\documentclass[preprint2]{proto}
\usepackage{times}
\usepackage{xcolor}
\usepackage{graphicx}
\usepackage{wasysym}
\usepackage{float}
\usepackage{dblfloatfix}
\usepackage{subfigure}
\usepackage{graphicx}
\usepackage{pdflscape}
\usepackage{afterpage}
\usepackage{capt-of}

\usepackage{sidecap}  

\graphicspath{{./}{Figures/}}

\newcommand{\mission}[1]{\textit{#1}}  
\newcommand{\DI}{\mission{Deep Impact}}
\newcommand{\Rosetta}{\mission{Rosetta}}
\newcommand{\CONTOUR}{\mission{CONTOUR}}
\newcommand{\Stardust}{\mission{Stardust}}
\newcommand{\EPOXI}{\mission{EPOXI}}
\newcommand{\StardustNExT}{\mission{Stardust-NExT}} 

\newcommand{\Philae}{\mission{Philae}}
\newcommand{\CI}{\mission{Comet Interceptor}}
\newcommand{\DSone}{\mission{Deep Space 1}}
\newcommand{\Giotto}{\mission{Giotto}}
\newcommand{\ZhengHe}{\mission{ZhengHe}}

\begin{document}

\title{\textbf{\LARGE Past and Future Comet Missions}}

\author {\textbf{\large C. Snodgrass}}
\affil{\small\em University of Edinburgh}

\author {\textbf{\large L. Feaga}}
\affil{\small\em University of Maryland}

\author {\textbf{\large G. H. Jones}}
\affil{\small\em Mullard Space Science Laboratory, University College London}

\author {\textbf{\large M. K\"uppers}}
\affil{\small\em European Space Agency}

\author {\textbf{\large C. Tubiana}}
\affil{\small\em Istituto di Astrofisica e Planetologia Spaziali}

\begin{abstract}

\begin{list}{ } {\rightmargin 1in}
\baselineskip = 11pt
\parindent=1pc
{\small 
We review the history of spacecraft encounters with comets, concentrating on those that took place in the recent past, since the publication of the \textit{Comets II} book. {This includes} the flyby missions \Stardust{} and \DI{}, and their respective extended missions,  the \Rosetta{} rendezvous mission, {and serendipitous encounters}. While results from all of these missions can be found throughout this book, this chapter focuses on the questions that motivated each mission, the technologies that were required to answer these questions, and where each mission opened new areas to investigate. There remain a large number of questions that will require future technologies and space missions to answer; we also describe planned next steps and routes forward that may be pursued by missions that have yet to be selected, and eventually lead to cryogenic sample return of nucleus ices for laboratory study.
\\~\\~\\~}
\end{list}
\end{abstract}  


\section{\textbf{INTRODUCTION}}
\label{sec:intro}

\begin{figure*}
    \centering
    \includegraphics[width=\textwidth]{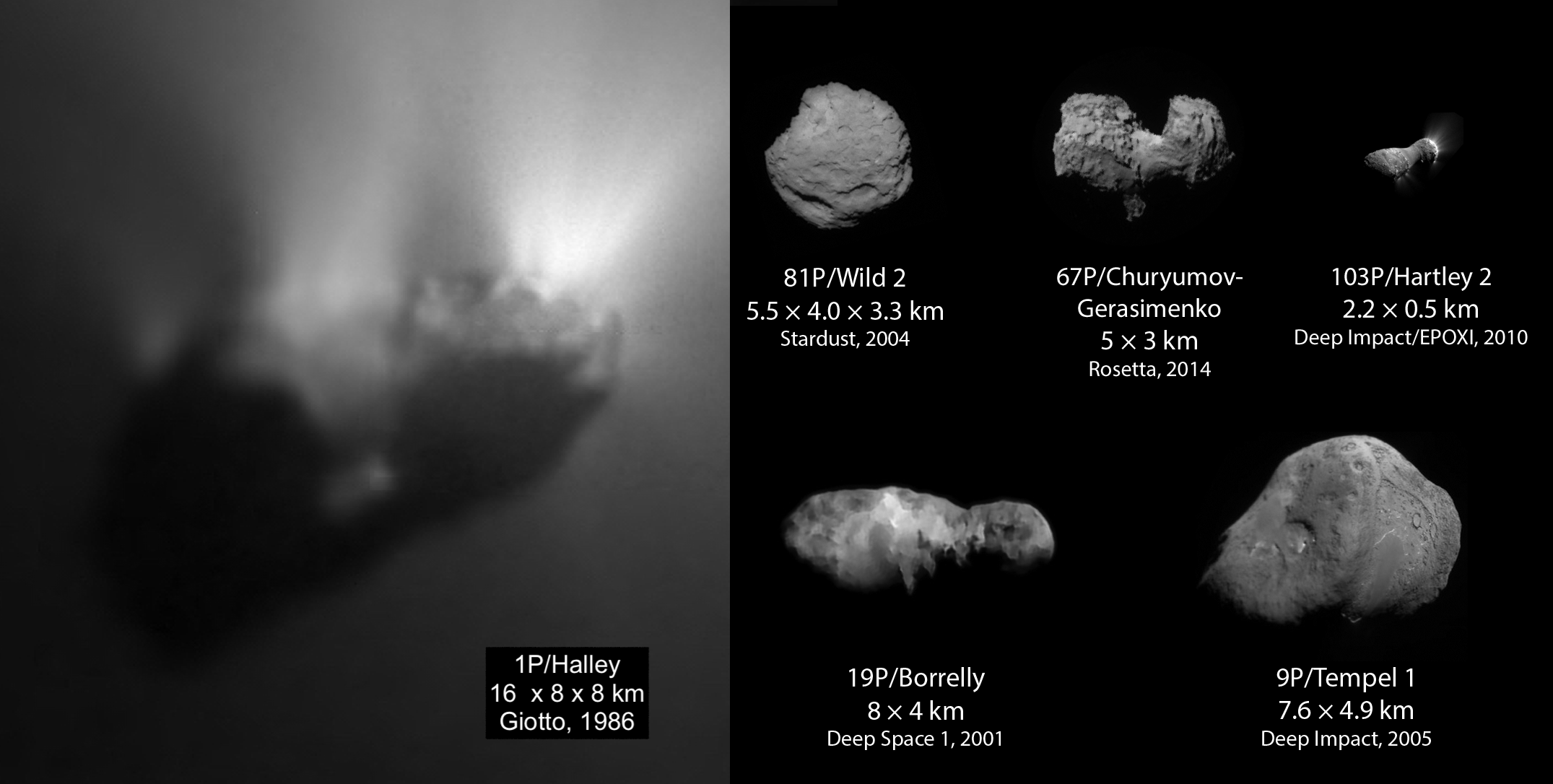}
    \caption{All comet nuclei imaged by spacecraft, at approximately the same spatial scale. Adapted from a graphic by E.~Lakdawalla, with permission. Credit for individual images: 1P -- ESA/MPS; 9P -- NASA/JPL/UMD; 19P -- NASA/JPL/Ted Stryk; 67P -- ESA/Rosetta/NavCam/Emily Lakdawalla; 81P -- NASA/JPL; 103P -- NASA/JPL/UMD.}
    \label{fig:nucleus_images}
\end{figure*}

In the period since the publication of \textit{Comets II} \citep{CometsIIbook} we have enjoyed a golden age of space exploration of comets. In that book, \citet{Keller-cometsII} reviewed results from the relatively low-resolution images returned by the missions to comet 1P/Halley (hereafter 1P) in the 1980s and \DSone's flyby of comet 19P/Borrelly (hereafter 19P). At the time, the next generation of missions had not yet returned any results, even though they were already on their journeys. \Stardust{}, \DI{}, and \Rosetta{} would completely change our view of comets, with more detailed imaging (Fig. \ref{fig:nucleus_images}), spectroscopy, and \textit{in situ} measurements, and {making} technological leaps forward (impacting or landing on the nucleus itself, and returning coma samples to Earth). This chapter reviews the motivations for each mission and the instruments and technologies that each required to answer their science questions{. Major results are highlighted from each mission}, although the detailed scientific results are described elsewhere in this book; nearly every chapter touches on results from space missions at some point, due to their significant impact on the field. 
{For more details on the results of the \DI{} / \EPOXI{} and \Stardust{} missions, see comprehensive reviews by \citet{AHearn+Johnson} and \citet{stardust-review}, respectively. \Rosetta{} results are both too extensive and too recent to have been captured in single reviews, but, alongside the many relevant chapters in this book, the reader could begin with the summary of key early results by \citet{Taylor-Rosetta}.}

\begin{table*}[t]
    \begin{tabular}{|l|c|c|c|c|c|c|c|c|c|}
   \hline
           Mission & Comet & \multicolumn{2}{c|}{Dates} & Agency$^*$ & Type$^{\P}$ & Mass$^{\diamond}$ & Speed & CA$^{\dag}$ & $r^{\ddag}$ \\
        &   & Launch & Encounter &  & & (kg)  & (kms$^{-1}$) & (km) & (au)  \\
        \hline
    \mission{ICE}       & 21P & 1978 Aug 12 & 1985 Sep 11  & NASA/ & F & 479 & 20.7  & 7862 & +1.03 \\
        & & &  & ESA &  & &  &  &  \\
    \mission{Vega 1}    & 1P  & 1984 Dec 15 & 1986 Mar 06  & SAS & F & 4920$^{\triangleleft}$ &79.2 & 8889 & +0.79  \\
    \mission{Vega 2}    & 1P  & 1984 Dec 21 & 1986 Mar 09  & SAS & F & 4920$^{\triangleleft}$ & 76.8 & 8030 & +0.83  \\
    \mission{Sakigake}  & 1P  & 1985 Jan 07 & 1986 Mar 11  & ISAS & F & 138 & 75.3 & $7\times10^6$ & +0.86  \\
    \Giotto             & 1P  & 1985 Jul 02 & 1986 Mar 14  & ESA & F & 583 & 68.4 & 605 & +0.90  \\
    \emph{GEM}             & 26P &      & 1992 Jul 10  & ESA & F & & 14.0 & $<$200 & -1.01 \\
    \mission{Suisei}    & 1P  & 1985 Aug 19 & 1986 Mar 08  & ISAS & F & 139 & 73.0 & 152400 & +0.82  \\
    \DSone              & 19P & 1998 Oct 24 & 2001 Sep 22  & NASA & F & 373 & 16.6 & 2171 & +1.36  \\
    \Stardust           & 81P & 1999 Feb 07 & 2004 Jan 02 & NASA & CSR & 300 & 6.1 & 237 & +1.86  \\
    \StardustNExT       & 9P  &     & 2011 Feb 15 & NASA & F & & 10.9 & 181 & +1.55  \\
    \CONTOUR            & --  & 2002 Jul 03 & -- & NASA & MF & 328 & -- & -- & -- \\
    \Rosetta            & 67P & 2004 Mar 02 & 2014 Aug 06 & ESA/ & R & 1230 & 0 & 0 & $^{\S}$  \\
    & & & & NASA & & & & & \\
    \Philae             & 67P & 2004 Mar 02 & 2014 Nov 12 & ESA & L & 100 & 0 & 0 & -2.99 \\     
    \DI                 & 9P  & 2005 Jan 12 & 2005 Jul 04 & NASA & F+I & 879 & 10.2 & 575 & -1.51 \\
    \EPOXI              & 103P&     & 2010 Nov 04 & NASA & F & & 12.3 & 694 & +1.06  \\
          \hline
    \end{tabular}
    \smallskip\\
    Notes:
    $^{*}$Agencies: NASA -- National Aeronautics and Space Administration (USA); ESA -- European Space Agency; SAS -- Soviet Academy of Sciences (USSR); ISAS -- Institute of Space and Astronautical Science (Japan);  
    $^{\P}$Mission type: F -- Flyby; CSR -- Coma sample return; MF -- Multiple flybys; R -- Rendezvous; L -- Lander; I -- Impactor; $^{\diamond}$Dry mass in kg at launch;
    $^{\dag}$Closest approach distance to the nucleus; 
    $^{\ddag}$Heliocentric distance at encounter (negative indicates pre-perihelion, positive post-perihelion);
    $^{\S}$\Rosetta's extended operations are detailed in Table \ref{tab:rosetta-timeline}. 
    $^{\triangleleft}$Includes Venus lander \& balloon. 
    \caption{Summary of all past 
    comet missions. See Fig. \ref{fig:tailcrossings}, sections \ref{sec:other-flybys} and \ref{sec:mars} for unplanned encounters and tail crossings. }
    \label{tab:all_missions}
\end{table*}

All comet missions share some common goals: cross-cutting themes that motivated past missions and continue to be key questions today, as each subsequent mission has shed some light on the topic but also opened new avenues to explore. In very broad terms, these motivations can be described as wanting to understand: 
\begin{enumerate}
    \item the formation and evolution of comets;
    \item the diversity of material in the Solar System and how it has been distributed and transported;
    \item how the physical processes that drive cometary activity work and influence the first two topics.
\end{enumerate}  
Individual missions also explore specific and focused questions, but at their heart all are built around these topics. Reviews of our current understanding of the three themes can be found in the chapters by 
Simon et al., Guilbert-Lepoutre et al., Pajola et al., Filacchione et al., Knight et al., Marschall et al., Biver et al., Engrand et al., Agarwal et al., and Marty et al. 
elsewhere in this book. Missions have made significant contributions to the first topic by returning resolved nucleus images of increasing resolution, revealing morphological features, and by providing clues to interior structure (density, porosity) through indirect (\DI; {\citealt{Ahearn_DI_Science}}) or direct (\Rosetta; {\citealt{Kofman2020}}) measurements. {Regarding the second topic,} sampling of cometary dust and gas by \textit{in situ} mass spectroscopy by \Giotto{} and especially \Rosetta{} has revealed a `zoo’ of chemicals that cannot be detected by remote observation \citep{Rubin2020-SSR}, while the return of a coma dust sample by \Stardust{} has enabled incredibly detailed measurements of composition using laboratory equipment, and revealed large-scale transport of material {in the protoplanetary disk} \citep{Brownlee2006}.  Finally, the question of how cometary activity works remains enigmatic, despite successive missions identifying various sources with increasing precision, from  discrete areas (\Giotto: \citealt{Keller1986}) to smooth flows and retreating scarps (\DI: \citealt{Thomas_DI_shape}). \DI{} and \EPOXI{} were able to associate dust and ice outflows with gas species  \citep{Ahearn_Hartley_Science}, and finally \Rosetta{} followed the changing patterns of activity with diurnal and seasonal cycles \citep{DeSanctis_67PVIRTISH2O,Fornasier2016}. The approved future missions will further advance our understanding of all these topics by visiting comets with potentially {more diverse regions of} origin, and at different evolutionary states. Key technological steps required to make major further progress in each topic include, respectively, {using ground-penetrating radar} to understand the interior structure of nuclei; (cryogenic) sample return of nucleus material; and \textit{in situ} investigation of the (micro-)physics of the near surface layers.

In this chapter we begin, in section \ref{sec:ground-based}, by highlighting the synergy between missions and supporting observations. Our review of past missions includes descriptions of all those with a recognized comet as a named target, {covering flyby missions in section \ref{sec:flyby} and \Rosetta{} in more detail in section \ref{sec:rosetta}}. 
We list all these missions in Table \ref{tab:all_missions}, and summarise their key goals, technologies and results in Table \ref{tab:goals_tech_results}. It is worth noting that the most important results of these exploration missions are often the unexpected ones, and not necessarily the ones described in the initial mission goals (\citealt{AHearn+Johnson} call this the `Harwit principle').

We include a first review of distant \textit{in situ} encounters by spacecraft with comets and their tails, whether targeted or not, in section \ref{sec:other-flybys}, and briefly summarise the serendipitous observations of the close approach of comet C/2013 A1 Siding Spring to Mars by the various probes in orbit (or on the surface) of that planet in section \ref{sec:mars}, but do not include remote observations of comets by space telescopes (these are described elsewhere in this book by Bauer et al.). 
We also look ahead, in section \ref{sec:future}, to mission concepts under development: the Japanese \mission{DESTINY$^{+}$} mission to Phaethon, the Chinese mission \ZhengHe, which is expected to visit an active asteroid; and the European Space Agency (ESA) mission \CI{}, whose objective is to visit an Oort cloud comet. We also suggest future possibilities of where and how we will continue to explore comets, in missions yet to be proposed or selected. 


\section{GROUND-BASED CAMPAIGNS}
\label{sec:ground-based}

Most missions are accompanied by significant international campaigns of supporting observations from ground-based (or remote space-based) telescopes. These serve multiple purposes: mission planning, providing the most complete and up-to-date reconnaissance on the comet before the spacecraft arrives; context, allowing comparison of spacecraft results to the comet on different spatial and time scales; and relation of the target comet to the wider population. The latter is particularly important as the number of comets visited by spacecraft will always be small compared to the number observed with telescopes, and it is necessary to identify what features of a particular comet are unique, versus characteristic of a class or comets in general.  

The scale of various ground-based campaigns varied: The \mission{International Halley Watch} that accompanied the 1980s missions defined, purchased, and distributed narrowband filters to isolate gas species in the coma that would be widely used for comet photometry for a decade \citep{Schleicher+Farnham2004}. The \DI{} observing campaign \citep{Meech2005} was an integral part of the mission, relied on to characterize the evolution of the ejecta plume after the spacecraft flyby, and as such is described in more detail in section \ref{sec:deepimpact}. The \Rosetta{} observing campaign continued throughout the 2.5 years that the spacecraft operated at the comet \citep{Rosetta-campaign}. 

A common goal of mission-supporting observation campaigns was monitoring  
the total brightness of the comet as a proxy for its activity levels -- typically of dust brightness (or $Af\rho$; {\citealt{Ahearn1984}}) via broadband imaging, but also including spectroscopy or narrowband photometry to constrain gas production rates when the comets were bright enough. One key result from the \Rosetta{} campaign was confirmation that the water production rates measured over months at the comet correlated well with the ground-based measurements of total dust brightness \citep{KC-waterproduction}. {This resulted in increased} confidence in models that derive gas production from broadband photometry \citep[e.g.,][]{Meech-cometsII}, which were used to estimate long-term evolution of gas production rates around previous flybys \citep[e.g., \EPOXI;][]{Meech-EPOXI}. 

Intensive monitoring of 9P/Tempel 1 (hereafter 9P) ahead of the \DI{} encounter revealed regular `mini-outbursts', although comparison between ground-based data and images taken by the spacecraft on approach showed many of these events were too small to affect the large scale brightness of the comet \citep{Lara2006,Ahearn_DI_Science}. Similarly, the small-scale outbursts that punctuate 67P/Churyumov-Gerasimenko's (hereafter 67P) activity were seen by \Rosetta{} but not the accompanying ground-based observations \citep{JB-summer-fireworks,Rosetta-campaign}. The long-term monitoring before and after spacecraft flybys possible from the ground was essential to understanding the rotation pole and period. The timing of the \StardustNExT{} encounter with respect to rotational phasing and illumination of 9P was critical to meet mission objectives (see section \ref{sec:stardust-next}; \citealt{Veverka2013,Belton2011}). {By monitoring} 103P/Hartley 2 (hereafter 103P), significant changes in rotation rate through the encounter perihelion passage {were revealed} \citep{103P-rotation}. For 67P, ground-based measurements of the rotation rate together with Rosetta approach data revealed long term spin-up, with the rotation period reducing by $\sim$20 minutes per orbit, over a timescale much longer than the \Rosetta{} mission \citep{Mottola67Pspinup}. Combining these results, estimates of the non-gravitational acceleration of the comet from ground-based astrometry and orbit fitting, and the detailed shape model from the spacecraft data, has enabled quite detailed modeling of the forces on 67P from outgassing \citep{Attree2019,Kramer2019}. {Furthermore, findings from \Rosetta{} confirmed that ground-based predictions of the pole position of 67P, via three independent methods, were correct \citep{67Ppole1,67Ppole2,67Ppole3}. The resulting seasonal effects were used to explain the temporal differences in CN gas production rates measured from ground-based spectroscopy \citep{Opitom67Pspec}. However, \Rosetta{} results also demonstrated the limitations of techniques for estimating nuclei shapes from ground-based observation (see section \ref{sec:rosetta-tech} and figure \ref{fig:67P_shape} for details).}

In addition to complementing the temporal context around missions, ground-based observations are key to understanding the wider spatial scale of comets. Apart from the more distant encounters described in section \ref{sec:other-flybys}, missions target the near-nucleus region, while the coma and tails extend over a region orders of magnitude larger. A prime example of this was on display at 103P where, prior to the \EPOXI{} mission, the comet was classified as  hyperactive  
\citep{Groussin_103P_hyper}{, implying $\sim$100\% of its nucleus surface was active compared to a more typical few percent level}.  Earth-based observations during the 2010 apparition indicated extended and asymmetric distributions of secondary species (e.g., OH and H) enhanced in the anti-sunward direction \citep{Meech-EPOXI}. Spacecraft visible and spectral images  revealed that the true source of the hyperactivity and extended water coma was an icy grain source that was being actively expelled from the nucleus during the encounter \citep{Ahearn_Hartley_Science,Protopapa_H2_ice,Sunshine_Feaga_hyper}. The color, polarization, and radial brightness profile of the dust coma of 67P indicated evolution of particle properties as dust moved outward, interpreted as fragmentation of these particles at distances beyond where \Rosetta{} was measuring their properties \textit{in situ} \citep{Boehnhardt2016,Rosenbush2017}. A combination of these large-scale observations and inner-coma spacecraft measurements is required to build a full model of how the cometary coma develops from the nucleus \citep{Marschall2020}. On the largest scales, cometary tails reveal both comet dust and gas properties and how their induced magnetospheres interact with the solar wind (see chapter by Goetz et al. in this volume). The interpretation of remote ion tail observations combined with spacecraft measurements of magnetic fields would be particularly illuminating. Unfortunately, the majority of spacecraft targets to date (low-inclination Jupiter family comets; JFCs) do not typically present strong ion tails with a favorable geometry; for example, any ion tail from 67P would have been projected behind the comet's dust coma as seen from Earth, and was not definitively detected. 
With favorable geometry, {dust and ion} tails can be well characterized by amateur astronomer observations, typically made with smaller telescopes with wide fields of view. 
There have been significant amateur contributions to mission-supporting observation campaigns, which, in addition to their  direct scientific value, have been hugely successful in engaging a wider community and the general public in cometary exploration \citep{McFadden2005,Usher2020}.

Finally, ground-based observations are key to understanding {how spacecraft targets relate} to the wider comet population.  Establishing a firm taxonomy of comets, and {linking} observed differences in composition to either formation or evolutionary processes, is an ongoing endeavour (see the chapter by Biver et al. in this volume), but observations have been able to broadly group comets into `typical' or `carbon-chain-depleted' classes \citep{AHearn1995}. Ground-based observations show that three of the eight comets visited by missions to date (Table \ref{tab:all_missions}) fall into the typical range (1P, 9P, and 103P), with the others counting as depleted, as would be expected when we consider that most have a source region in the Kuiper belt, and the majority of comets from there were seen to be carbon-chain depleted in the original taxonomy \citep{AHearn1995}. {More recently, a compilation of infrared spectroscopic data shows that both dynamical classes of comets are represented throughout the spread of compositional abundances, however, there is an overall emerging trend that JFCs are depleted in volatile species with respect to water as compared to Oort cloud comets \citep{DelloRusso2016}.}  Smaller nuclei are often seen to be `hyperactive', another often-used classification of comets, including 103P and the original \Rosetta{} target, 46P/Wirtanen (hereafter 46P). Of the spacecraft target comets,  21P/Giacobini-Zinner (hereafter 21P) and 103P have shown some evidence of hyperactivity (active fraction $>$ 50\%) in remote observations with the SOHO/SWAN instrument \citep{Combi2019}. Recent work suggests a continuum of activity levels \citep{Sunshine_Feaga_hyper} {and that the degree of hyperactivity may correlate with the D/H isotopic ratio \citep{Lis2019}.} 


\afterpage{%
    \thispagestyle{empty}
    \begin{landscape}
    \hspace{-0.5cm}
    \begin{tabular}{| p{1.2cm} | p{5.8cm} | p{4cm}  | p{10.2cm} | p{1cm} |}
   \hline
     Mission & 
        Goals & 
        Technologies & 
        Results &
        Ch.$^{*}$\\
        \hline
    \Stardust           & 
    	Return $\sim$1 mg of coma dust sample to Earth; measure dust flux; image nucleus and inner coma & 
	    Aerogel dust capture; sample return (high-speed capsule, sample recovery, handling, curation); laboratory investigations & 
	    Highly pitted nucleus; fragmentation of coma particles; solid and fragile particles; composition of dust (mineralogy, petrology, atomic and isotopic composition of components), including high temperature minerals  &
	    1, 9, 10, 18\\
    \DI                  & 
    	Excavate a $\sim$100 m diameter x $\sim$25~m depth crater; study interior composition, layering, heterogeneities; determine porosity,  density, and strength of the nucleus; identify vertical/lateral structure; characterize nucleus and coma environment prior to impact & 
	    Releasable sub-spacecraft; autonomous navigation and targeting & 
	    Frequent mini-outbursts; morphological diversity, flows and layers, but overall homogeneous nucleus color and albedo; low strength, density and thermal inertia; limited surface ice, possible frosts; compositional homogeneity with depth, but variation across surface (differing source regions of H$_2$O and CO$_2$); localized activity; presence of small subsurface H$_2$O ice grains, a hot ($\sim 1,000$ K) ejecta plume immediately following impact  &
	    9, 10, 11, 14 \\
    \EPOXI               & 
    	Explore diversity of comets; {determine the origin and significance of hyperactivity} & 
	    Repurposed spacecraft; test of the deep space communications network & 
	    Bilobed morphology; rough terrain on lobes, a smooth `waist', and tall spires; dm-scale chunks of icy material in coma on bound orbits; {spatially and temporally resolved  distribution of gasses;} $\mu$m-sized ice associated with CO$_2$ driven activity providing explanation for `hyperactivity' &
	    10, 11, 14, 20\\
    \StardustNExT        & 
    	Image \DI{} impact site; search for surface changes over one orbit; increase observed surface area of 9P& 
	    Repurposed spacecraft; precise prediction of rotational phase; targeting with very little fuel; recalibration of flying camera & 
	    Nucleus layering; localized erosion; scarp retreat by tens of meters; identification of crater and confirmation of low density and strength nucleus  &
	    10 \\
    \Rosetta             & 
    	Study changes in nucleus, gas, dust, and plasma around orbit; understand activity processes; provide detailed composition and structure measurements at unprecedented resolution & 
	    Rendezvous (inc. spacecraft hibernation, long-term operations at an active comet); lander targeting and release; extensive \& novel payload (see Table \ref{tab:rosetta-instruments} and section \ref{sec:rosetta-instruments}) & 
	    Complex and varied nucleus morphology, with clues on interior structure; mass, volume, density and porosity of the nucleus; surface variation with time; frosts, ice revealed by cliff collapse; significant fallback of dust; detailed inventory of minor gas species, including complex organics and isotopic measurements; variation of activity and coma composition with time and location; strong seasonal effects; physical structure of dust particles; importance of electron impact dissociation; better understanding of solar wind interactions at low/ intermediate activity 	    &
	    1, 9, 10, 11, 13, 14, 15, 16, 17, 18, 20\\    
    \Philae              & 
    	Investigate nucleus surface composition, mechanical, electrical and physical properties; interior structure via sounding  & 
	    Landing equipment for unknown surface; highly miniaturized instrumentation & 
	    Surface images and dust composition measurements; (debated) local surface strength assessment &
	    9, 11, 17\\    
          \hline
    \end{tabular}
    \smallskip\\
    Notes:
    $^{*}$ Relevant chapters elsewhere in this book for more details on the scientific results of each mission: 
    1: Bergin et al.;
    9: Guilbert-Lepoutre et al.;
    10: Pajola et al.;
    11: Filacchione et al.;
    13: Bodewits et al.; 
    14: Marschall et al.;
    15: Biver et al.;
    16: Beth et al.;
    17: Goetz et al.;
    18: Engrand et al.; 
    20: Agarwal et al.
       \captionof{table}{Summary of key science goals, enabling technologies, and results for the missions described in detail in sections \ref{sec:flyby} and \ref{sec:rosetta}. Here references are given to relevant chapters elsewhere in this book that discuss the results in more detail; see the relevant sections in the text for additional information and references.}
    \label{tab:goals_tech_results}
    \end{landscape}
}


\section{\textbf{FLYBY MISSIONS}}
\label{sec:flyby}

\subsection{Early missions (pre-{\it Comets II})}
\label{sec:halley-missions}

{The first targeted cometary spacecraft encounter was an ion tail crossing, performed by the NASA/ESA \mission{ISEE-3} spacecraft, which had been monitoring the solar wind upstream of Earth since 1978. The spacecraft was renamed as the \mission{International Cometary Explorer} (\mission{ICE}), and repurposed to encounter comet 21P, which it achieved by crossing the comet's ion tail on September 11, 1985, as planned, 7,800~km downstream. \mission{ICE}'s payload comprised instruments designed to study the solar wind only and had no cameras. The data returned were however extremely valuable: they confirmed the long-expected pattern of draped heliospheric magnetic field lines \citep{alfven57}, forming an induced magnetotail $\sim$10,000~km wide. Outside this region lay the ionosheath (or cometosheath) of weaker, loosely-draped magnetic fields; its outer boundary marked by a bow wave/shock \citep{smith1986}. \mission{ICE} returned to Earth's vicinity in 2014, but attempts to adjust its trajectory were largely unsuccessful and any further targeted comet encounters are not possible \citep{Dunham2015}. }

The first missions dedicated to cometary science were the spacecraft encountering comet 1P, the most famous comet, during its perihelion passage in 1986.
It was an opportunity to study the nucleus and the innermost coma, point sources in Earth-based observations, in spatial detail. A total of 5 spacecraft visited 1P within $\sim$2 weeks, see Table \ref{tab:all_missions}. 
Due to 1P's retrograde orbit, all of the spacecraft flew by the comet at very high velocities of 68--79~km~s$^{-1}$. 
While \mission{ICE} and the two Japanese missions \mission{Suisei} and \mission{Sakigake} measured the plasma environment of 1P, the two \mission{Vega} missions and \Giotto{} {also} provided the first detailed investigation of a comet nucleus, with most detail of the nucleus surface provided by \Giotto . 
A composite  of 68 images taken during the approach to 1P is shown in Fig.~\ref{fig:nucleus_images}.

The missions to 1P and their results are described in detail in {\it Comets II} and will not be repeated here. However, after the rendezvous of \Rosetta{} with comet 67P, several authors revisited the 1P data. Notably, the detection of molecular oxygen in comet 67P prompted a reanalysis of mass spectrometer data on \Giotto{}. If 3.7\,$\pm$ 1.7\,\% of O$_2$ relative to water is included, similar to the abundance detected in 67P \citep{Rubin2015}, the fit to the data is improved. This suggests that 67P is not special in terms of O$_2$ abundance and that molecular oxygen is a parent molecule with a typical abundance of a few percent relative to water in  comets. Furthermore, plasma measurements from the \Giotto{} and \mission{Vega} missions were revisited to compare with \Rosetta{} results on ion composition \citep{Heritier2017} and plasma processes and boundaries 
(see the chapter by Goetz et al. in this volume, and references therein). The re-use of \Giotto{} data 30+ years after the mission illustrates the value of long-term archiving of scientific data.      

To change trajectories in order to intercept comet 26P/Grigg-Skjellerup (hereafter 26P) on July 10, 1992, \emph{Giotto} performed the first Earth flyby by a planetary spacecraft in July 1990. The spin-stabilized probe’s high-gain antenna was oriented specifically for the Halley encounter; therefore to maintain real-time communication with Earth, the spacecraft had to encounter 26P without full protection by its dust impact shield, but did so successfully \citep{GrensSch1993}.
Despite not carrying an operational camera, much was learned about the comet's coma structure \citep{ACLR1993}, the possible existence of nucleus fragments  \citep{McBride1997}, and the solar wind interaction of this relatively low-activity object \citep[e.g.,][]{CoatesJones2009}. \Giotto{} operations ceased on July 23, 1992, with insufficient fuel remaining for a further encounter \citep{GrensSch1993}.

\DSone{} was a NASA technology demonstration mission. Its extended mission included flybys of asteroid (9969) Braille and comet 19P \citep{Rayman2001, Rayman2002}, which provided the first opportunity to identify surface morphological structures  in sufficient detail to investigate a comet from a geological point of view \citep{Britt2004}.
\DSone{} paved the way for future comet and asteroid missions by pioneering new technologies \citep{Rayman2000}. In particular, solar electric propulsion, autonomous navigation (including autonomous spacecraft maneuver execution), and miniaturized scientific instruments are now used by several spacecraft.  


\subsection{CONTOUR}
\label{sec:contour}
{The \textit{COmet Nucleus TOUR} (\CONTOUR{}) was a cometary flyby mission selected under the NASA Discovery program led by Principal Investigator (PI) Joseph Veverka. It successfully launched on July 3, 2002, but suffered 
a catastrophic failure on August 15, 2002, during the solid rocket burn that would have propelled the spacecraft out of Earth’s orbit and on its trajectory to its first target \citep{CONTOUR-report}. The mission’s primary objective was to complete close flybys of two active cometary nuclei around 1 au from the Sun: 2P/Encke, a highly thermally evolved comet, and 73P/Schwassmann-Wachmann 3 (hereafter 73P), a recently fragmented comet \citep{Cochran2002}.} 

{\CONTOUR{}'s design was risk averse with a suite of heritage instruments (a visible and infrared imaging spectrograph, a visible imager, a mass spectrometer, and a dust analyzer) and a simple and compact spacecraft design with few movable parts \citep{Cochran2002}. \CONTOUR{} would have been one of the first missions to use only a tone to alert Earth that the spacecraft was still functioning during long periods of minimal power hibernation ({\it Cochran, A., 2021, private communication}). 
The mission would have doubled the number of cometary nuclei imaged by spacecraft at the time, advanced our scientific knowledge of the characteristics of comet nuclei, and most importantly explored the diversity and evolution of comets \citep{Cochran2002}. The most remarkable aspect of the \CONTOUR{} mission was the flexibility in its trajectory design, implemented with many Earth flybys, commonly used now, to allow retargeting maneuvers as desired \citep{Cochran2002}. This enabled options for multiple comet flybys, or even more scientifically compelling, rerouting the prime mission to a newly discovered long period comet (LPC) from the Oort cloud, a class of comet then yet to be imaged by spacecraft.}


\subsection{Stardust}
\label{sec:stardust}
\Stardust{}, a NASA Discovery mission led by PI Donald Brownlee, is the only mission to date to return cometary dust of known origin to Earth for detailed analysis. {The mission objectives included quantifying the nature and amount of dust released by the JFC 81P/Wild 2 (hereafter 81P), and relating cometary dust to meteoritic samples collected on Earth and interstellar dust particles also collected by \Stardust{} \citep{Stardust_Brownlee_JGR}. A major component in the success of this mission was the comprehensive state of the art analysis that was conducted in the lab once the $\sim$1 mg total sample was returned to Earth. These detailed measurements would not have been feasible \textit{in situ}. See \citet{stardust-review} for a detailed recent review of the mission results.} 

The biggest challenge of this mission was to capture a sample of dust particles without destroying them in the processes nor harming the spacecraft. This was achieved thanks to the design of a spacecraft trajectory that produced a low encounter relative velocity of 6.1 km~s$^{-1}$ and the use of an aerogel capture medium \citep{Stardust_Brownlee_JGR}. Aerogel is a porous silica foam with density comparable to that of air: its low density gradually slowed the impacting particles, avoiding melting and/or vaporization. 
{\Stardust{} was equipped with a $\sim40$ cm diameter tray of aerogel cells that was held outside the  spacecraft, and exposed to the dust flux, while a Whipple bumper protected the main body of the spacecraft \citep{Stardust_Brownlee_JGR}. Following the encounter, the whole tray was  placed within a return capsule.}

\begin{figure}[t]
\begin{center}
\includegraphics[angle=90,width=\columnwidth]{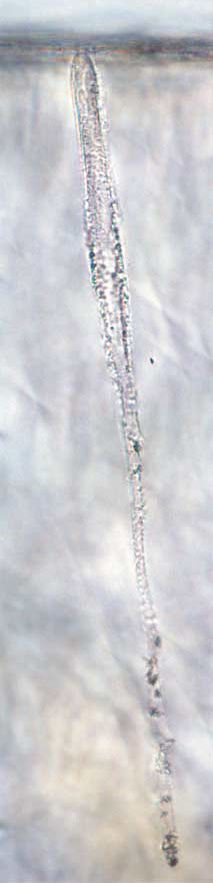}
\caption{Track created in the aerogel by a dust particle. Source: \citet{Burnett2006}}
\label{fig:81P_track_stardust}
\end{center}
\end{figure}

On January 2, 2004, the \Stardust{} spacecraft flew by 81P at a distance of 234 km from the surface \citep{Brownlee2004} and collected more than 10,000 dust particles in the size range between 1 and 300 $\mu$m, which are expected to be a representative sample of the non-volatile component of the interior of the comet \citep{Brownlee2006}. 
Particles impacting on the aerogel produced millimeter-sized tracks (Fig. \ref{fig:81P_track_stardust}); most of the material of the incoming particle was left as fragments on the wall of the track, but usually an intact terminal particle was found at the end of the track \citep{Burnett2006}. The shape of the deep tracks in the silica aerogel varied depending on the nature of the impacting particle. In addition, the aluminum frame used to hold the aerogel tiles showed bowl-shaped impact craters containing residues of the incoming particles \citep{Grahametal2006}.  

\begin{figure}[t]
\begin{center}
\includegraphics[width=\columnwidth]{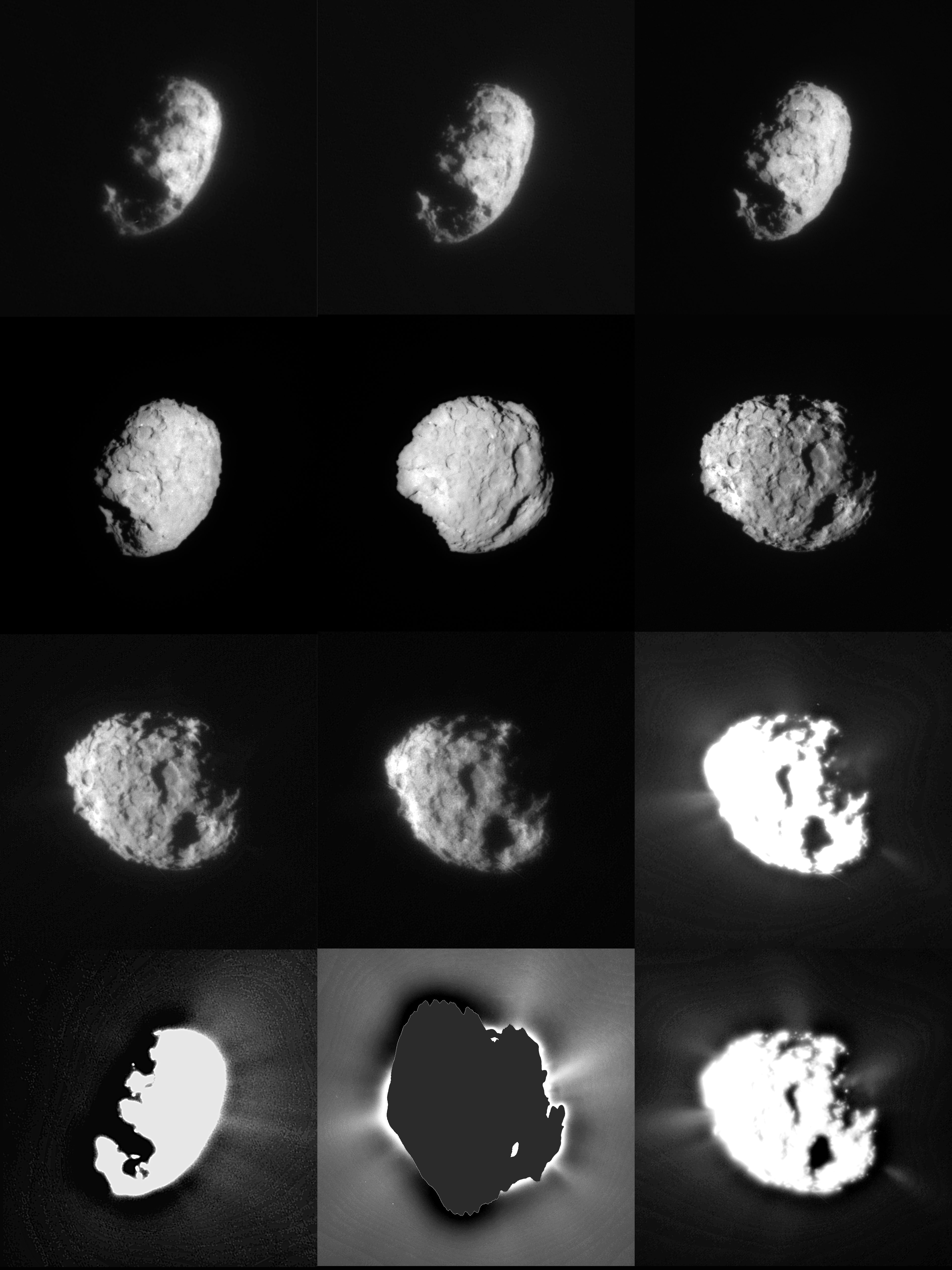}
\caption{{Images of the nucleus of comet 81P/Wild 2, acquired by the \Stardust{} navigation camera during the comet flyby. Source: NASA/JPL/Stardust Team}}
\label{fig:81P_collage}
\end{center}
\end{figure}

Another obstacle for sample return missions is to survive Earth atmospheric re-entry and the environment of the Utah Test and Training Range (UTTR), or other such landscape, until recovery. \Stardust{}'s \textit{Sample Return Capsule} withstood peak temperatures from aerodynamic friction near 3,000 K during the fastest Earth entry of a manmade object in history, while a mission requirement was levied to keep the temperature of the sample canister inside the capsule below 343 K until retrieval from the UTTR \citep{Stardust_Brownlee_JGR}. \Stardust{} employed a heat shield constructed from a lightweight ceramic ablating material to meet these necessary conditions and to stay within mission mass and cost constraints. The 0.05 m thick phenol impregnated carbon ablator, PICA, was an enabling technology with its first flight application on \Stardust{} \citep{Stardust_heatshield}.

The benefit of bringing samples back to Earth is to characterize and quantify small amounts of even smaller particles with instrumentation too large and expensive to send into space \citep{Zolenskyetal2000}. The \Stardust{} samples underwent a wide range of analysis including, but not limited to, laser mass spectrometry, liquid chromatography, time of flight mass spectrometry, scanning transmission x-ray microscopy,  infrared and Raman spectroscopy, ion chromatography, secondary ion mass spectrometry, and scanning and transmission electron beam microscopy \citep{Sandfordetal2006}. A key implication afforded by the detailed analysis of the returned 81P samples is the importance of large-scale mixing in the early Solar System, based on the direct evidence for high temperature phase materials like chondrules accreted within a cold icy comet \citep{Brownlee2006}. Other significant results include the detection of glycine and other amino acids, the presence of amorphous and crystalline silicates, the absence of hydrous material, a heterogeneous organic distribution similar to interplanetary dust particles, and an isotopic makeup suggestive of interstellar origin \citep{Glavinetal2008, Elsilaetal2009, Kelleretal2006, Zolenskyetal2006, Sandfordetal2006}. 
More details on the \Stardust{} dust composition results  are given in the chapter by Engrand et al. in this volume.

Additionally, while particles were collected during the flyby, the suite of onboard instruments also acquired data. 
Throughout the closest approach to 81P, the spacecraft maintained a constant attitude relative to the ram direction, to enable dust collection and to protect the spacecraft behind dust shields, while the navigation camera \mission{NAVCAM} tracked the comet via a rotating mirror, which was itself protected by a `periscope' of two 45$\degr$ mirrors to look forward around the shield \citep{Newburn2003}. \Stardust{} acquired 72 close up images of the nucleus (Fig. \ref{fig:81P_collage}) that allowed for the determination of the size and shape of the comet as well as characterization of its surface and near surface activity. \Stardust{} observed an oblate nucleus, with a landscape of numerous flat-floored depressions and rounded central pits (\citealt{Duxbury2004,Brownlee2004,Sunshine_comet_processes}; see also chapter by Pajola et al. in this volume) and active jets of material emanating from it. 
The range of viewing geometries enabled by the rotating mirror approach, and the significant improvement in spatial resolution of the nucleus imaging compared to previous missions, allowed the source regions of activity (the pits) to be unambiguously identified and resolved for the first time.
Although  planned, multi-color imaging was not possible as the camera filter wheel stuck in the white-light position before arrival at the comet. 81P remains distinct from other nuclei imaged by spacecraft in appearing to be much more heavily pitted, and is relatively unusual in being a single comparatively  round object, rather than bi-lobed.

The \Stardust{} spacecraft also had two {\it in situ} dust experiments onboard, the \mission{Dust Flux Monitoring Instrument} (\mission{DFMI}) that measured the dust particle flux and size distribution \citep{Tuzzolino2003} and the \mission{Comet and Interstellar Dust Analyzer} (\mission{CIDA}) that measured dust grain composition \citep{Kissel2003} during various phases of the mission. {Measurements from \mission{CIDA} unambiguously identified organic compounds within the dust {\citep{Kissel2004}}. The \mission{DFMI} measurements of non-uniform bursts of dust hits, with few impacts between, indicate that the nucleus of 81P releases large aggregate dust clumps that subsequently disintegrate once in the coma \citep{Tsouetal2004}. 
\mission{DFMI} had two different sensor subsystems:  thin films of polyvinylidene fluoride sensitive to impacts by small particles, and acoustic detectors mounted on the dust shield to measure larger impacts.
By carefully calibrating these, \citet{Green2004} showed that \Stardust{} was much more sensitive than the dust detectors on \Giotto, and that such dust fragmentation could be an important effect in all comets that had previously been missed; {it would subsequently be seen again by \StardustNExT{} and \Rosetta{} (see sections \ref{sec:stardust-next} and \ref{sec:rosetta}).}} 

{Although the aerogel capture was highly successful, it} modified all collected particles to some degree. Particles larger than 1 $\mu$m were generally well preserved due to their higher thermal inertia. Sub-micron dust particles, instead, were strongly modified and survived only when shielded by a larger particle \citep{Brownlee2006}. A good understanding of this alteration process is important for understanding the properties of the cometary dust. Many laboratory calibration efforts were carried out to determine impactor properties from track characteristics (e.g., \citealt{Kearsley2012}). This nonetheless motivates the need for future sample collection with little to no particle alteration, {i.e., at essentially zero relative velocity to the comet, which can only realistically be achieved with a rendezvous mission}. 


\subsection{Deep Impact}
\label{sec:deepimpact}
\begin{figure}[t]
\begin{center}
\includegraphics[width=\columnwidth]{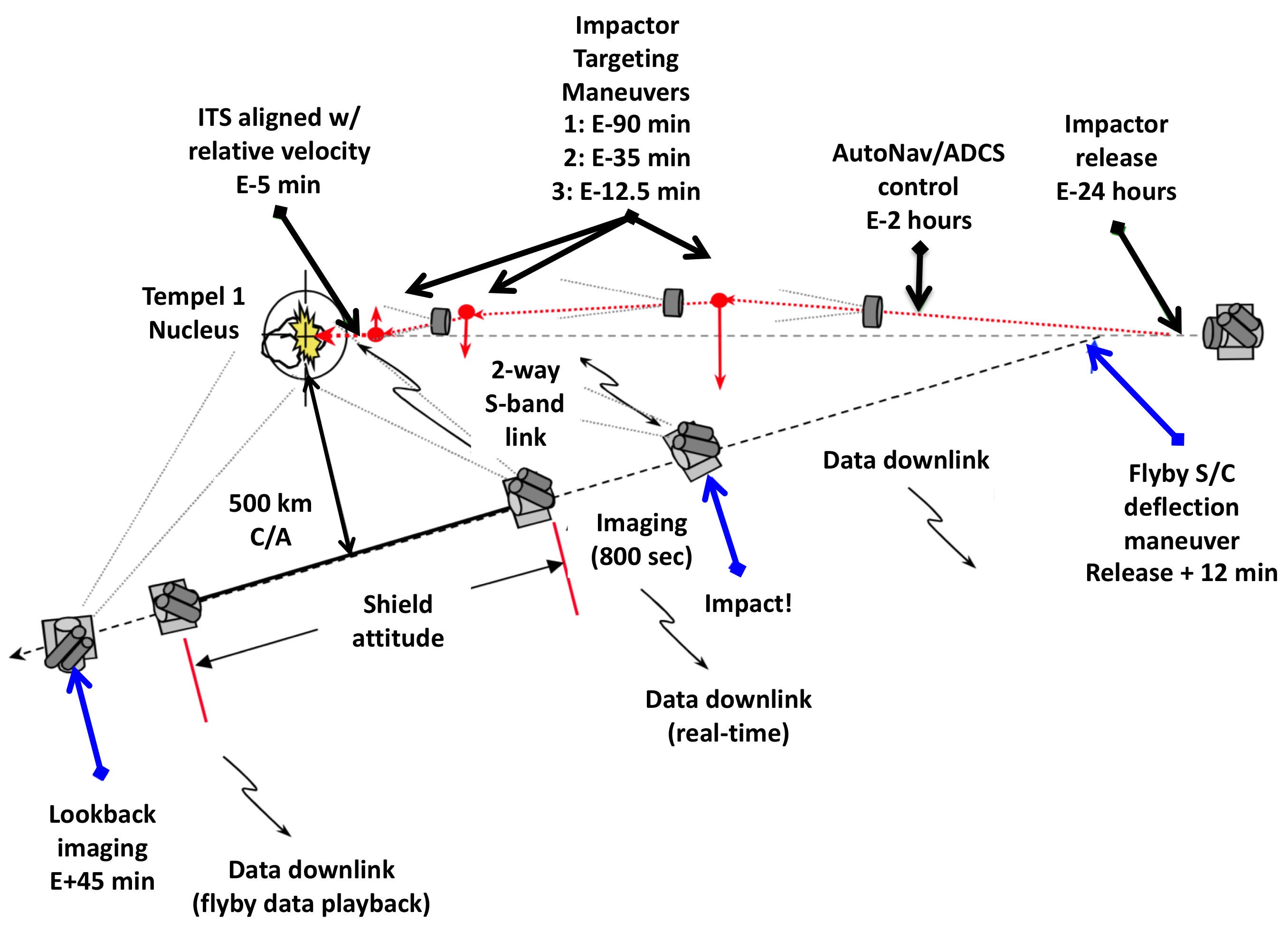}
\caption{Graphic adapted from \citet{Mastrodemos_DI_SSR} depicting the timeline of the impactor separation from the \DI{} flyby spacecraft and targeting maneuvers, deflection maneuver and shield mode of the flyby spacecraft, and various communication links.} 
\label{fig:DI_autonav_graphic}
\end{center}
\end{figure}

\DI{} elevated the complexity of cometary science missions by including an experiment onboard that would interact with the nucleus of comet 9P. Envisioned first by \citet{Belton_DI_concept} and  successfully proposed by PI Michael A’Hearn in NASA’s eighth round of \mission{Discovery} missions in 1998, \DI{} was the most risky cometary mission up to that time.  The project comprised two fully functional spacecraft mated together, a flyby and an impactor. The goal was to explore beyond the highly evolved upper meters of the nucleus, exposed to cosmic radiation and solar insolation, by excavating less altered and relatively pristine material at depth and comparing the properties and composition of the subsurface and coma components. This would be achieved via a hypervelocity impact with known kinetic parameters \citep{Ahearn_DI_SSR}. In order to be properly executed, the two spacecraft had to separate $\sim$24 hr before impact and the flyby spacecraft had to divert and slow down to safely miss the nucleus by 500 km (Fig. \ref{fig:DI_autonav_graphic}) while providing an 800 s viewing window of the cratering event \citep{Blume_DI_SSR}.  

Target considerations for cometary missions are often determined by launch window and orbit feasibility and not chosen simply for the science case. In the case of \DI{}, there were further constraints to consider. For purposes of both reliable targeting of the impactor during its autonomous navigation journey after release by the flyby spacecraft, and nucleus size and strength to sustain a crater of $\sim100$ m size, a comet nucleus with radius $>$2 km was required \citep{Ahearn_DI_SSR}. The approach phase angle was also restricted such that the nucleus would be illuminated as seen from the impactor. For successful autonomous navigation, the crater would be illuminated as seen from the flyby spacecraft before and after its creation, and the comet could be studied prior to the impact experiment for broader context \citep{Mastrodemos_DI_SSR}. The mission was to deliver a $>$350 kg impactor at a velocity $>$10 km~s$^{-1}$ in order to create a large-scale cratering event observable from the flyby spacecraft and the Earth \citep{Blume_DI_SSR}.
An offset from the center of brightness, determined from in-flight scene analysis, was programmed into the autonavigation system for both spacecraft for greatest probability of impacting the comet in an observable and illuminated location and to avoid pointing confusion from a bright limb of the nucleus or the ejecta plume \citep{Mastrodemos_DI_SSR}. The event timing was also crucial as it needed to occur in darkness from at least one major astronomical observatory because of the important role of remote sensing \citep{Meech_DI_SSR}. 9P was chosen because it met or exceeded all of these constraints \citep{Ahearn_DI_SSR}.

\begin{figure}[t]
\begin{center}
\includegraphics[width=\columnwidth]{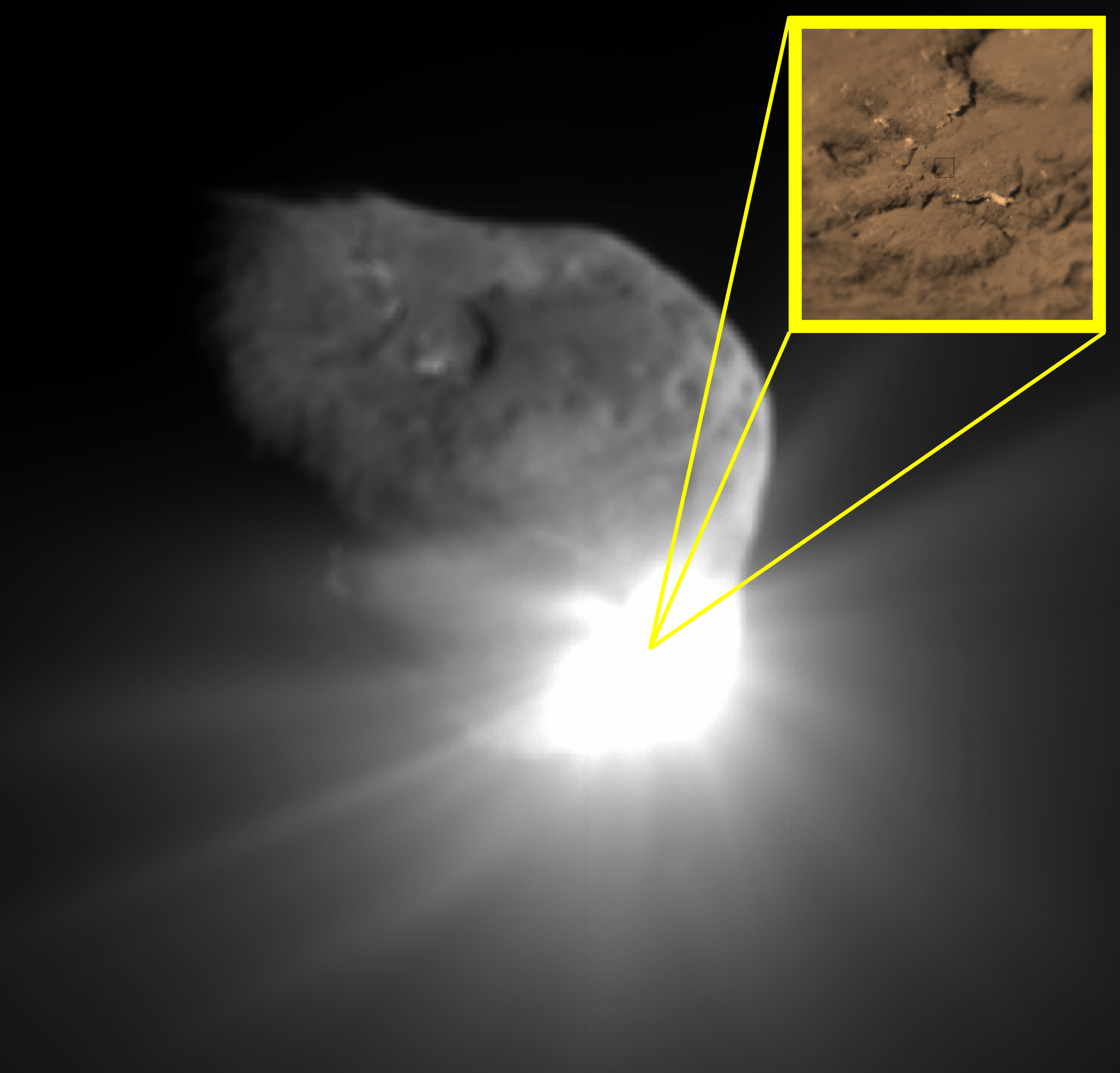}
\caption{Clear filter image of comet 9P acquired with the \DI{} \mission{HRI-VIS} $\sim$1 min after the impact experiment while the flyby spacecraft was $\sim$8,000 km away from the nucleus (image ID: 9000937, \citet{McLaughlinPDS}). The growing ejecta cloud is obvious in the image. The spatial resolution is $\sim$15 m pixel$^{-1}$. The inset composite image is adapted from \citet{Wellnitz_DI_impact} and was derived after geometric and pointing corrections were performed on nested \mission{ITS} images from its final approach before impacting the nucleus. The processed images were then carefully registered to preceding images to a small fraction of a pixel resulting in an image with scale of 1 m~pixel$^{-1}$.}
\label{fig:tempel_DI_impact_ejecta}
\end{center}
\end{figure}

The highly capable instrument suite on the flyby spacecraft included a high-resolution camera (\mission{HRI-VIS}), a wide-field medium-resolution camera (\mission{MRI}), and an infrared imaging spectrometer (\mission{HRI-IR}) that were body mounted and co-aligned \citep{Hampton_DI_SSR}. The \mission{HRI-VIS} and \mission{HRI-IR} shared a single $f/35$ telescope with 10.5 m focal length, {which was discovered to be out of focus after launch due to flawed pre-launch calibrations, resulting in spatial resolution degradation and blurred images that were improved with a deconvolution algorithm \citep{Lindler_deconvolve}.} The $f/17.5$ telescope feeding the \mission{MRI} had a focal length of 2.1 m. Both visible cameras utilized filter wheels with 9 optimized filters for cometary science studies. Aside from two broadband clear filters for each camera, the \mission{HRI-VIS} geology filters had a bandpass of $\sim$100 nm for pre- and post-impact context imaging, observing the progression of the ejecta curtain and crater growth (Fig. \ref{fig:tempel_DI_impact_ejecta}), monitoring photometric variation with rotation, analyzing the surface color and reflectivity, and examining surface layers. The narrowband \mission{MRI} filters were $\sim$5-10 nm in bandwidth and were designed for detailed coma gas (e.g., OH, CN, and C$_2$) and dust studies, including morphology of faint structures and overall coma production. The \mission{HRI-VIS} achieved 10 m~pixel$^{-1}$ or better spatial resolution over 25$\%$ of the nucleus during the flyby and the \mission{MRI} provided stereoscopic information for the nucleus shape reconstruction \citep{Ahearn_DI_Science}. The \mission{HRI-IR} {double prism} spectrometer covered a wavelength region of 1.05-4.85 $\mu$m with variable spectral resolution, peaking at the short and long wavelength ends of the spectrometer and having a minimum $R\sim$200 at 2.5 $\mu$m. This wavelength region afforded the simultaneous detection and mapping of the three primary volatiles known to constitute the frozen interior of comets, namely H$_2$O, CO$_2$, and CO, as well as other important compositional components like carbon-based organics \citep{Ahearn_DI_Science}. The spectral range also encompassed short and long wavelength continuum used to parameterize the reflective and thermal properties of the nucleus and ejected material. 

In addition to the flyby spacecraft instrumentation, the impactor, made predominantly of copper to reduce contamination in the measurements of the cometary composition, had a complete attitude control system, propulsion system, and an S-band radio link to the flyby spacecraft \citep{Blume_DI_SSR}. It carried a wide-field camera (\mission{ITS}) identical to the \mission{MRI} but without a filter wheel. The \mission{ITS} captured images of the intended impact site on 9P up until a few seconds before impact (Fig. \ref{fig:tempel_DI_impact_ejecta}) and achieved spatial resolutions of $\sim$1 m~pixel$^{-1}$ \citep{Ahearn_DI_Science}, the highest resolution ever obtained on a cometary nucleus until \Rosetta. 

As previous missions encountered, in order to successfully study the diverse aspects of a comet, whose  nucleus is comparatively bright with respect to the dim diffuse coma, the instruments were designed to handle a large dynamic range. Before launch, a spectral attenuator that covered one third of the slit length was added at the entrance slit of the \mission{HRI-IR} to reduce emission from the nucleus to account for the potential 10,000:1 brightness ratio between the nucleus and coma \citep{Hampton_DI_SSR}. In addition, observing sequences, exposure times, and data compression lookup tables for all of the instruments were created to span the nucleus and coma signal, so as not to compromise  low signal details in the coma jets and ejecta. In the end, the ejecta was quite bright and optically thick and overwhelmed some of the images. The ejecta material was so bright that it  biased the autonavigation and instrument pointing of the flyby spacecraft, which misinterpreted the ejecta as part of the nucleus and thus pointed further from the impact site than the planned center of brightness offset \citep{Klaasen_DI_cal}.

As with most cometary missions, 9P’s nucleus properties were not well {constrained} at the time of mission design. The mission team had the challenge of designing to various physical parameters (e.g., composition, ice-to-silicate ratio, gas-to-dust ratio, mass, density, porosity, material strength, thermal properties), which often spanned an order of magnitude or more. Additionally, although modeled by \citet{Richardson_DI_SSR} and \citet{Schultz_DI_SSR}, the exact impact scenario could not be assumed prior to the event and best guess boundary conditions of low strength and low density were used to {limit} the expected situation in order to design the full mission. Regardless of the resulting cratering scenario, e.g., burrowing into very porous material or fragmenting a piece off the nucleus, the data from the impact experiment would lead to fundamental conclusions about cometary structure. 

The \DI{} flyby spacecraft {implemented a 45$\degr$ rotation} with respect to the velocity vector 800~s after impact to protect the instruments and other critical subsystems from dust particle hits during closest approach and the comet’s orbital plane crossing (Fig. \ref{fig:DI_autonav_graphic}). As an additional safety measure, the spacecraft design included dust shielding on the leading edge of the solar panels and instruments in this flyby orientation. Particle hits were detected and measured by the impactor's {navigation system once released by the flyby spacecraft (neither the impactor nor flyby had dedicated dust impact monitoring instruments)}, and while some of the larger hits did cause a slew in pointing, none were fatal to the mission and the attitude control system always recovered the impactor pointing. This result has since been used to inform the assessment of dust hazards for subsequent mission proposals.

Major science results from the \DI{} mission were first presented in \citet{Ahearn_DI_Science} and led to advances in our understanding of cometary nuclei (see also chapters in this book by Filacchione et al. and Pajola et al.). For example, the results from pre-impact monitoring and mapping of 9P are extensive. {The mini-outbursts observed on approach were nearly instantaneous, with many being cyclic and originating from regions on the limb coming into daybreak \citep{Farnham_DI_outbursts, Belton_DI_outbursts}. The diverse examples of surface terrain are suggestive of numerous geologic processes at work on 9P: the smooth surface flows are hypothesized to be cryogenic in nature; the rough regions have an assortment of circular features, where some may be attributed to cratering events but others are pits likely formed by sublimation and material collapse; the active scarps are suggestive of backwasting; and the thick layered ridge may be a relic of formation or a result of erosion \citep{Thomas_DI_shape}. The single-scattering albedo was 0.039 at 550 nm wavelength and the photometric variations on 9P, showing little variegation, were relatively small compared to other comets \citep{Li_DI_photometry}. A low thermal inertia was sufficient to explain the temperature map of 9P’s nucleus, which had a maximum of 336 K near the subsolar point, and fell off toward the terminator, following the topography, and was indicative of cold temperatures just below the surface that may harbor ices \citep{Groussin_DI_thermal}. The surface ice was comprised of loose aggregates 10’s of microns in size and, like the mini-outbursts, appeared to be confined to small patches along the morning terminator \citep{Sunshine_DI_surfaceice}. {Heterogeneous gas emission was observed for the first time:} the H$_2$O outgassing was solar driven with enhancements in the sunlit hemisphere, while the CO$_2$ was released in nighttime jets and may be correlated with the smooth flows and eroding scarps \citep{Feaga_DI_coma}.} 

On July 4, 2005, at 05:44:36 UTC, \DI{} delivered 19 GJ of kinetic energy to the nucleus of 9P as designed, fulfilling a major mission objective and furthering the scientific return of the mission. {The volatile and silicate composition of the ejecta was similar between the interior and exterior of the comet, suggesting that the primordial inventory may be reflected in ambient activity. The exception was that the first meter of excavated material was devoid of H$_2$O ice, while  micron-sized H$_2$O-ice grains were present at depths as shallow as 1 m and extending 10’s of meters into the nucleus \citep{Sunshine_DI_ejectaice}. The impact  heated and vaporized silicates, volatiles, and organics in the nucleus and created a self-luminous impact flash. The properties of the flash were indicative of a highly-porous substrate \citep{Ernst_DI_flash}. Material properties of the nucleus were derived from analysis and modeling of the ejecta rays and curtain and their behavior. The results conclude that the surface of 9P is layered and has a high porosity \citep{Schultz_DI_ejecta}.  Additionally, the nucleus has low strength on the order of 1-10 kPa, bulk density of 400 kg~m$^{-3}$, and the cratering event was controlled by gravity rather than material strength \citep{Richardson_DI_ejecta}.}


\subsection{EPOXI}
\label{sec:epoxi}

After a successful 7-month mission and impact experiment, the \DI{} flyby spacecraft remained healthy, with all three scientific instruments functioning nominally, and had adequate consumables onboard to continue on for an extended mission \citep{Ahearn_Hartley_Science}. A second cometary mission, one to explore the diversity of JFCs, was proposed for the extension and selected as a Mission of Opportunity in the 2006 round of Discovery proposals with the caveat that the \mission{Deep Impact eXtended Investigation} (\mission{DIXI}) mission (PI: A’Hearn) would be merged with the \mission{Extrasolar Planetary Observation and Characterization} (\mission{EPOCh}) mission (PI: Drake Deming) resulting in the \mission{EPOXI} mission. 
The out-of-focus \mission{HRI} telescope was advantageous to fulfilling the EPOCh objective of using the \DI{} flyby spacecraft as a remote observatory for transiting extrasolar planet science during its long cruise to another comet \citep{EPOCh_Ballard}. Photometric observations acquired by the \mission{HRI-VIS} did not saturate the CCD, but rather the \mission{HRI} focus issue spread the light from observations over many pixels generating better data for the purpose of the study. The extended mission also incorporated a technology demonstration for \mission{NASA}, testing the deep space communication network \citep{DI_tech_demo}.

The scientific value of repurposing the \DI{} spacecraft was enormous. The extended mission utilized three Earth gravity assists to put it on a trajectory bringing the spacecraft within 700 km from comet 103P on November 4, 2010. During the Earth-Moon system flybys, hydration on the lunar surface was investigated \citep{Sunshine_Moon} and the Earth-Moon system was observed from afar as an analog extrasolar planet \citep{EPOCh_Robinson,EPOCh_Crow}. In addition, \DI{} had a short six-month commissioning, calibration, and cruise phase from launch to approach of 9P during the primary mission, leaving much to improve upon for the extended mission, including extensive calibration campaigns throughout cruise and designing more effective observing  sequences to be executed for the 103P encounter \citep{Klaasen_DIXI_cal_paper}.

The \EPOXI{} team was faced with a challenge leading up to the 103P encounter as a result of the spacecraft design for the primary mission, where the instruments were to point at the comet while the antenna was directed at Earth. During the 103P encounter, the flyby geometry was such that the instruments and antenna could not simultaneously point to their respective targets. The mitigation strategy was to manage data acquisition, data storage, data downlink, and thermal control of the spacecraft using a {\it do-si-do} maneuver where data were taken and stored, then the spacecraft slewed to point the antenna to Earth for data downlink, during which the spacecraft heated up, and then slewed back to the comet and cooled before data acquisition resumed.  This programmed {`dance'} transpired daily on approach and caused spikes in the instrument temperatures but resulted in successful observing sequences. 

\begin{figure}[t]
\begin{center}
\includegraphics[width=\columnwidth]{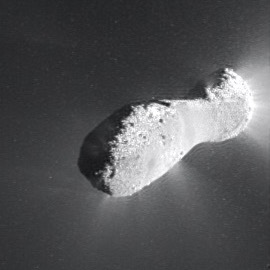}
\caption{Deconvolved \mission{HRI-VIS} image of 103P/Hartley 2 acquired November 4, 2010, when \DI{} was {$\sim$1,950} km away from the nucleus. There are large icy chunks surrounding the nucleus as well as small icy grains being expelled from the nucleus in the image. Every speck in the image is a particle. The image was downloaded from the PDS-SBN (\citet{LindlerPDS}, image ID: {5004008}).}
\label{fig:H2_stretch}
\end{center}
\end{figure}

103P was the fifth comet imaged up close by a spacecraft, but was not the original target of the extended mission.  Comet 85D/Boethin was the proposed target for \EPOXI, but was abandoned when it was not recovered in an extensive Earth-based campaign in 2005-2007 \citep{Meech_Boethin}, prior to a {planned} trajectory correction maneuver to put the spacecraft on the path to intercept the comet, again emphasizing the importance of mission support observational campaigns (Section \ref{sec:ground-based}). The primary target became 103P, which resulted in paradigm changing science due in part to its intrinsic hyperactivity {\citep{Groussin_103P_hyper}}.  The cometary objectives of the \EPOXI{} mission were to systematically determine the degree of diversity among comets by comparing comets of similar age and orbit, using the same suite of instruments and to disentangle evolutionary versus primordial aspects of cometary behavior and morphology. The resolved images of 103P showed {an elongated bilobed nucleus with a smooth `waist', rough terrain on both lobes, and several tall spires ($>$ 40 m; \citealt{Thomas2013b_H2})}. The nucleus, {which was in a state of complex rotation \citep{Belton_2013_H2spin}}, was  engulfed in icy particles (Fig. \ref{fig:H2_stretch}; {\citealt{Hermalyn2013}}). {Reminiscent of 9P's heterogeneous coma, non-correlated discrete concentrations of H$_2$O and CO$_2$ above the ubiquitous ambient coma were also detected at 103P \citep{Ahearn_Hartley_Science}}. It was clear that the hyperactivity of 103P was due to the sublimation of H$_2$O-ice grains, including a population of small grains that were expelled from the small lobe of the nucleus in a jet of CO$_2$ gas \citep{Ahearn_Hartley_Science,Protopapa_H2_ice}. These particles did not cause damage to the spacecraft.

The \DI{} flyby spacecraft was subsequently used for the remote study of distant comets such as C/2009 P1 (Garradd) in early 2012, determining that the comet had a very high and increasing CO to H$_2$O abundance as it receded from the Sun \citep{Feaga_Garradd}, and C/2012 S1 (ISON) in early 2013, detecting several small spontaneous outbursts when it was approaching the Sun but still beyond the H$_2$O snowline \citep{Farnham_ISON}. Communication with the \DI{} spacecraft was lost in August 2013 after it was determined that the spacecraft computer entered an infinite fault protection loop due to a 64-bit time stamp being converted to and not fitting into a 32-bit value \citep{Farnham_ISON}, thus bringing a close to the mission.

The combination of the \DI{} and the \EPOXI{} missions unveiled certain details about several extreme cometary behaviors and their manifestations, 
but generated just as many questions. In particular, it is still uncertain how deep {below the surface} the ice reservoirs are;  what mechanism causes outbursts {\citep{Belton_Melosh_outbursts}}; whether the surface flows are cryogenic as suggested 
\citep{Belton_DI_outbursts}; what the true nature of the small versus large ice particle populations is {(\citealt{Sunshine_DI_ejectaice}; \citealt{Protopapa_H2_ice}; \citealt{Kelley_H2_particles})}; and, most importantly, which morphological and compositional cometary properties reflect evolutionary versus formational conditions {\citep{Sunshine_comet_processes}}. \Rosetta{} would shed light on many of these topics, but not completely close these knowledge gaps. 


\subsection{Stardust-NExT}
\label{sec:stardust-next}

Also selected as a Mission of Opportunity in the 2006 round of NASA Discovery proposals {and led by PI Joseph Veverka}, \StardustNExT{} (New Exploration of Tempel) repurposed the healthy \Stardust{} spacecraft to fly by 9P one perihelion passage after the \DI{} experiment \citep{Veverka2013}.  
The \mission{NAVCAM}, which was built from spare parts from previous flight projects \citep{Newburn2003}, and not well calibrated during the primary mission because of its supporting role during the \Stardust{} sample collection, became the priority instrument for the extended \StardustNExT{} mission. As such, substantial calibration improvements were made to the geometric correction, spatial resolution, and radiometric calibration during the extended mission \citep{Klaasen2013}. In addition, for optimal imaging quality, recurring camera contamination from dust was minimized by periodically heating the instrument via internal heaters or direct sunlight. The dust instruments, \mission{DFMI} and \mission{CIDA}, both collected data during the 9P encounter. 

\begin{figure}[t]
\begin{center}
\includegraphics[width=\columnwidth]{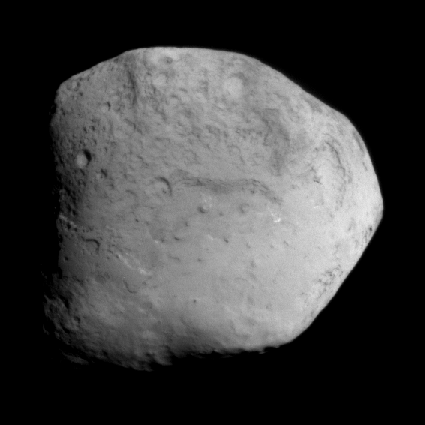}
\caption{Clear NAVCAM image of comet 9P acquired by  \StardustNExT{} on February 15, 2011, from $\sim$200 km away from the nucleus. The illuminated face of the comet displayed here is territory not imaged by \DI{} and shows a diversity in morphology including examples of layering, ridges, and circular depressions on the surface at a resolution of $\sim$12 m pixel$^{-1}$. Downloaded from the PDS-SBN (\citet{VeverkaPDS}, image ID: 30039).}
\label{fig:tempel_SDN}
\end{center}
\end{figure}

For the first time, \StardustNExT{} imaged a comet previously visited by a spacecraft  and documented surface changes after a full orbital period had passed, and after its surface was altered by an artificial impact. \StardustNExT{} extended the nucleus surface coverage (Fig. \ref{fig:tempel_SDN}), and thus knowledge, of 9P from $\sim$30\% to $\sim$70\% and measured the \DI{} experimental crater site \citep{Veverka2013}. Precise targeting and timing of the flyby were essential in order to realize these goals because the crater specifically, and other regions studied by \DI{} more generally, needed to be illuminated and in the field of view of the spacecraft at the time of closest approach to make the intended measurements and comparisons. For this, 9P’s spin state and spin rate had to be accurately predicted before the encounter, and thus the importance of synergistic observations was recognized by the community yet again. A worldwide observing campaign of 9P \citep{Meech2005,Meech2011}, supporting both the \DI{} and \StardustNExT{} missions, was used to acquire enough data from which the evolution and current state of the rotational period and pole orientation could be derived from the lightcurve to feed to the navigation team a full year before closest approach \citep{Belton2011}. A trajectory correction maneuver in 2010 was made to delay arrival by $\sim$8.5 hr. Upon arrival, it was clear that the team met the challenge of accurately determining the rotation rate and spin axis orientation as the encounter occurred within $\sim$20$^{\circ}$ of the targeted longitude \citep{Veverka2013,Belton2011,Chesley2013}. This was the first time that a change in rotation rate of a comet was unambiguously measured from one perihelion to another.

At a speed of 10.9 km~s$^{-1}$ and closest approach distance of 178 km, \StardustNExT{} arrived at comet 9P on February 15, 2011 and took 72 images with the \mission{NAVCAM} \citep{Veverka2013}. The highest quality image of 9P and the artificial crater had a resolution of 11 m~pixel$^{-1}$. The imaging frequency was constrained to {one image every} 6 s due to the maximum rate supported by the data system. During the flyby, the change in geometry and lighting was amenable to taking and creating stereo pairs of images. The data collected during closest approach allowed for the 
\DI{} crater to be detected and measured: {the images reveal a 50~m diameter bowl-shaped crater within a larger ($\sim180$~m diameter) depression, hinting at either a layered nucleus structure and/or a complex cratering process \citep{Schultz2013,Vincent2015}}. 
Even more smooth terrain was discovered on 9P than had been seen with \DI{}, amounting to $\sim$30\% of the surface, most likely new material that had erupted from the subsurface well after formation of the comet \citep{Veverka2013,Thomas2013}. There was also additional {evidence of nightside activity, jets emanating from steep slopes,} extensive layering, and pitted terrain {on 9P \citep{Veverka2013,Farnham_SDN}}.  Contrary to some expectations, the active regions of the surface identified by temporal variability comprised only $\sim$10\% of the surface, especially evident in scarp retreat, with 1--10~m of material lost in one apparition \citep{Veverka2013,Thomas2013}. Additional images were acquired on approach and departure, but the faint comet was not detected in the maximum commandable exposure time of 20 s until 27 days prior to encounter and was not resolved in the latter. The jet activity in 2011 was weaker than in 2005, which may be due to the pre- versus post-perihelion behavior of the comet and the timing of the encounters. \DI{} was 1 day pre-perihelion while \StardustNExT{} was 34 days post-perihelion. There was also no evidence in the photometry of the periodic mini-outbursts seen by \DI{} \citep{Belton_DI_outbursts,Belton2013}.

The dust instruments were only operational for a very short time ($\sim$hours) surrounding closest approach \citep{Veverka2013}. Similar to the dust environment at 81P, the \mission{DFMI} data were not uniform and were dominated by detections of clusters of particles fragmenting from aggregates rather than a steady stream of small particle hits originating from active areas of the nucleus. Up to 1,000 particles were measured over kilometer scales followed by voids with no particle impacts \citep{Economou2013}. Dust mass spectra collected with \mission{CIDA} showed prominent peaks at mass 1 and 26, H and CN respectively, and long tails at high mass numbers suggesting complex molecules \citep{Veverka2013}.

Although extended missions are challenging, as enumerated here and in section  \ref{sec:epoxi}, recycled spacecraft have led to scientifically enlightening outcomes at an order of magnitude less cost than the primary mission. 

\begin{figure*}[th]
  \begin{minipage}[c]{0.65\textwidth}
\includegraphics[width=\columnwidth]{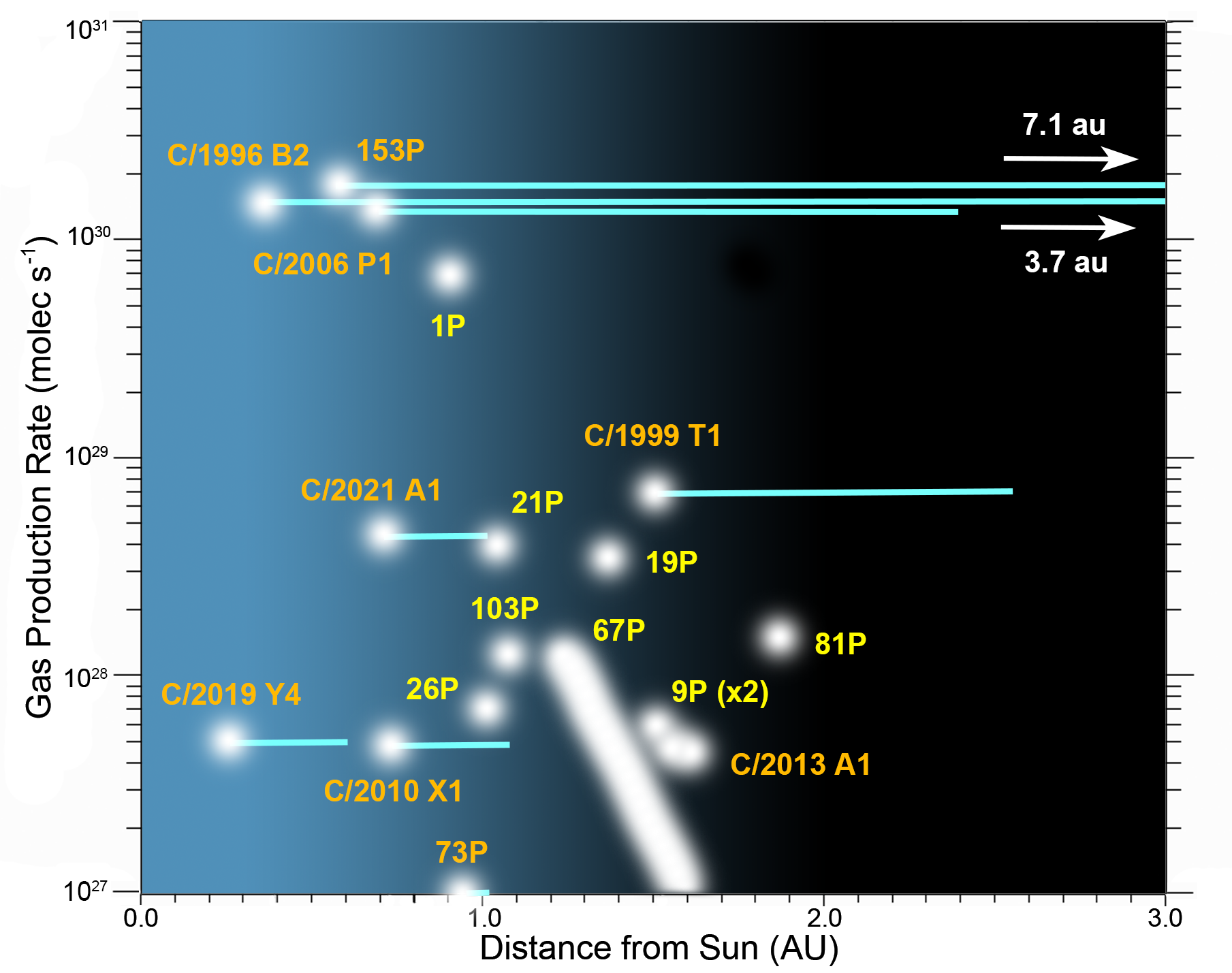}
  \end{minipage}\hfill
  \begin{minipage}[c]{0.33\textwidth}

\caption{
 All targeted comet encounters (yellow), and serendipitous ion tail crossings (orange). 
   Positions are plotted to show the heliocentric distance of each comet's nucleus against its gas production rate at the time. Tail crossings occurred at the end of each blue line; those outside the plot's range are as shown by the arrows. Sources of production rates, for H$_{2}$O unless indicated, if not cited in the text: 1P: (total) \citet{krankowsky1986}; 9P: \citet{schleicher2006}; 19P: \citet{Young2004}; 21P: \citet{neugebauer07}; 26P: \citet{johnstone1993}; 81P: \citet{fink1999}; 103P: \citet{dellorusso2011}; 153P, C/1999~T1: \citet{Combi2008}; C/2019~Y4: \citet{Combi2021}.
   Continuum of 67P values: \citet{KC-waterproduction}; {C/2021 A1: \emph{Combi, personal communication}}. 
}
\label{fig:tailcrossings}
  \end{minipage}
\end{figure*}


\subsection{Other encounters}
\label{sec:other-flybys}

In addition to planned ion tail crossings, spacecraft can also serendipitously traverse tails if they are 
at the right place and time to be in the flow of ions downstream of a comet. 
As many comets' ion tails have a significant width, this 
alignment does not need to be perfect for cometary ions to be detectable at the spacecraft. 
The first such tail crossing, found unexpectedly, was the \mission{Ulysses} spacecraft's encounter with the ion tail of C/1996 B2 Hyakutake in 1996: \citet{jones00} recognized the magnetic field structure 
to be consistent with {the} comet's ion tail, and \citet{gloeckler00} reported the observation of cometary pickup ions in the craft's \mission{SWICS} instrument data.

Several such unplanned comet tail crossings are now known (Fig. \ref{fig:tailcrossings}). 
 There were two more by \mission{Ulysses}:  C/1999 T1 McNaught-Hartley on October 19-20, 2000 \citep{gloeckler2004}, and C/2006 P1 McNaught over several days centered on February 7, 2007 \citep{neugebauer07}. Pick-up ions from 73P were detected by the near-Earth \mission{ACE} and \mission{Wind} spacecraft in May and June 2006 \citep{gilbert2015},  and ions from C/2010~X1 Elenin  by \mission{STEREO-B}  \citep{galvin2015}. 
    
These unplanned comet encounters complement data from targeted comet missions, and provide: \textit{in situ} information on the solar wind's response to a comet's presence, including flow speed and direction; the magnetic field structure of a comet's ion tail; and the {wind's} composition and that of pickup ions that are carried by it. Despite not being designed for cometary encounters, general advancements in the sensitivity of \textit{in situ} instruments have improved the quality of the serendipitous observations obtained.

As demonstrated by the case of C/2013~A1 (see section \ref{sec:mars}), the advance warning of a tail crossing can allow mission teams to select favorable instrument modes to maximize the science return of these serendipitous events. Such  warnings were provided for ESA's \mission{Solar Orbiter} crossing  the tails of C/2019 Y4 ATLAS \citep{jones2020} in June 2020 {and C/2020 A1 Leonard in December 2021}. \textit{In situ} instruments detected {these} {tails'} draped magnetic field, pick-up ions, and the presence of ion-scale waves excited by associated instabilities, { e.g.,
\citet{Matteini2021}}.

NASA’s \mission{Parker Solar Probe} (\mission{PSP}) passed as close as 0.025 au, or 3.7 million km, to  
322P/SOHO 1 on September 2, 2019. This encounter is particularly intriguing, given this object's possible asteroidal nature \citep{Knight2016}. However, there was little unambiguous evidence of the comet in data gathered then by \mission{PSP} \citep{Jiansen2021}.

A remarkable aspect of several of the serendipitous tail crossings is the immense tail lengths sometimes involved. Pickup protons from 153P/Ikeya-Zhang were detected by the NASA-led \mission{Cassini} spacecraft at $\sim$6.5 au downwind of the nucleus \citep{JONES2022115199}. Detections at such large distances imply that ion tails, although ultimately originating at bodies of a few km across, can potentially remain as detectable structures all the way to the heliosphere's edge.

We also note the phenomenon of Interplanetary Field Enhancements, IFEs. These are  uncommon solar wind events, characterized by a heliospheric magnetic field magnitude increase, often with a thorn-shaped profile, that are usually accompanied by a discontinuity in the magnetic field direction. The first reported event of this kind \citep{russell1983} has been followed by several surveys and case studies, e.g. \citet{russell1984, jones2003}. Although {these signatures'}
exact cause is  still to be widely accepted, it appears clear from several associations between IFEs and cometary orbital plane crossings (e.g., \cite{lai2015,jones2003a}) that they are somehow associated with interactions between the solar wind and dust trails that lie along the orbits of some comets and asteroids. Once the formation process for IFEs is better understood, routine solar wind observations by spacecraft may add to our understanding of dust trails.


\subsection{C/2013 A1 at Mars}
\label{sec:mars}

By far the closest known unplanned approach to a comet is the case of C/2013 A1 Siding Spring. This dynamically-new {LPC}'s orbit took it to within 140,751${\pm}$175~km of the centre of Mars on October 19, 2014 \citep{Farnocchia2016}. All spacecraft then operating in the planet’s orbit -- NASA’s \mission{Mars Odyssey}, \mission{Mars Reconnaissance Orbiter (MRO)}, and \mission{MAVEN}, ESA’s \mission{Mars Express}, and India’s \mission{Mangalyaan} -- effectively performed a comet flyby. Pre-encounter studies of the comet suggested a not-insignificant dust risk for spacecraft near Mars, so these satellites’ orbits were phased such that they were largely protected from dust impacts by Mars itself during the the comet’s closest approach to the planet. Instruments on several of the spacecraft were also purposefully turned off for safety reasons.

The comet's $\sim$0.5~km-wide nucleus was successfully resolved by the \mission{MRO HiRISE} camera at a resolution of 138~m~pixel$^{-1}$ \citep{Farnham2015,Lisse2015}.  Images of the inner coma were also obtained using NASA's \mission{Curiosity} and \mission{Opportunity} rovers.
Lyman-$\alpha$ observations by \textit{MAVEN} indicated a water production rate of 0.5$\times$10$^{28}$~s$^{-1}$ \citep{Mayyasi2020}.
Interpretations of the solar wind-comet-planet interactions were complicated by the arrival of the interplanetary counterpart of a coronal mass ejection shortly before the encounter.
Nevertheless, the influence of the comet's own induced magnetosphere on the near-Mars environment was clear: Mars's much smaller induced magnetosphere underwent a rotation in magnetic field orientation as the comet approached, with particularly strong distortion of the planet's ionospheric magnetic field around closest approach, and increased planetary magnetosheath turbulence post-encounter \citep{Espley2015}. 
A significant increase in energetic particle fluxes was observed during the comet's passage; these displaying similarities to features observed at other comets \citep{SanchezCano2018}.

Of particular interest was the interaction between Mars's atmosphere and the {comet coma's} dust and gas. The energetic particles mentioned above would have been deposited in the planet's atmosphere, as well as pickup ions such as O$^{+}$ that were detected for $\sim$10~hours near the planet \citep{SanchezCano2020}. A transient ionized layer at altitudes of 80-100~km was present in the hours after closest approach, resulting from {the shower of cometary dust}
\citep{Gurnett2015}. An increase in the abundance of metallic ion species, consistent with cometary dust deposition, was also detected at altitudes of $\sim$185~km \citep{Benna2015}.


\section{\textbf{THE ROSETTA RENDEZVOUS MISSION}}
\label{sec:rosetta}

\subsection{Introduction}
\label{sec:rosetta-intro}

The original concept that would become  \Rosetta{}  began life in the 1980s as a hugely ambitious comet nucleus sample return mission. Despite the fact that the study began before anyone had even seen a nucleus (pre-Halley encounters), it would conclude that a mission should drill a meters-long core and return the sample for analysis in laboratories on Earth, maintaining the ices at cryogenic temperatures throughout (see \citealt{grumpy-Keller} and \citealt{Thomas2019SSRv} for short histories on the early evolution of the mission concept). As described in section \ref{sec:future-mission-types}, the technology required for such a concept appears to be some way off, even today. However, the motivation for a sample return mission was, and is, compelling and obvious: by collecting ice from the interior of a comet, we would sample more-or-less completely unaltered material laid down during the formation of our Solar System, and be able to subject it to the full range of investigative techniques possible with modern laboratory equipment. This would be revolutionary not only in the study of comets, but also in understanding the wider topic of planet formation processes. Motivated by a desire to understand the physics of how cometary activity really works, \Rosetta{} would also study the changes on the surface of the comet over an extended period as it approached the Sun. This is important within the field in its own right, and would also be essential to determine just how `pristine' (or not) the returned sample really is, by measuring the effects of activity on the immediate sub-surface layers.

Buoyed by the success of \Giotto, ESA selected \Rosetta{} as the only planetary probe among the four cornerstone missions for its \mission{Horizon 2000} program. An advanced set of \textit{in situ} instruments was proposed as a more realistic alternative to sample return, including separate landers (originally two), but the key advance from previous missions was retained: \Rosetta{} would rendezvous with a comet and follow its evolving behavior as it approached the Sun, instead of seeing it at only a single moment in time with a flyby.  This 2-year period of operation was the key to the \Rosetta{} mission, enabling much more data to be collected on the comet and allowing for repeated measurements to be made, to see how the nucleus surface, coma, and interactions with the solar wind changed over time. 

\begin{table}[t]
    \begin{tabular}{|c|c|l|}
   \hline
    Dates &  $r^*$ &  Event\\
        \hline
2004 Mar 02 & 1.0	& Launch	\\        
2005 Mar 04	& 1.0	& 1st Earth flyby	\\        
2007 Feb 25	& 1.5 & Mars flyby	\\        
2007 Nov 13	& 1.0	& 2nd Earth flyby	\\        
2008 Sep 05 & 2.1 & {\v S}teins flyby (800 km)	\\        
2009 Nov 13	& 1.0	& 3rd Earth flyby	\\        
2010 Jul 10 & 2.7 & Lutetia flyby (3200 km)\\        
2011 Jun 08	& 4.5 &  Hibernation entry \\
2012 Oct 03 & 5.3 & Aphelion\\        
2014 Jan 20 & 4.5 & Hibernation exit \\        
2014 Aug 06 & 3.6 & Arrival at comet \\        
2014 Nov 12 & 3.0& \Philae{} landing\\        
2014 Nov 15 & 3.0 & End of \Philae{} science operations\\        
2015 Feb 14 & 2.3 & 1st close flyby (6 km) \\
2015 Mar 28 & 2.0 & 2nd close flyby (14 km) \\
2015 May 10 & 1.7 & Equinox \\
2015 Jul 09 & 1.3 & Last signal from \Philae\\         
2015 Aug 13 & 1.2 & Perihelion\\   
2015 Sep 30	& 1.4 & Sunward excursion to 1,500 km\\
2016 Mar 21 & 2.6 & Equinox \\
2016 Mar 30	& 2.7 & Tailward excursion to 1,000 km\\  
2016 Sep 02 & 3.7 & \Philae{} found \\
2016 Sep 30 & 3.8 & End of mission\\        
           \hline
    \end{tabular}
    \smallskip\\
    Notes:
    $^*$ Heliocentric distance at event (au)
    \caption{Timeline of key events in the \Rosetta{} mission.}
    \label{tab:rosetta-timeline}
\end{table}

\Rosetta's original comet was 46P, but a year-long delay following the failure of a previous Ariane 5 rocket meant that a new target was required. The mission launched toward 67P on March 2, 2004. To match orbits with the comet, \Rosetta{} took a looping 10-year trajectory through the inner Solar System (Table \ref{tab:rosetta-timeline}). This trajectory included two asteroid flybys en route as \Rosetta{} passed through the main belt, the 5 km diameter (2867) {\v S}teins and the 100 km (21) Lutetia \citep{ast-flybys}. Additionally, a number of `bonus' cruise phase science investigations were completed utilizing the \mission{OSIRIS} cameras for remote observations of comets 
C/2002 T7 LINEAR, C/2004 Q2 Machholz, and
9P \citep{Keller2006-OSIRIS-comets,Rengel2007-OSIRIS-comets,Kueppers2005,Keller2005, Keller2007} 
and active asteroid 354P/LINEAR \citep{2010a2}. 
Finally, while \Rosetta{} was too far from the Sun to generate sufficient power from its solar panels, it entered a 31 month-long `hibernation' through its final aphelion before arrival. 

\begin{figure}[t]
\begin{center}
\includegraphics[width=\columnwidth,trim=0.2cm 0.1cm 0.2cm 1.2cm,clip=true]{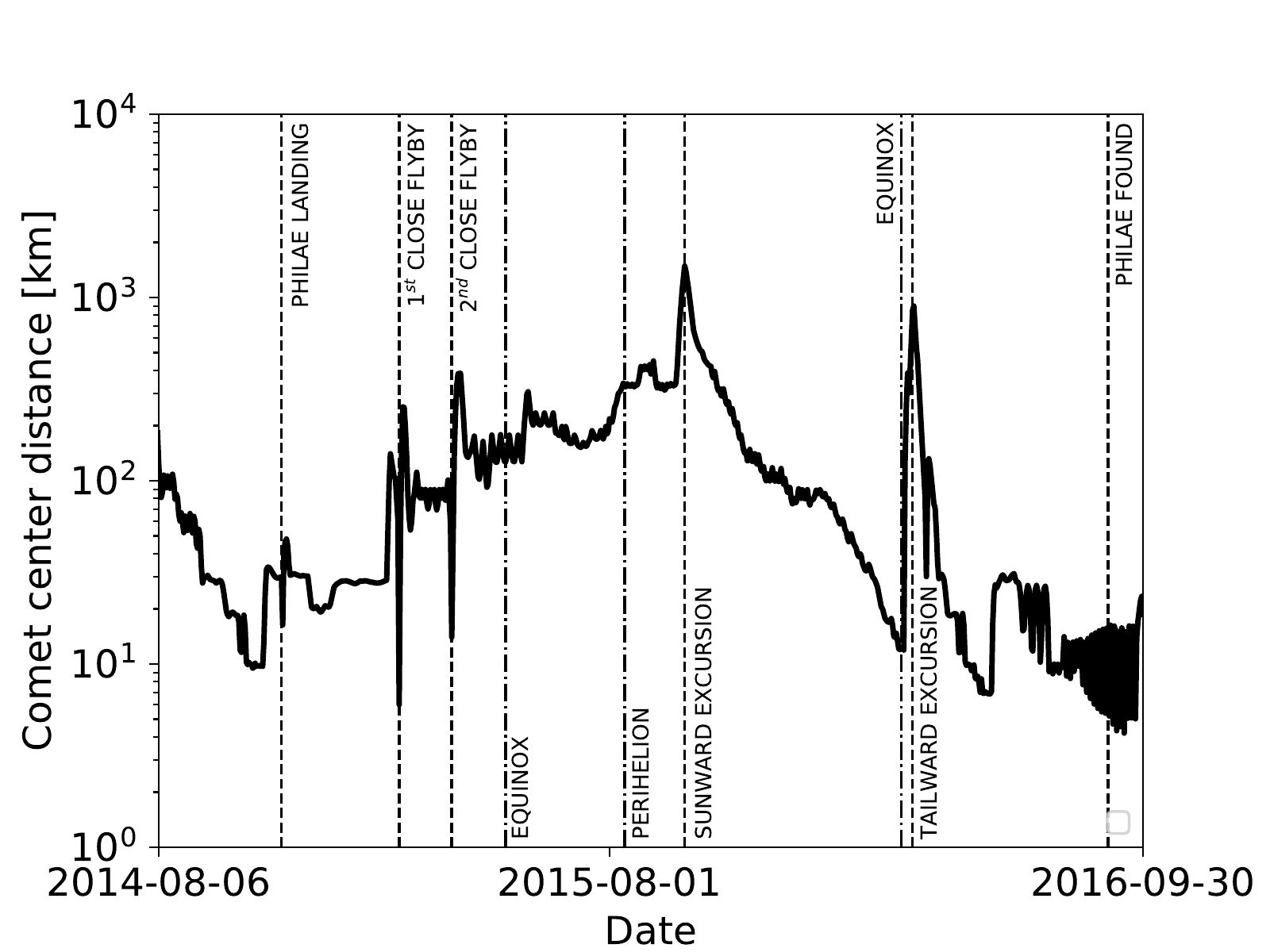}
\caption{Distance between \Rosetta{} and the comet center from 
{arrival at the comet} until end of mission. {The vertical dashed lines indicate spacecraft activities and the vertical dash-dotted lines indicate comet orbital milestones}.}
\label{fig:rosetta-distance}
\end{center}
\end{figure}

\Rosetta{} woke from hibernation in January 2014 and began recommissioning of its payload, with the first images of 67P being acquired by \mission{OSIRIS} on March 23, 2014, from a distance of $\sim 5 \times 10^6$ km \citep{Tubiana-67Papproach}. 
The next five months saw a careful approach to the comet, followed by three months of mapping and preparation for the \Philae{} landing, which occurred on November 12, 2014 (see section \ref{sec:landing} for details). \Rosetta{} then entered the `escort phase' of the mission, continuing through 67P's perihelion in August 2015, and an extended phase beyond the nominal operations period into 2016. During the escort phase the distance between \Rosetta{} and the nucleus was typically between 10 and 200 km (increasing to more than 300 km with higher activity levels near perihelion; Fig. \ref{fig:rosetta-distance}). `Excursions' to larger distances (1,500 km toward the Sun, and 1,000 km in the ion tail direction) were made, primarily to better understand the plasma environment around the comet.

\begin{figure}[t]
\begin{center}
\includegraphics[width=\columnwidth]{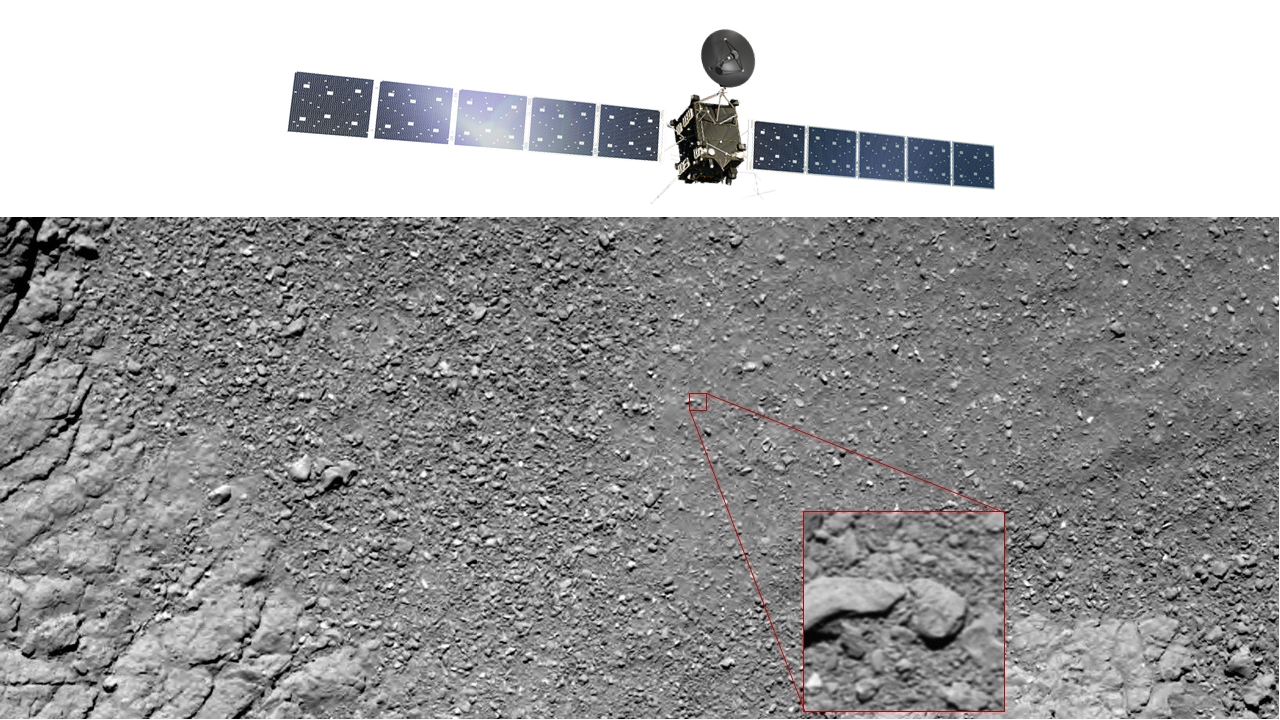}
\caption{Image from the \mission{OSIRIS} 
Wide Angle Camera on \Rosetta, showing the landing area, with the last image acquired by \mission{OSIRIS} before the crash landing superimposed. The image of the \Rosetta{} spacecraft is to scale. {Credit: OSIRIS\protect\footnotemark[1]{}.}}
\label{fig:Rosetta_landing}
\end{center}
\end{figure}

The \Rosetta{} mission ended on September 30, 2016, with the controlled impact of the orbiter onto the smaller lobe of the nucleus. Figure \ref{fig:Rosetta_landing} shows the approximate landing site. The spacecraft most likely bounced, 
but as contact was lost at impact time, we may never know its final location. 


\subsection{Instrumentation}
\label{sec:rosetta-instruments}

\begin{table*}[thp]
    \centering
    \begin{tabular}{|p{6cm}|p{6cm}|l|}
        \hline
        Instrument & Description & Reference \\
        \hline
\mission{Alice} (Ultraviolet Imaging Spectrometer) & $R\sim100$, 70-210 nm FUV long-slit spectrograph & \citet{RosIns-Alice} \\
\mission{CONSERT} (Comet Nucleus Sounding Experiment by Radio wave Transmission) & Radio signal timing experiment between orbiter and lander & \citet{RosIns-CONSERT} \\
\mission{COSIMA} (Cometary Secondary Ion Mass Analyser) & SIMS dust composition measurements over 1-3500 amu mass range, and microscope camera & \citet{RosIns-COSIMA} \\
\mission{GIADA} (Grain Impact Analyser and Dust Accumulator) & Microbalance plate and laser curtain to measure mass/momentum of dust particles larger than $\sim$ 15 $\mu$m & \citet{RosIns-GIADA} \\
\mission{MIDAS} (Micro-Imaging Dust Analysis System) & Atomic force microscope to image dust in 3D, on nm-$\mu$m scales & \citet{RosIns-MIDAS} \\
\mission{MIRO} (Microwave Instrument for the Rosetta Orbiter) & Microwave temperature measurements at 190 and 562 GHz, with 44 kHz resolution spectroscopy around the latter band & \citet{RosIns-MIRO} \\
\mission{OSIRIS} (Optical, Spectroscopic, and Infrared Remote Imaging System) & Narrow-angle (2$\degr$) and Wide-angle (11$\degr$) CCD cameras, with a total of 26 filters covering 0.25-1 $\mu$m & \citet{RosIns-OSIRIS} \\
\mission{ROSINA} (Rosetta Orbiter Spectrometer for Ion and Neutral Analysis) & Separate high-resolution ($m/\Delta m > 3000$) ion and neutral mass spectrometers and pressure gauge & \citet{RosIns-ROSINA} \\
\mission{RPC} (Rosetta Plasma Consortium) & Sensors for the plasma environment: ion mass spectrometer, ion and electron sensors, Langmuir probe, magnetometer, and mutual impedance probe & \citet{RosIns-RPC} \\
\mission{RSI} (Radio Science Investigation) & Ultrastable oscillator to allow accurate ranging of Rosetta via radio signals & \citet{RosIns-RSI} \\
\mission{VIRTIS} (Visible and Infrared Thermal Imaging Spectrometer) & Medium ($R\sim200$) and High ($R\sim2000$) resolution spectrometers, covering 0.22-5 $\mu$m and 1.9-5 $\mu$m, respectively & \citet{RosIns-VIRTIS} \\
\hline
\mission{APXS} (Alpha Proton X-ray Spectrometer) & Surface elemental composition measurements & \citet{RosIns-APXS} \\
\mission{CIVA} (Comet Infrared and Visible Analyser) &	Panoramic cameras (7 monochromatic visible cameras) and visible and infrared spectroscopic microscopes  & \citet{RosIns-CIVA} \\
\mission{CONSERT} (Comet Nucleus Sounding Experiment by Radio wave Transmission) & 	Radio signal timing experiment between orbiter and lander & \citet{RosIns-CONSERT} \\
\mission{COSAC} (COmetary SAmpling and Composition) & Gas chromatograph and time of flight mass spectrometer focussed on elemental and molecular composition  & \citet{RosIns-COSAC} \\
\mission{Ptolemy} (Mass spectrometer) &	Gas chromatograph and ion trap mass spectrometer focussed on isotopic composition & \citet{RosIns-Ptolemy} \\
\mission{MUPUS} (MUlti-PUrpose Sensors for Surface and Sub-Surface Science) & Thermal and mechanical surface and subsurface properties probe	 & \citet{RosIns-MUPUS} \\
\mission{ROLIS} (Rosetta Lander Imaging System) &	Descent camera system, monochromatic visible camera  & \citet{RosIns-ROLIS} \\
\mission{ROMAP} (Rosetta Lander Magnetometer and Plasma Monitor) &	Fluxgate magnetometer, ion, electron and pressure sensors 	 & \citet{RosIns-ROMAP} \\
\mission{SD2} (Sampling Drilling and Distribution) & Drill and system to pass samples to other instruments & \citet{RosIns-SD2} \\
\mission{SESAME} (Surface Electric Sounding and Acoustic Monitoring Experiment) & Combination of three probes on landing legs; acoustic sounder, permittivity probe, and dust impact monitor & \citet{RosIns-SESAME} \\
         \hline
    \end{tabular}
    \caption{\Rosetta{} instruments on the orbiter (top) and lander (bottom).}
    \label{tab:rosetta-instruments}
\end{table*}

The instrument suite on \Rosetta{} was chosen, via a separate call for proposals after the mission was selected, to achieve as much of the original sample-return science as possible with \textit{in situ} measurements. This led to a large collection of 11 instruments on the orbiter and 10 on the lander (Table \ref{tab:rosetta-instruments}), with the latter in particular focused on local composition measurements and including a drill to sample the sub-surface. 
{Some of the most important results from the mission would come from composition measurements, particularly via the \mission{ROSINA} mass spectrometer that sampled neutral and ionised gas in the coma with high sensitivity and mass resolution \citep{RosIns-ROSINA}, while the \mission{COSIMA} instrument revealed both composition and structure of dust particles \citep{RosIns-COSIMA}.}
\Rosetta's instruments included a number of novel elements, including the first atomic force microscope in space (\mission{MIDAS}; \citealt{RosIns-MIDAS}), to measure dust grain structure at nanometer scales, and an experiment to use the communication transmission between the orbiter and lander to explore the nucleus interior structure (\mission{CONSERT}; \citealt{RosIns-CONSERT}). \Rosetta{} also carried remote sensing instruments covering ultraviolet through to microwave wavelength ranges, establishing a legacy that would see some of these instrument designs fly on other missions throughout the Solar System. The UV spectrograph \mission{Alice} \citep{RosIns-Alice} would form the basis of the \mission{New Horizons} instrument of the same name and the \mission{Lunar Reconnaissance Orbiter LAMP} instrument. The {visible and infrared spectrometer} \mission{VIRTIS} \citep{RosIns-VIRTIS} would also be flown on \mission{Dawn} and \mission{Venus Express}.  
The two cameras of \mission{OSIRIS}, the visible scientific imaging system \citep{RosIns-OSIRIS}, were equipped with 16-bit 2k x 2k CCD detectors and mechanical shutters, unique features that enabled the return of more than 80 thousand detailed images. 
Two of the instruments were actually made up of suites of independent sensors provided by different teams but operated via a common interface to the spacecraft: \mission{RPC} and \mission{SESAME}. The \mission{RPC} plasma package \citep{RosIns-RPC} was located on the orbiter and contained five sensors to study the comet's interaction with the solar wind (\mission{Ion Composition Analyser}, \mission{Ion and Electron Sensor},  \mission{Langmuir Probe}, \mission{Fluxgate Magnetometer}, and \mission{Mutual Impedance Probe}), while the \mission{SESAME} collection of sensors \citep{RosIns-SESAME} was on the lander (\mission{Cometary Acoustic Surface Sounding Experiment}, \mission{Permittivity Probe}, and \mission{Dust Impact Monitor}).

The selection of the \Rosetta{} payload after the mission and its primary science goals had already been confirmed had some key advantages, primarily in allowing an `optimized' collection that achieved the best fit to the science goals based on what ESA member states and the international partner agency, NASA, would support. 
However, this progression led to some problems, as requirements for different instruments were often mutually exclusive with each other or with spacecraft limitations. For example, \mission{RPC} measurements of the comet's interaction with the solar wind ideally needed to sample a wide range of distances from the comet, while \textit{in situ} instruments had to be as close to the nucleus as possible for sensitivity considerations. Also, there were frequent conflicts between remote sensing and \textit{in situ} payloads in pointing requirements. The ESA approach of independent instrument teams encouraged an (occasionally fraught) atmosphere of competition over spacecraft resources rather than cooperation. It was only relatively late in the mission that coordinated inter-instrument measurement campaigns were organized, and successful joint analyses of data from multiple instruments were carried out, e.g., in the analysis of serendipitous simultaneous observations of an outburst \citep{Gruen-outburst}, or in joint remote sensing and \textit{in situ} study of dust \citep{Guettler-dust}.

\footnotetext[1]{All \mission{OSIRIS} images shown are \copyright{} ESA/Rosetta/MPS for OSIRIS Team MPS/UPD/LAM/IAA/SSO/INTA/UPM/DASP/IDA.}


\subsection{Challenges for a comet rendezvous and solutions}
\label{sec:rosetta-tech}
As the first mission to rendezvous with, and make a soft landing on, a comet, \Rosetta{}  faced several challenges. In what follows we describe how \Rosetta{} coped with some of them. The treatment partially follows \citet{Kueppers2017}. 

\subsubsection{Matching a comet orbit}
Due to the high eccentricity of the orbit of 67P, matching the orbit of \Rosetta{} with that of the comet required 10 years of interplanetary cruise, including 4 planetary flybys as gravity assists (three at Earth and one at Mars: Table \ref{tab:rosetta-timeline}). The propellant required by \Rosetta, most of it for the rendezvous maneuvers at comet arrival, constituted more than half of the spacecraft mass at launch. 
The orbit matching also required a heliocentric distance range of 0.89 - 5.3 au between \Rosetta{} and the Sun. At low heliocentric distance, stringent attitude constraints, determined through in-flight tests, had to be applied to avoid overheating of thrusters and radiators. At large heliocentric distance, \Rosetta{}, with its 64 m$^2$ of solar panels, could only collect sufficient power to operate out to approximately 4.5 au, the largest heliocentric distance for a solar powered spacecraft at the time. For the 2.5 years spent outside 4.5 au, the spacecraft had to be commanded into hibernation, meaning that only essential elements, in particular heaters and thermostats, were kept on to keep all spacecraft and payload components within their designed temperature range. All other elements were switched off and needed to survive 2.5 years without activation or communication.
Power-driven constraints on simultaneous operation of payloads prevailed {from hibernation exit} until summer 2014 and again in the last weeks before \Rosetta{}  landed on 67P at a heliocentric distance of 3.83 au.  

\subsubsection{Rendezvous with a little known target}
When 67P was chosen as \Rosetta's new target in 2003, its orbit was certain but little else was known about it. An observing program was initiated to prepare for the \Rosetta{} mission implementation by collecting as much information as possible about 67P. Here we summarize the results obtained and compare them to the ground truth from \Rosetta.

\paragraph{Orbit and spin}
67P was discovered by Klim Ivanovich Churyumov and Svetlana Ivanova Gerasimenko in 1969. The comet was observed over 45 years between first detection and the arrival of \Rosetta{}, meaning that the orbit was known well enough that it could be used to define the spacecraft trajectory until the onboard cameras detected the comet and images taken by \Rosetta{} could be used for further trajectory refinements and navigation.

In a synthesis of data acquired between 2003 and 2007, \citet{67Ppole2} determined a rotation rate of 12.76137 $\pm$ 0.00006 hours. Somewhat surprisingly at the time, when \Rosetta{} approached 67P in 2014, several lightcurves taken of the unresolved nucleus resulted in a spin period of 12.4043 $\pm$ 0.0007 hours \citep{Mottola67Pspinup}. The difference was explained by a change in the rotation period during the perihelion passage in 2009. It was caused by the reaction torque from the non-radial component of cometary outgassing. This explanation was confirmed by monitoring the spin period of the comet during the perihelion passage in 2015, when the period was again observed to decrease by about 20 minutes \citep{Keller2015}. 

The spin axis direction of the comet as determined by \citet{67Ppole2} agreed with the solution obtained \textit{in situ} by \Rosetta{} to within approximately $10\degr$. No deviation from principal axis rotation was found in ground-based data.  \Rosetta{} detected an excited rotational state, but with a {very} small amplitude ($\approx$ 0.15$\degr$, \citealt{Jorda2016}).

The spin period of 67P was known well enough to plan for onboard navigation. The only adaptation was a change of the frequency of navigation slots to once every 5 hours, from 4 or 6 hours, to avoid the resonance with the spin period of 67P of $\approx$ 12 hours after perihelion. 

\begin{figure}[t]
\begin{center}
\includegraphics[width=\columnwidth]{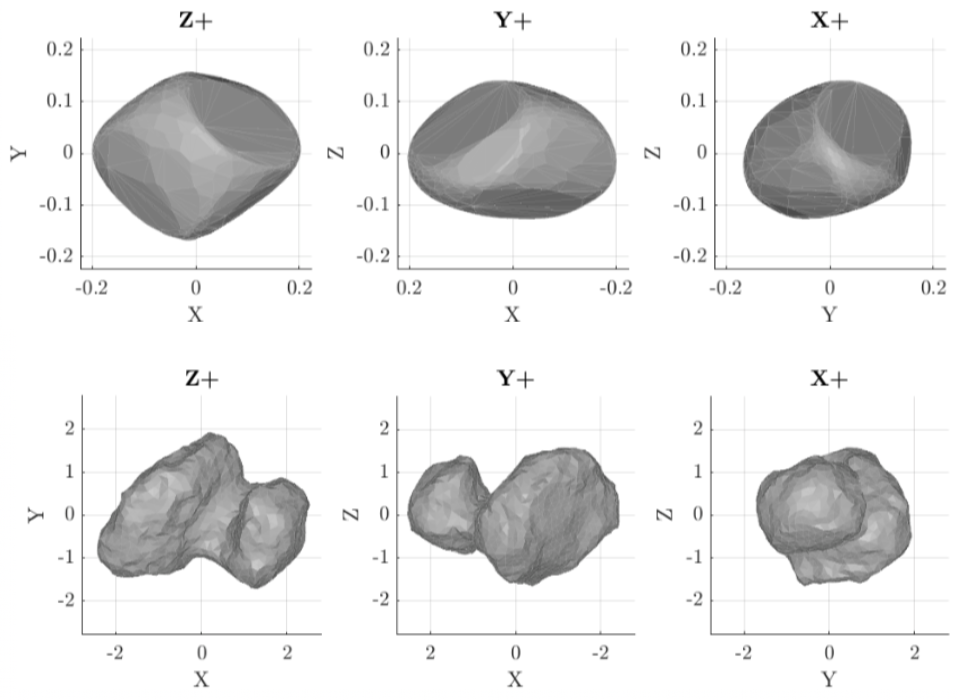}
\caption{Comparison between the shape of comet 67P  derived from ground-based observations (top, \citet{67Ppole2}) with a shape model determined by \Rosetta{} (bottom, \citet{Preusker2017})}
\label{fig:67P_shape}
\end{center}
\end{figure}

\paragraph{Size, Shape, Volume, Mass and Density}
Ground-based and \mission{Hubble Space Telescope} observations of 67P arrived at an effective radius of ~2 km and an albedo of 0.04-0.06 \citep{Lamy2006,67Ppole2}. Those values were later confirmed by the images acquired by \Rosetta.

The determination of small body shapes is difficult if only lightcurve data are available, in particular in the case of an object like 67P, as the technique is not sensitive to concavities. Cometary activity complicates the determination of a nucleus lightcurve close to the Sun, while the faintness of the nucleus and corresponding limitations in signal-to-noise, together with contamination of the data by any faint coma, complicate the analysis far from the Sun. Furthermore, 67P orbits the Sun close to the ecliptic, implying that lightcurves cannot cover the whole surface and the shape reconstruction is non-unique. Therefore it was not surprising that the shape models from Earth-based lightcurves did not resemble the very peculiar shape of 67P (see Fig. \ref{fig:67P_shape}), although the presence of large flat facets in these models can be taken as an indication of significant concavities \citep{Devogele2015}. The highly concave shape of the comet caused some initial problems for shape modeling from resolved \Rosetta{} imaging. However, the routines were adapted to reach accurate representations of the nucleus (e.g., \citet{Jorda2016, Preusker2017}).

The mass of 67P was estimated from the modification of the cometary orbit by non-gravitational forces (the reaction force to outgassing) by \citet{67Ppole1}. Their models provided nucleus masses in the range of 0.35 – 1.22 $\times$ 10$^{13}$\,kg. The measured mass of 67P gleaned from spacecraft data, 0.998 $\times$ 10$^{13}$\,kg \citep{Paetzold2016}, is within this range. However, even with the size of 67P being approximately known from ground-based observations, the volume of the rather regular shape models is larger than that of the highly concave, bilobed nucleus. Therefore \citet{67Ppole1} overestimated the volume of the nucleus, underestimating its density. Nevertheless, their correct determination of a mass range for 67P provides some confidence in applying this method to other comets.

\paragraph{Activity}
67P's activity profile turned out to be relatively repeatable from one perihelion passage to the next; the global dust and gas production could be {well} predicted  \citep{Snodgrass2013}. However, the possibility of predicting conditions in the innermost coma, where \Rosetta{} would spend most of its mission, was extremely limited due to the comparably low spatial resolution of Earth-based data. One valuable correct prediction from Earth-based data was that 67P would already be active at the time of \Rosetta{}'s arrival and the \Philae{} landing.

\subsubsection{Choosing a landing site and delivering a lander}
\label{sec:landing}

The relative velocity between \Rosetta{} and its target was successively reduced from 775\,m s$^{-1}$ to less than 1\,m s$^{-1}$ in a series of maneuvers taking place between May and August 2014. 
On August 6, 2014, \Rosetta{} was inserted into the first of a series of hyperbolic arcs around 67P, representing the first comet rendezvous. As the \Philae{} lander had to be delivered at a heliocentric distance not much below 3\,au, due to thermal reasons, this left only 3 months for nucleus and gravity field characterization, moving to closed orbits, global mapping, landing site selection, and planning of the \Philae{} delivery.

The characterization of the nucleus was achieved from hyperbolic orbit arcs, first at distances of 80-100\,km, then moving in to 50-70\,km. During this phase, five landing site candidates for \Philae{} were selected by August 24, 2014. The main information sources for landing site selection were images by \mission{OSIRIS} and the navigation cameras (for shape, surface roughness, etc.) and temperature information provided by \mission{MIRO} and \mission{VIRTIS}. Selection criteria included the scientific interest of the site and the following engineering criteria \citep{Ulamec2015}: 
\begin{itemize}
    \item Illumination of the landing site at a Sun angle $<$ 60$\degr$ from the time of landing until at least 40 minutes later.
    \item Impact velocity between 0.3 m s$^{-1}$ and 1.2 m s$^{-1}$.
    \item The vertical axis of the lander, lander velocity vector, and surface normal must be aligned within 30$\degr$, implying a constraint on average surface roughness. 
    \item Nominal separation velocity from the orbiter was commandable between 5 and 50 cm s$^{-1}$. A backup (emergency) release mechanism would eject the lander with a separation velocity of $\sim$18 cm s$^{-1}$. Preferred landing sites could be reached with a nominal separation velocity of $\sim$18 cm s$^{-1}$, so that the nominal descent trajectory would be approximately followed using the backup mechanism. Accordingly, a longer than necessary descent time was acceptable.
\end{itemize}

On September 9, \Rosetta{} entered a 30 km circular orbit around 67P, becoming the first comet orbiter. A few days later, the five landing sites were ranked and the prime and backup landing sites were determined. In subsequent weeks, \Rosetta{} moved to orbits of 20 km and 9 km distance from the comet, and, as no showstopper (e.g., very high surface roughness at dm - m scale) was found in the close observations, the landing site was confirmed on October 14, 2014. For the orbit changes during the mapping and close observation phases, the feasibility of getting closer was decided only a few days before the maneuvers were executed, requiring the maintenance of three different payload operations plans. More details of the operational aspects of landing preparation and landing are given in section \ref{sec:rosetta-lander} and by \citet{Accomazzo_landing} and \citet{Ashman_landing}.  

\subsubsection{Operating a spacecraft around an active comet}
  
Due to the low gravity and irregular shape of the nucleus, the gas drag from cometary outgassing, dust particles in the field of view of cameras resembling stars, and the changes of cometary activity over time, the spacecraft was exposed to an environment of limited predictability and strong variability. The resulting challenges for spacecraft operations were unprecedented.  
 
\begin{figure}[ht]
\begin{center}
\includegraphics[width=\columnwidth]{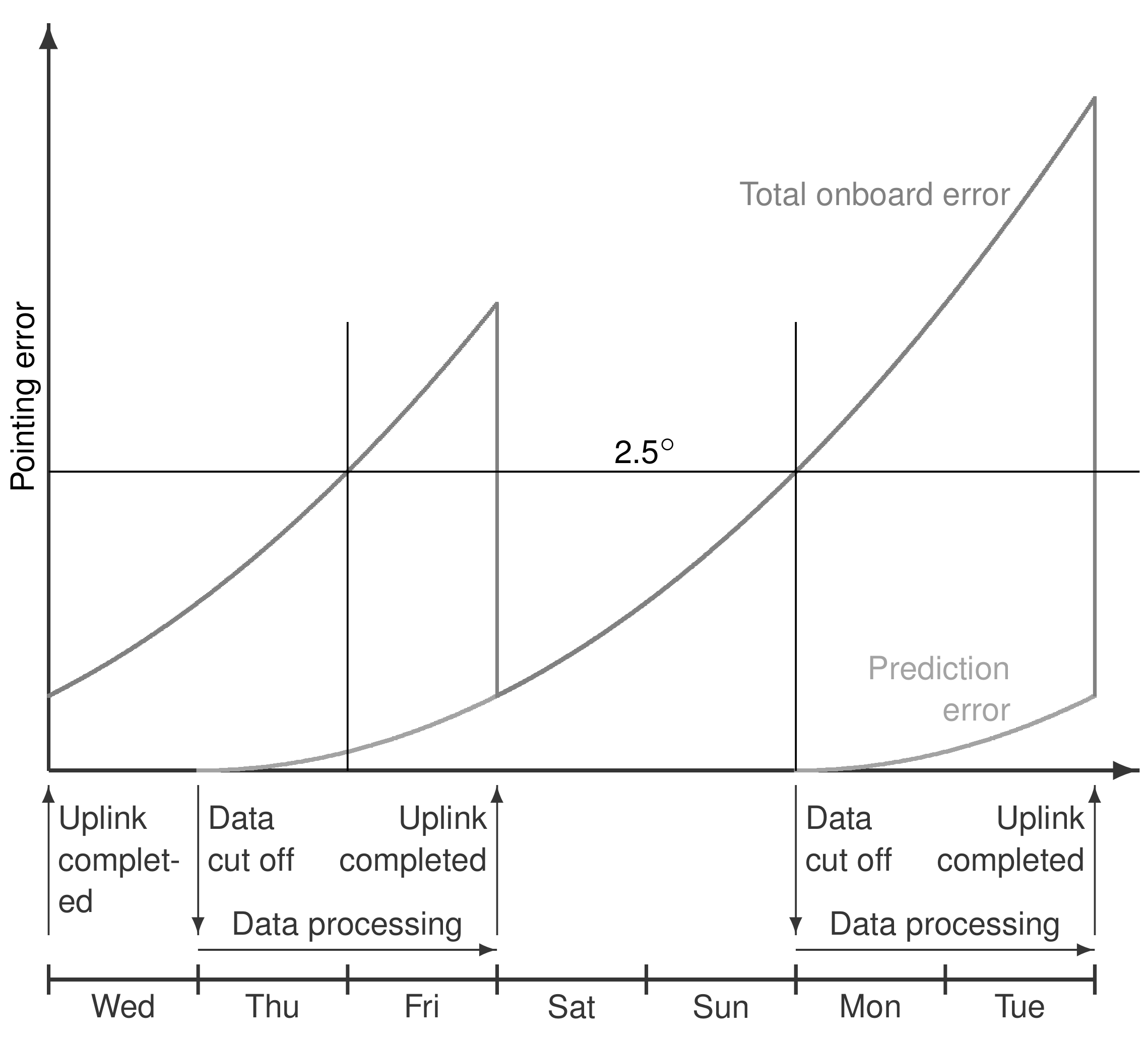}
\caption{The navigation cycle during the \Rosetta{} mission. Navigation images were continuously taken every 4-6 hours. All data up to a cutoff (usually 2 days before uplink of the new commands to the spacecraft) were used for the next navigation cycle. At the time of uplink, the pointing error was reduced to the prediction error from the last data cutoff. Then, the error propagated to the next data uplink. The requirement for a flight trajectory was that the pointing error at data cutoff was not larger than 2.5$\degr$. The plot shows the limiting case.  Source: Björn Grieger }
\label{fig:navigation}
\end{center}
\end{figure}

Before lander delivery, the \Rosetta{} mission focused largely on the preparations for landing. After \Philae’s first science sequence, the focus shifted to the needs of the orbiter instruments for scientific observations. This involved coordination of requests from 11 different instruments and adaptation to the operational constraints,
{such as the} limitation on trajectories while navigating the spacecraft around the comet. The gas drag  from the expanding cometary coma accelerated the spacecraft, and the amount of the drag (comet activity) was not fully predictable. In addition, inaccuracies of spacecraft maneuvers (maneuver error) contributed to an uncertainty in the motion of \Rosetta{} relative to 67P. The uncertainty of the position of \Rosetta{} translated into an error for spacecraft pointing  (see Fig. \ref{fig:navigation}). As navigation around the comet required regular images to be taken to confirm the relative position, a trajectory was defined to be possible if the pointing error did not exceed 2.5$\degr$ during navigation periods, {dictated by the field-of-view of the navigation cameras}.

The feasibility of a trajectory therefore depended on the short-term predictability (a few days ahead) of the gas drag on the spacecraft, and therefore on the predictability of cometary activity. The problem was attacked by defining two cases. A preferred case, a trajectory defined by the science team based on a best guess on the activity (and its predictability), and a high activity case, based on very conservative assumptions.
\Rosetta{} flew trajectories according to the preferred case trajectory, and if that trajectory at some point was not feasible in terms of navigation, the spacecraft was commanded to change to the high activity case trajectory. Notably, as most of the cross section of the spacecraft was its 64\,m$^2$ solar panels, gas drag on \Rosetta{} was minimized by staying close to the terminator, where the gas flow was directed toward the edge of the solar panel. This trajectory design was leveraged for a large part of the mission, although it limited viewing geometry and sampling locations relative to the nucleus and its illuminated regions.   

The remaining problem for the \Rosetta{} Science Working Team (SWT) was to define a preferred case activity level that allowed trajectories close to the  nucleus without violating navigational constraints. Observations from previous perihelion passages, measurements of the current conditions at 67P by \Rosetta{} instruments, and theoretical models were used to propose an activity level that was then agreed on by the SWT and used for trajectory design. A comprehensive summary of the science planning activities in this phase is given by \citet{Vallat2017}.

After the planning scheme had been executed for a few months, the concern was raised that the resulting trajectories were quite conservative in that \Rosetta{} was significantly more distant from 67P than necessary, as suggested by the gas pressure measured by the \mission{COmet Pressure Sensor} (\mission{COPS}), part of the \mission{ROSINA} instrument. Steps were prepared to improve the planning, but before implementation, events at \Rosetta{} enforced a more dramatic change.

The possibility of star trackers being confused between dust particles and stars, and thus not fully functional, was discussed in preparation for the \Rosetta{} mission, but not considered to be a major constraint. However, the number of individually visible dust particles in the coma of 67P was higher than expected. During a close flyby on February 14, 2015, the \Rosetta{} star trackers temporarily lost track due to such confusion. Operational measures were taken to mitigate the problem during the execution of a second close flyby on March 28, 2015. However, in spite of those measures, the star trackers again lost tracking over an extended period of time during that flyby, and while the spacecraft narrowly avoided a loss of attitude control, it ended up in safe mode. It became clear that neither the planned (preferred case) trajectory nor the high activity trajectory, designed to be robust against navigation error, were safe in this situation and thus could not be flown.

Prior to this safing event, \Rosetta{} had been flown with the typical planning scheme of {an ESA} planetary mission: in iterations between the Mission Operations Centre (MOC), the Science Ground Segment (SGS), and the PI teams, the trajectory was fixed in a long-term planning process 4-8 months in advance, the spacecraft pointing in a medium-term process 1-2 months in advance, and the details of resource (data volume, power) distribution and commanding were fixed 1-2 weeks in advance. As a consequence of the star tracker problems, the planning of the \Rosetta{} mission had to be redesigned. The new scheme was used for most of the remaining mission operations. Some of the main changes in the planning process were:
\begin{itemize}
    \item 	A large reduction in all turn-around times, so that planning could better consider the comet conditions.
    \item   The SWT and SGS requested trajectories from MOC that met certain criteria, rather than specific trajectory implementations.
    \item   The long-term trajectory was indicative only; the actual trajectory was defined 1 day before execution.
\end{itemize}

For the final $\sim$7 weeks of the mission, a still more demanding and less risk averse scheme was designed. To get \Rosetta{} even closer to the comet, elliptical orbits with a constant duration of 3 days were designed (corresponding to a  semi-major axis of 10.5 km), implying an even shorter turnaround. In order to support this effort, a pre-defined attitude profile was flown, thereby somewhat reducing the flexibility in pointing. Starting with 13 $\times$ 8 km ellipses, the pericenter distances were successively reduced, reaching less than 2 km, or one nucleus radius, from the surface. 

Overall, \Rosetta{} was operated successfully over its more than two-year rendezvous with 67P. However, some of the mission goals were not fully achieved. With the benefit of hindsight, what could be improved for the future?  
In reviews of the \Rosetta{} mission \citep{Thomas2019SSRv, grumpy-Keller}, the main science objective that was identified as not being fully achieved is understanding the mechanism of cometary activity. High resolution observations of the nucleus during the most active phase of 67P, around perihelion, would have been required to obtain the relevant data. The main obstacle to being close to the active nucleus was the attitude uncertainty, first due to maneuver error and uncertainties in the prediction of gas drag, then because of the star tracker failing when in the inner coma during high comet activity. While details of the operations scheme, chosen trajectories, and risk philosophy during \Rosetta{} operations may be debated, the most straightforward way to overcome those constraints is autonomous attitude navigation of the spacecraft. It would eliminate the build-up of `pointing error' as shown in Fig. \ref{fig:navigation} and therefore the main constraint preventing the spacecraft from getting closer to the nucleus. Even the star tracker failure, while unlikely to reoccur in a future mission, may have been at least partly overcome with such a scheme. The position of the comet plus that of the Sun would provide at least a rough attitude solution. While a closed loop between the attitude control system and the navigation cameras was available to keep the \Rosetta{} target asteroids in the field of view during the flybys, the algorithm was unfortunately not adequate for operations close to 67P, as it required the target object to underfill the field of view of the \mission{NAVCAM}s.

Another shortcoming pointed out in the reviews is that there was no systematic approach to revisit, within a short time-frame, locations where interesting events were detected. The example most frequently mentioned is the fracture in the Aswan cliff that resulted in a (predictable) collapse \citep{Pajola2017}. While most of those locations were observed repeatedly over the mission, in most cases those revisits were random during either comet centre pointing attitude or pointings designed for other objectives, as opposed to the outcome of a {specific targeted} observing strategy. This may have allowed several objectives to be met in a single go, but it was generally not optimal for following the development of changing features. The long lead-times in the early comet phase of \Rosetta{} were not ideal for such dedicated follow-up observations. In the later phases, the main issue was again the pointing uncertainty, which complicated targeted observations from close distances. Again, autonomous attitude guidance would have helped to increase the flexibility.  


\subsection{The \Philae{} lander}
\label{sec:rosetta-lander}

During the \Rosetta{} mission study, the original two lander concepts for \Rosetta{} were merged into the \Philae{} lander. As some of  the properties of the cometary surface, such as Young's modulus and strength, were unknown prior to \Rosetta, \Philae{} was designed against a large range of those properties, covering 3-4 orders of magnitude. This included several devices to keep it on the surface at touchdown \citep{Philae-summary}. During descent, the Active Descent System (ADS) was foreseen to fire thrusters in an upward direction to avoid rebound. Furthermore, anchoring harpoons were to be fired into the ground at landing. Finally, the feet of the lander contained screws to secure it on the surface after touch down.  The overall tensile and compressive strength of the surface of 67P turned out to be lower than the minimum assumed when \Philae{} was built, although analysis of the \Philae{} landing showed that the compressive strength at the original landing site Agilkia was within the \Philae{} design range \citep{Groussin2019,Biele2015,Moehlmann2018}. Furthermore, when the \Rosetta{} target was changed from 46P to 67P, an update to the \Philae{} landing gear was required to cope with the increased gravity of the new mission target, which is about 3 times larger and therefore $\approx$ 30 times more massive \citep{Ulamec2006}. 

\begin{figure}[t]
\begin{center}
\includegraphics[width=\columnwidth]{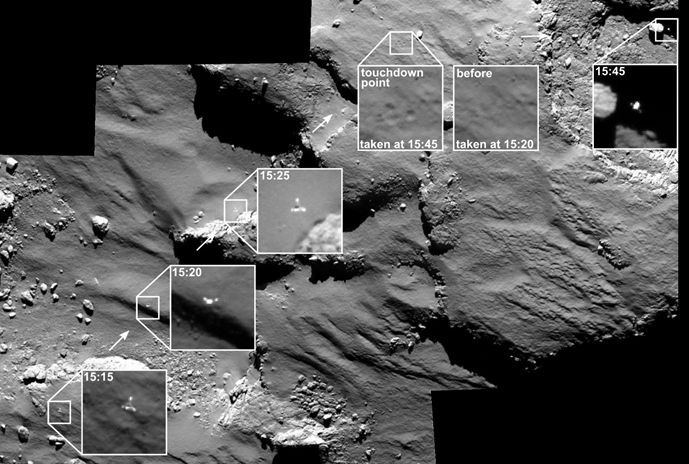}
\caption{Images of \Philae{} flying over the surface of the comet on its way to the touch down location on November 12, 2014. {Credit: OSIRIS\protect\footnotemark[1]{}.}}
\label{Fig:Philae_over_surface}
\end{center}
\end{figure}

\begin{figure}[t]
\begin{center}
\includegraphics[width=\columnwidth]{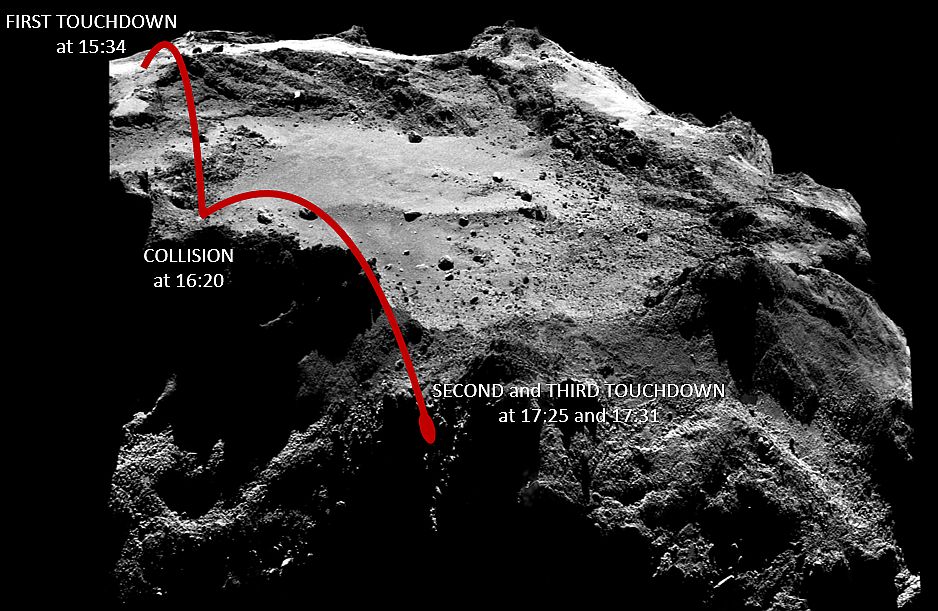}
\caption{\Philae{}'s flight to its final touch down. {Credit: OSIRIS\protect\footnotemark[1]{}.} }
\label{fig:Philae_touchdowns}
\end{center}
\end{figure}

On November 12, 2014 \Philae{} separated from \Rosetta{} at an altitude of 20.5 km and with a relative velocity of 19 cm~s$^{-1}$ \citep{Biele2015}. After 7 hours of ballistic descent, \Philae{} touched down on the Agilkia landing site at 15:34:03 UTC, only 120 m from the planned position, after having been imaged while flying over the surface of the comet (Fig. \ref{Fig:Philae_over_surface}). 
At touchdown, the anchor harpoons failed to fire and the hold-down thrust of the cold gas system did not work, leading to \Philae{} bouncing and rising again from the surface {before the screws could activate}. \Philae{} traveled across the nucleus for 2 hours, touching the surface twice before reaching its final resting position in Abydos, a shaded rocky area in the southern hemisphere  (Fig. \ref{fig:Philae_touchdowns}) at 17:31:17 UTC. It took almost two years to {successfully find and} undoubtedly image the lander in its final location.

\Philae{} continued communicating with \Rosetta{} during its entire flight.This provided an unexpected opportunity for some of the payload instruments to acquire data in two substantially different locations (Agilkia and Abydos), and the magnetometer \mission{ROMAP} to collect data during most of \Philae's voyage over the surface. The communication link was interrupted, for geometrical reasons, about 27 minutes after the final landing. It was re-established four more times until the primary batteries ran out of power on November 15, 2014 \citep{Biele2015}.
In addition to \mission{OSIRIS} and \mission{NAVCAM} images, \mission{CONSERT} played a key role in narrowing down the location of the final landing site to an area of 22 $\times$ 106 m$^2$ \citep{Herique2015}.

The originally planned science activities of the \Philae{} instruments had to be rapidly adapted to the illumination conditions of the final landing site and to the unknown attitude of the lander, but eventually they were all activated and collected invaluable \textit{in situ} data. Unfortunately, the lander attitude (`hanging' {sideways} at a cliff) did not allow sample collection and drilling, and the \mission{MUPUS} hammer failed to penetrate into the surface. {It is still debated whether this was because of a harder than expected surface layer, or because the hammer did not deploy as expected due to the final lander attitude \citep{Spohn2015,Heinisch2019}}.
The \mission{MUPUS} surface strength measurement should therefore be treated with caution, and may not be a useful constraint in future mission design; it is orders of magnitude higher than all other observational, experimental, and theoretical results (see \citealt{Groussin2019}, in particular their figure 10, and the chapter in this book by Guilbert-Lepoutre et al.).
Communication between \Rosetta{} and \Philae{} was briefly re-established on June 15, 2015, when \Rosetta{} flew over the Abydos region at a distance of 200 km. Communications were irregular and unstable and ended on July 9, 2015 \citep{2019ORourke}. The intermittent communications did not allow for a restart of science observations. Therefore, the \Philae{} long-term science, {expected to last weeks} based on power from rechargeable batteries, could not be recovered.  

\begin{figure}[t]
\begin{center}
\includegraphics[width=\columnwidth]{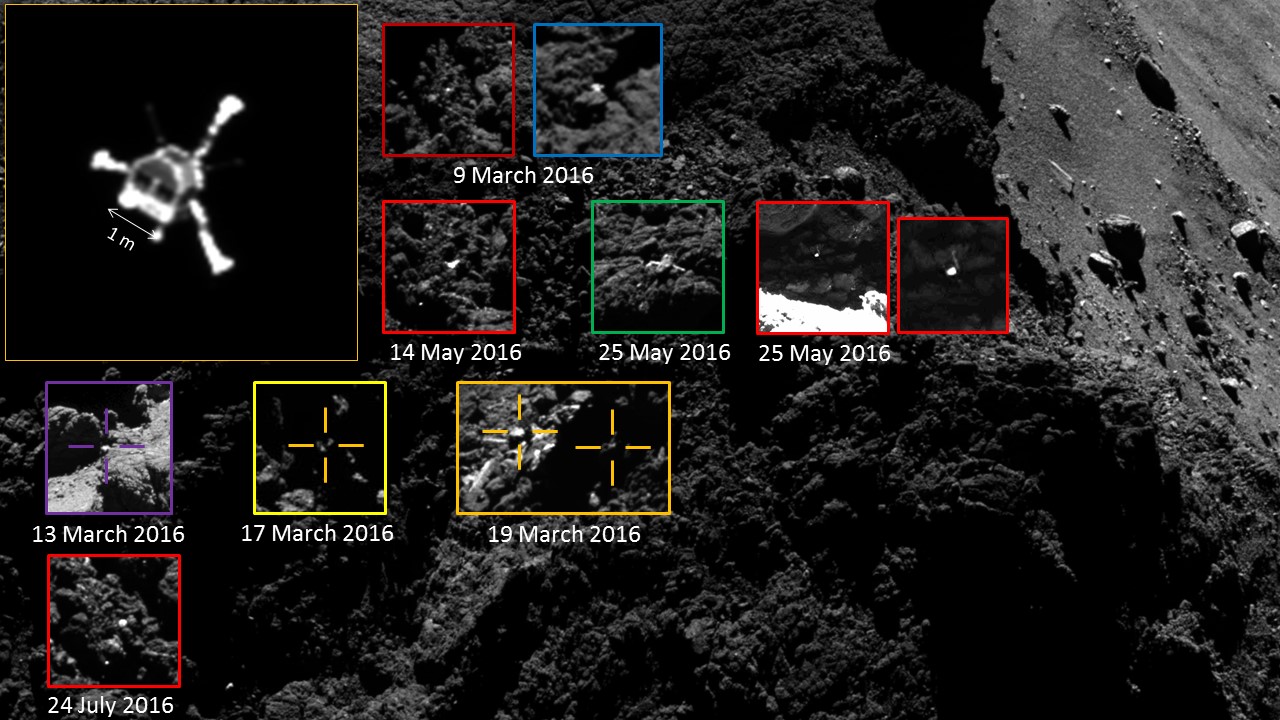}
\caption{Examples of \mission{OSIRIS} images acquired during the search campaign for \Philae{}, showing how difficult it was to unambiguously identify \Philae{} in the images. {Credit: OSIRIS\protect\footnotemark[1]{}.} }
\label{Fig:Philae_search_campaign}
\end{center}
\end{figure}

From March 2016 the \Rosetta-comet distance was regularly below 20 km, after several months spent at large distances  due to the high perihelion activity levels (Fig. \ref{fig:rosetta-distance}), and the lander search restarted. 28 dedicated {observing campaigns} were carried out. 
A major difficulty was to be able to unambiguously identify the `bright dot' in the \mission{OSIRIS} images as \Philae{} (Fig. \ref{Fig:Philae_search_campaign}). \Rosetta{} needed to be at a distance $<$ 5 km from the comet for \Philae{} to be at least 10 pixels {in diameter} in \mission{OSIRIS} images. 

\begin{figure}[ht]
\begin{center}
\includegraphics[width=\columnwidth]{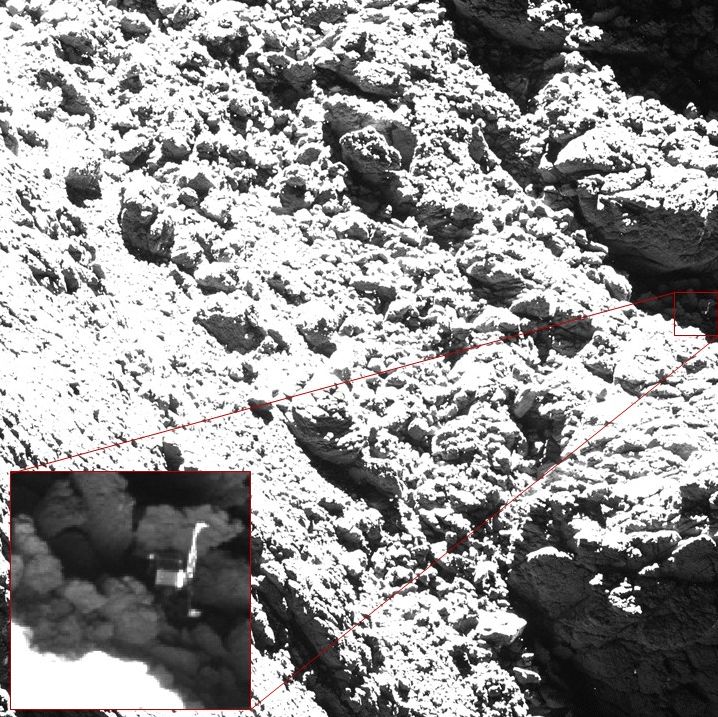}
\caption{\Philae{} in its final resting position in Abydos in an \mission{OSIRIS} image acquired on September 2, 2016, at 19:59 UTC. The full field of view is approximately 100 m across. {Credit: OSIRIS\protect\footnotemark[1]{}.}}
\label{fig:Philae_found}
\end{center}
\end{figure}

Despite the many challenges (e.g., limited observation slots at small enough distance, no pointing flexibility, non-optimal viewing geometries and illumination conditions, large pointing error, small projected fields of view) \Philae{} was successfully and unmistakably imaged on September 2, 2016, at 19:59 UTC from a distance of 2.7 km from the surface.
Imaging \Philae{} in its final resting position (Fig. \ref{fig:Philae_found}) was important for several reasons: it provided context for the measurements performed by the \Philae{} instruments; it confirmed the location identified by \mission{CONSERT} and allowed its measurements to be properly interpreted in terms of the nucleus interior \citep{Kofman2020}; and it brought to a close the story, begun almost two years before, of the first spacecraft to land on a comet.
The landing of \Philae{} marked a milestone in  space history.

\begin{figure}[t]
\begin{center}
\includegraphics[width=\columnwidth]{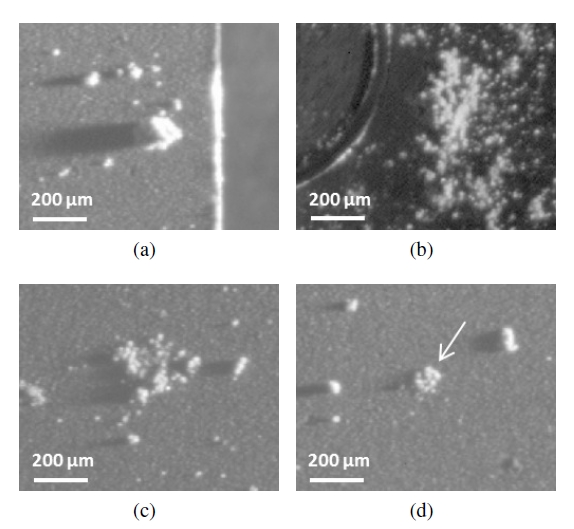}
\caption{{Selection of dust particles imaged by the \Rosetta/\mission{COSIMA}. From \citet{Merouane2016}.}  
}
\label{Fig:cosima-dust-image}
\end{center}
\end{figure}


\subsection{{Results and} open questions for the future}
\label{sec:rosetta-future}

The most important results from \Rosetta{} were, arguably, those on the composition of the comet, including the discovery of abundant O$_2$ and its correlation with H$_2$O \citep{Bieler2015,Keeney_O2,Luspaykuti_O2}, complex organics (including the amino acid glycine; \citealt{Altwegg2016}), salts \citep{Altwegg2020}, noble gasses \citep{Rubin2018}, and the role of electron impact dissociation in coma chemistry \citep{Feldman2015}. The \mission{ROSINA} and \mission{COSIMA} instruments were also able to probe the comet's composition at an isotopic level, with implications for comet formation (including incorporation of pre-solar material), activity models, and delivery of various species to the early Earth \citep[e.g.,][]{Altwegg2015,Marty2017,Hoppe2018}.  
The remote sensing instruments revealed, in exquisite detail, the morphology and composition of the nucleus, and changes in these with time. Meanwhile microscopic imaging of coma dust grains showed their structure and fragility on nm-$\mu$m scales (Fig. \ref{Fig:cosima-dust-image}; \citealt{Guettler-dust}). 

\Rosetta's contribution to cometary science is undoubted; {in addition to the highlights summarised here (and  listed in Table \ref{tab:goals_tech_results}),} results from the mission are quoted in nearly every chapter in this book. 
 However, \Rosetta{}  is also not the final word in the field; some important questions remain, and some new ones that  arose following \Rosetta{}'s findings. 
\citet{Ahearn_after_Rosetta} and \citet{Thomas2019SSRv} both consider what \Rosetta{} achieved, and what future missions could do to answer remaining questions.
The highly critical review by \citet{grumpy-Keller} also raises some valid points; the most important being that our understanding of the processes driving cometary activity is still far from complete. Since \Rosetta{} followed a comet from approximately the onset of activity through perihelion and then outbound again, this does feel like a missed opportunity. 

A key piece of missing information appears to be the temperature of the (near-)surface, which was constrained  by near-IR and microwave observations by \mission{VIRTIS} and \mission{MIRO}, respectively, and locally from \Philae{} measurements \citep[][and references therein]{Groussin2019}, but would have been most effectively mapped by thermal IR (e.g., 5-20 $\mu$m) imaging at the equilibrium surface temperatures expected at 1-3 au from the Sun. Such an instrument was proposed for \Rosetta, but not selected for budget reasons \citep{Thomas2019SSRv}. The limited measurements from \Philae{}, which should have provided details on the microscopic structure of the surface and range of temperatures in the immediate sub-surface layers, also mean that there are still significant assumptions in the interpretation of the existing remote sensing data and uncertainties in where sub-surface ice layers may exist (see the chapter by Guilbert-Lepoutre et al. in this volume). 

In addition, the navigation issues described in section \ref{sec:rosetta-tech} meant that \Rosetta{} was not at its closest to the nucleus during the more active phases, and therefore missed opportunities to directly sample the acceleration region \textit{in situ}, and resolve cm-scale surface changes or lifted particles, when these effects were most significant. Improved onboard autonomy, which would allow safe spacecraft operation for extended periods at closer distances, as well as the ability to react and sample certain phenomena with a real-time `joystick' control options, seems a necessary technology for future comet rendezvous missions.

 The fine-scale structure and composition of the nucleus surface remains a key missing piece of information, relevant not only for understanding activity, but also the formation processes of comets, and for interpreting unresolved data (e.g., photometric or spectroscopic behavior seen from ground-based telescopes). Remote sensing can only give broad assessments of nucleus composition (e.g., that it contains organics; \citealt{Capaccioni2015}), and \Philae's composition measurements were unable to give any further detail to compare with measurements of dust particle composition in the inner coma (from \mission{COSIMA} measurements; \citealt{Hilchenbach2016}). One key question that remains the topic of energetic debate is what the volatile-to-refractory ratio is, both within particles lifted into the coma, and originally within the nucleus \citep[see][for a review]{Choukroun2020}. While models of comet formation that use coma measurements of this parameter have been proposed \citep{Blum-model}, the lack of surface (or ideally sub-surface) measurements limits our current knowledge. As well as how much ice is present, we would still like to know what and where it is, both locally and globally; e.g., changes with depth from the surface due to heating, or on the scale of the whole nucleus to test ideas of primordial (in)homogeneity.  \mission{CONSERT} could only operate for a short time and the detailed interior structure of 67P remains largely unknown \citep{Kofman2020}. Ground-penetrating radar, operating from an orbiter without the need for a lander, could enable significant advances in future comet (or asteroid) missions \citep{CORE,Herique2018}.

Finally, one of the surprising results from \Rosetta{} was the revelation that much more dust falls back to the nucleus than was previously thought \citep{Thomas2015-fallback,Keller2017,Hu2017-fallback}. Together with the array of surface morphology features seen on 67P (e.g., pits, terraces, fractures; \citealt{Thomas2015}), it has become clear that the nucleus surface has been significantly eroded or covered by cometary activity. To access {more} pristine material, and really test planet formation models, will require either reaching sub-surface layers (possibly at considerable depth), or visiting a comet that has not had previous close perihelion passes, and whose surface remains nearly unaltered. 


\section{\textbf{FUTURE COMET MISSIONS}}
\label{sec:future}

\subsection{DESTINY$^{+}$}
\label{sec:destinyplus}

We include the {Japanese Aerospace Exploration Agency's (JAXA)} \mission{DESTINY$^{+}$} (Demonstration and Experiment of Space Technology for INterplanetary voYage with Phaethon fLyby and dUst Science; \citealt{Ozaki2022}) as a future comet mission as its primary target, near-Earth asteroid (3200) Phaethon, has been demonstrated to have repeated activity at perihelion and is responsible for the Geminid meteor stream \citep{Jewitt2010-Phaethon,Jewitt2013-Phaethon,Li2013-Phaethon}. Phaethon has been dubbed a `rock comet' as its orbit is such that water ice is not expected to survive, even in the interior, but at perihelion (0.14 au) it is close enough to the Sun for thermal fracturing and electrostatic lifting of rock to form a dust coma without ice sublimation \citep{Jewitt2010-Phaethon}. \mission{DESTINY$^{+}$} is primarily a technology demonstration mission that will show that a medium-sized ($\sim500$ kg wet mass) satellite launched into low Earth orbit by a modest rocket can perform interplanetary missions, through patient use of ion engines to slowly raise the orbit to reach a lunar flyby, and then escape from the Earth-Moon system through a lunar gravity assist maneuver. It promises low-cost deep-space exploration via the enabling technologies of ion engines; large and very lightweight solar arrays; and advanced, compact, thermal and attitude control systems. It has a lot of flexibility in its trajectory and launch window \citep{Sarli2018}. \mission{DESTINY$^{+}$} is currently expected to launch in 2024 and perform a fast (30 - 40 km s$^{-1}$) flyby of Phaethon in 2028 \citep{Ozaki2022}. Its primary science goal is to measure dust composition; both interplanetary / interstellar dust during the long cruise, and dust released from Phaethon. It will have an instrument based on the \mission{Cassini Cosmic Dust Analyzer}, which relies on   high-speed destructive particle impact with the back of the instrument, and mass spectroscopy of the resulting plasma \citep{CDA}, similar to \mission{CIDA} on \Stardust. The mission will also carry narrow-angle panchromatic and wide-angle multi-color cameras to image the nucleus from an expected minimum distance of 500 km \citep{Ishibashi2020}.


\subsection{ZhengHe}
\label{sec:ZhengHe}

The coming decade {should} also see the first {dedicated small bodies mission from China,} expected to be called \ZhengHe, although the name and final approval for the mission remain to be confirmed at the time of writing \citep{ZhengHe1,ZhengHe2}. \ZhengHe{} is ambitious and will demonstrate a wide range of technologies by first performing an autonomous touch-and-go surface sampling of a small near-Earth asteroid (2016 HO$_3$) {in the late 2020s}, returning the sample to Earth, and then redirecting the spacecraft to rendezvous with 311P/Pan-STARRS (hereafter 311P), {an active asteroid.} This is an interesting object that showed multiple tails at its discovery epoch in 2013, but has not yet shown the repeated activity from orbit-to-orbit that is associated with sublimation of ices and therefore marks main belt objects as true comets (see the chapter by Jewitt \& Hsieh in this volume). In any case, \ZhengHe{} is expected to arrive at 311P in 2034 and operate there for a year \citep{ZhengHe2}, covering the approach to aphelion phase of the orbit, and not the true anomaly range in which activity was previously reported. The mission may therefore not encounter an active object, but in any case should provide detailed characterization of the nucleus, and help solve the puzzle of its strange activity pattern, which had been attributed to rotational disruption until lightcurve observations revealing a relatively slow rotation and possible binarity \citep{Jewitt2018}. The \ZhengHe{} payload will include cameras and spectroscopic capability covering the ultraviolet to thermal infrared range, radar (possibly with the capability to detect sub-surface ice), a magnetometer, and \textit{in situ} analysis of any gas \citep{ZhengHe2}. 
{A proposed dust analysis instrument for \ZhengHe{} presents some novel aspects; if flown and the mission encounters an active body - it will combine momentum and size measurements via piezoelectric microbalances and a laser curtain, as used by \mission{GIADA} on \Rosetta, with microscopic imaging and a first-order assessment of composition (volatile vs refractory fraction) by heating (evaporating) collected particles and measuring the resulting change in mass \citep{ZHAO2022}.} A sub-surface penetrator  has also been proposed for this mission (\citealt{ZhengHe-penetrator}; see section \ref{sec:penetrators}).


\subsection{Comet Interceptor}
\label{sec:comet-interceptor}

The primary goal of the ESA-led \CI{} mission is to visit a `pristine' comet \citep{Snodgrass+Jones-CI}. Due to launch in 2029, this will be the first of ESA's F-class missions, relatively short development time (for ESA) programs. Despite an order-of-magnitude smaller budget, and wet launch mass 
less than $\sim1000$ kg, \CI{} will go beyond what ESA achieved with \Giotto{} and \Rosetta{} in some ways, such as a first multi-point \textit{in situ} investigation of a cometary coma and by including thermal IR and more extensive
polarimetric imaging.

\CI{} will be a flyby mission, targeting a population that has yet to be visited: LPCs with a source region in the Oort cloud. If possible, it will target a dynamically new LPC (i.e., one that hasn't previously had a perihelion in the inner Solar System), but in all cases the target comet will have experienced very little insolation-driven evolution on past perihelion passages, and is expected to have a relatively pristine surface with little or no devolatized layer. The main challenge associated with visiting a LPC is that the mission cannot be designed around a well characterized target on a known orbit. Suitable LPCs have orbital periods of many thousands of years and are therefore only discovered months or years before their perihelion. The \CI{} concept involves launch before the target is known. \mission{ISEE-3/ICE} had, arguably, demonstrated this concept already, albeit with a spacecraft not specifically designed for comet characterization. \mission{ISEE-3} had operated at the Sun-Earth L1 for several years before being directed to 21P and then upstream of 1P.
{\citet{PrOVE2} have proposed a similar `waypoint' approach for a CubeSat-sized comet mission.}
\CI{} will  wait in a halo orbit around the Sun-Earth L2 point, a favored location for modern space telescopes, and will depart from there to meet its chosen comet once a suitable one is found and a transfer orbit calculated. The number of possible targets will depend on the capability of the spacecraft (specifically, the maximum possible change of velocity, $\Delta v$, achievable by its propulsion system and available fuel), the length of time that the mission is prepared to wait, and the amount of warning time available for optimized transfer trajectories. Simulations suggest a high probability of a suitable target being reached within the nominal 6-year lifetime of the mission \citep{Pau-CI}. Should no LPC target be found within an acceptable mission lifetime, \CI{} will instead be sent to encounter one of 8 possible backup targets (known JFCs; \citealt{Schwamb-CI}), which will not achieve the primary mission goal of exploring a pristine surface, but are interesting in their own right (including comets thought to be at the end of their lifetimes, and either almost dormant -- 289P/Blanpain -- or disintegrating via multiple fragmentation events -- 73P). Calculations have also been made that suggest \CI{} will have a low (but non-zero; $\sim 5\%$) chance of encountering an interstellar comet (\citealt{Hoover2021}; see section \ref{sec:future-target-types}). 

Aside from its new target type and launch-before-discovery approach, \CI{} will be novel as the first multi-point exploration of a coma, with three separate spacecraft: a `mothership' and two small-sat probes, one of which will be provided by JAXA. The probes (currently designated B1 and B2) will be released from the main spacecraft (A) $\sim$ 24  hours before closest approach, and will pass closer to the nucleus than spacecraft A, which will perform a relatively distant flyby at around 1000 km closest approach distance. This approach allows for sampling the coma on three different trajectories, and also combines a low-risk flyby for A with `high-risk/high-reward' measurements from the B-probes, which are designed to be short-lived and expendable, permitting a less risk-averse approach than is normally found in spacecraft operations. All three spacecraft will carry magnetometers to investigate the 3-dimensional structure of the comet's interaction with the solar wind, and cameras to obtain different viewing geometries of the nucleus and inner coma, while there will be dust detectors on both ESA-provided spacecraft (A and B2) and ion and/or neutral mass spectrometers for \textit{in situ} composition measurement on A and  B1. These multi-point measurements will be complemented by a high-resolution visible color camera, infrared imaging/spectroscopy in the 1-25 $\mu$m range, and fields and energetic particle measurements on spacecraft A; wide- and narrow-angle visible and far UV (Ly-$\alpha$) cameras on B1; and all-sky polarimetry and a forward looking visible camera on B2.  It is expected that the highest resolution images ($\sim$ 10 m pixel$^{-1}$) will come from spacecraft A, despite it being at the largest distance at closest approach, as the  mothership will carry larger instruments, including the main science camera, \mission{CoCa}, which is based on the \mission{CaSSIS} camera onboard the ESA Mars orbiter \mission{Trace Gas Orbiter} \citep{CaSSIS}. The probes instead rely on advances in instrument miniaturization from the small-sat and CubeSat developments of recent years, particularly on the JAXA B1 probe, which reuses the UV camera developed for the \mission{PROCYON} small-sat technology demonstration mission \citep{LAICA}.

\CI{} will encounter its comet near 1 au from the Sun and as it crosses the ecliptic plane; for both fuel and thermal design reasons, the spacecraft is not designed to travel very far from Earth's orbit. Since LPCs can have high inclinations (including retrograde orbits) and will be near their perihelion at the encounter, the relative velocity of the flyby could be high (up to 70 km s$^{-1}$), providing a number of technical challenges that the spacecraft design must meet, including adequate shielding and maintaining the nucleus in the fields of view of the various cameras. Spacecraft A is expected to employ a rotating mirror to track the nucleus while keeping shields towards the dust flow, as was done by \Stardust. Probe B2 will be spin-stabilized, akin to a mini-\Giotto, which it will take advantage of to scan the full sky as it rotates, with the \mission{EnVisS} polarimeter looking out of one side. There is expected to be limited communication with Earth during the flyby itself, depending on the geometry of the encounter. Data from all three spacecraft will be stored aboard A and transmitted in the weeks or months after encounter.


\subsection{Future concepts}
\label{sec:future-concepts}

\begin{figure}[t] 
   \centering
   \includegraphics[width=\columnwidth]{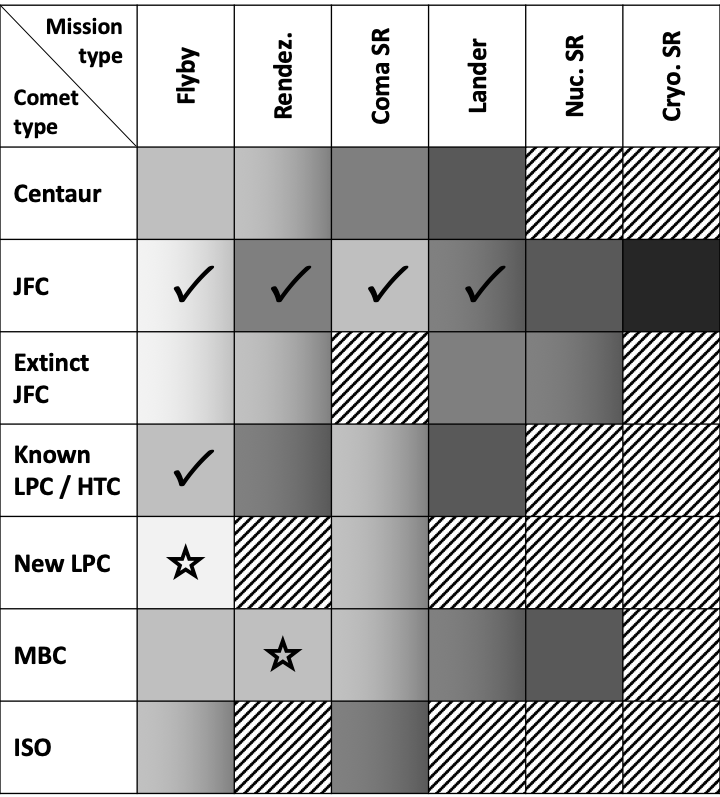} 
   \caption{
   Grid of mission type and comet class, indicating the approximate level of mission (cost/difficulty) needed to achieve the combination, from lighter shades (cheaper/easier)
   to darker (most challenging), {including gradients between shades where a variety of approaches or targets give a wide range of possibilities}. Hatched squares indicate that such a combination (e.g., nucleus sample return [SR] from an interstellar object) seems unfeasible with current or near future technology, or within the cost caps of the largest current interplanetary missions.
   Checks mark those combinations already tried (various JFC and Halley flybys described in section \ref{sec:flyby}, JFC coma sample return by \Stardust, and JFC rendezvous and landing by \Rosetta{} and \Philae) and {stars} indicate those selected for future missions (flyby of a new LPC and rendezvous with a MBC, by \CI{} and \ZhengHe, respectively). Based on a figure from \citet{Ambition}. }
   \label{fig:mission-types}
\end{figure}

Looking further into the future, there are two main routes to advancing the spacecraft exploration of comets; visiting different types of comets and performing different types of experiments, enabled by advances in spacecraft technology. Such missions will vary in cost and size depending on how challenging the targets are (mostly governed by the $\Delta v$ required to reach them) and the complexity of the spacecraft and instruments, progressing from `simple' single-spacecraft flybys to the significant challenges of cryogenic sample return of nucleus ices. The latter remains a distant prospect but is widely regarded as the holy grail of comet exploration. Both the \mission{Ambition} concept white paper presented to ESA's \mission{Voyage 2050} call \citep{Ambition} and the \mission{Cryogenic Comet Sample Return} white paper presented to NASA's 2023-2032 Planetary Science and Astrobiology Decadal Survey call \citep{Westphal_cryo} explored the current state of the field and the technological steps needed to reach cryogenic sample return (see section \ref{sec:future-mission-types}). The \mission{Ambition} paper also considered a wide range of potential mission and target types, identifying opportunities for different scales of mission that may be proposed in upcoming calls. In addition, over the past decade, NASA has funded two cometary Phase A concept studies, one in the Discovery class (\mission{Comet Hopper}; \citealt{CHopper}) and one in the New Frontiers class (\mission{CAESAR}; \citealt{CAESAR}), emphasizing the key role that comets play in planetary exploration and providing insight into the origin of life. Figure \ref{fig:mission-types} presents a grid of comet type and mission class, listing the past or planned missions described earlier in this chapter, and showing where the most promising future opportunities are. Some specific concepts in these areas are described in the following subsections.
{We note that the list of future missions we present is incomplete -- there will always be concepts studied and proposed that we are not aware of, and we do not try to list all missions, but rather a selection covering the range of possibilities. We mention proposed missions by name where there are published descriptions in the literature, but note that there are very different norms regarding advertising competition-sensitive details of pre-selected mission concepts in different communities, due to differences in the selection processes of the various space agencies (e.g., NASA vs ESA), and related commercial sensitivities. Inclusion or exclusion of any particular mission concept should not be read as endorsement or criticism.}

\subsection{Target types}
\label{sec:future-target-types}

\subsubsection{Jupiter Family Comets and Centaurs}

Of the two major comet classes, JFCs with orbits near the ecliptic and a source region in the Kuiper belt and scattered disc, and nearly-isotropic LPCs or Halley-types (HTCs) from the Oort cloud\footnote[2]{The distinction between LPCs and HTCs is on orbital period (longer or shorter than 200 years, respectively), but from a mission point of view they present similar challenges (potentially high inclinations), so we group them together for the purposes of this chapter. We note that HTCs may not have an Oort cloud source after all \citep{Levison2006,Fernandez2016}; see chapters by Kaib \& Volk and Fraser et al. in this volume.}, the former present much easier spacecraft targets. This is due to their typically low inclinations and short periods, with aphelia typically within the orbit of Jupiter. As demonstrated by \Rosetta, it is feasible with current launchers to match these orbits and, with modern solar panels such as those used by \mission{Juno} or \mission{Lucy} {(\citealt{Juno_spacecraft}; \citealt{Lucy_spacecraft}; respectively)}, perform scientific operations all the way to aphelion. JFCs have well characterized orbits and all classes of mission can be planned well in advance. This is reflected in the dominance of these comets in the list of previous missions. However, there remain interesting cases for future missions to JFCs: different types of mission (see section \ref{sec:future-mission-types}), or exploring the long-term evolution of these bodies by visiting split or fragmented comets or related populations, e.g.,  Centaurs or  extinct comets. 

Centaurs are thought to be precursors to JFCs as objects transitioning from outer Solar System orbits. They have a wide range of orbits and sizes (see chapter by Fraser et al. in this volume.), and consequently vary considerably in how feasible a mission to one would be. Feasible medium-sized missions have been proposed \citep[e.g., ][]{chimera} for flyby or rendezvous with 29P/Schwassmann-Wachmann 1, which has a relatively circular orbit just beyond that of Jupiter, and is particularly interesting as it is likely in the process of transitioning from a Centaur to a JFC \citep{29P}. More distant Centaurs, e.g., the large ($\sim$200 km diameter) ringed worlds Chariklo or Chiron, with  orbits between 8 and 18 au from the Sun, have been proposed as mission targets as more accessible Kuiper Belt Objects, rather than due to their relationship with comets as such, but would be more costly: e.g., NASA Frontiers class for a \DI-style flyby and impactor combination \citep{chariklo-mission} {or multi-Centaur tour \citep{centaurus}}. 

Comets at the other end of their active lifetimes can present more accessible targets. A large number of asteroids in cometary orbits (ACOs) have been identified, including many in near-Earth space,  some of which may have low $\Delta v$ transfers for flyby, rendezvous, or even sample return (see chapter by Jewitt \& Hsieh in this volume).
As these objects are thought to be dormant or extinct comets, with their near-surface ices either exhausted or buried at sufficient depth to extinguish activity, they offer an opportunity to conduct a \emph{relatively} easy comet mission, without the concerns for spacecraft safety caused by activity, but a scientifically compelling one, to understand what it is that causes comet activity to cease. However, the most significant uncertainty with such a proposal is in being sure of the identification of any individual ACO as definitely a previously active comet, rather than a body with an asteroidal origin but somehow scattered onto a comet-like orbit.

\subsubsection{Long Period Comets}

As noted in section \ref{sec:comet-interceptor}, LPCs {or HTCs} are much more challenging targets to reach, as they can have high inclinations (including retrograde orbits) and relatively high speeds relative to Earth and any spacecraft as they pass through the inner Solar System. For this reason, only flyby missions to LPCs seem reasonably feasible in the short term. The Halley missions took advantage of a long-predicted return of a rare well known comet in this class, while \CI{} has the chance to visit a new comet by implementing the use of a waypoint.
Further sampling the diversity of LPCs may be rewarding, especially given the recent discoveries of activity at large heliocentric distances well beyond the water snow line (e.g., C/2017 K2; \citet{MeechK2}) and low activity or inactive `Manx' or `Damocloid' objects in LPC orbits (see chapter by Jewitt \& Hsieh in this volume). Discussions of a return to 1P at its 2061 perihelion, to rendezvous or even return a sample, highlight the current need for significant technology development to achieve this step \citep{Halley2061}. While another mission at the next return of this most famous comet has  strong cultural and historic appeal, and its predictability decades in advance inspires some of the long-term planning needed to develop the necessary technology, there are perhaps easier LPCs/HTCs to consider. Although it comes close to Earth, 1P's highly retrograde orbit ($i = 162\degr$) will always mean very high energy requirements to match its velocity.
{A concept was investigated to address this issue for 1P's 1986 apparition through the use of a solar sail \citep{friedman1976}; this technology deserves to be considered further.}

\subsubsection{Main Belt Comets}

In the decades since the last comet mission was launched, it has become apparent that there may be a third class of comet orbit, and source reservoir, within the Solar System: the Main Belt Comets (MBCs), which suggest an unexpected significant population of ice-rich bodies in the outer asteroid belt (see chapter by Jewitt \& Hsieh in this volume). With very low activity levels and relatively circular orbits, MBCs are reasonably accessible for rendezvous missions, particularly for missions propelled by ion engines or other low thrust technology, as employed by the \mission{Dawn} mission to Vesta and Ceres \citep{Dawn}. As described in section \ref{sec:ZhengHe}, an upcoming Chinese mission may make the first visit to a MBC as a secondary target, {but may not encounter an active body}. A number of other missions have been proposed to other agencies, but not yet selected. {As a potential guide to future proposals in this area, we identify some of the approaches and technologies proposed in those with publicly available information. These include, for example,} the NASA Discovery class \mission{Proteus} rendezvous mission concept \citep{Proteus}; the \mission{Castalia} rendezvous concept proposed as an ESA medium-sized (M-class) mission and described in detail by \citet{Castalia}; and 
a \Stardust-like coma sample return mission, \mission{Caroline}, which was also previously proposed as an ESA M-class \citep{Caroline}. The main science drivers for most MBC missions rely on \textit{in situ} sampling of their volatiles, to confirm that MBC activity is indeed driven by sublimating water ice, which has not yet been possible via remote observation \citep{Snodgrass-MBCs}, and to reveal their composition and thus test Solar System formation and evolution models, and whether or not the asteroid belt could have been a significant source of Earth's water. The very low activity levels of MBCs make this challenging -- to build up the necessary signal-to-noise to make a sensitive detection via mass spectroscopy, \mission{Castalia} proposed extended periods of automated close proximity station keeping. This would have been a significant technological advance from \Rosetta, where navigation with ground control in the loop meant reaction times of a few days, but automated on-board navigation is now a mature spacecraft technology, to the point of enabling precision landings on asteroids or Mars \citep{auto-landing}. MBCs also differ from other comets in that their ices may be buried deeper under a refractory crust, or exposed in only a small and recently uncovered active area, and locating volatiles must {constitute} an important part of the mission. Surface (or sub-surface) exploration with landers, drills, or penetrators would also be very interesting, if they can be targeted to volatile-rich areas, although estimates of the depth of buried ice vary from cm to tens of metres, depending on location and the individual MBC's age, orbit, and pole orientation \citep{Snodgrass-MBCs}.

\subsubsection{Interstellar Objects}

Finally, recent years have seen the discovery of the first interstellar comets, which would naturally be very high priority spacecraft targets, as they represent an opportunity to sample extra-solar material without the massive technological challenges or centuries-timescales of an interstellar probe. Our knowledge of interstellar objects (ISOs) and the motivations and concepts for a mission to one are discussed in the chapter by Fitzsimmons et al. in this volume, and were explored by a recent NASA Planetary Science Summer Seminar cohort \citep{Moore2021}, but we briefly discuss the technology requirements for such a probe. In terms of mission design, ISOs are similar to new LPCs, with high velocities and potentially high inclinations, and consequently present similar challenges. Fast flyby missions, perhaps combined with a \DI-style impactor \citep{Moore2021}, appear to be the limit of what is feasible. A waiting spacecraft, similar to \CI{} in concept, seems the most likely way to reach an ISO, but the current poor constraints on the total population size mean that it is hard to judge the necessary waiting times or $\Delta v$ requirements. Given the limited heliocentric distance range in which \CI{} will be able to operate, it seems that it will have to be exceedingly lucky to be able to encounter an ISO within its planned 6-year lifetime and $\Delta v <$ 1 km s$^{-1}$. 
A more capable version of a similar concept could be imagined for a larger-budget mission class (e.g. ESA M-class or NASA Discovery), as the true size of the ISO population becomes clearer following the expected decade of discovery with the Vera C. Rubin Observatory's Legacy Survey of Space and Time (LSST) \citep{Seligman2018}. Alternatively, `launch on discovery' concepts, rather than waiting in space, have been proposed to encounter ISOs \citep{Seligman2018,Moore2021}, although these also require quite rapid reaction (see section \ref{sec:rapid}). Clearly cometary ISOs like 2I/Borisov, if discovered inbound, present the best chances of success for interception, while the available warning times for another small 1I/`Oumuamua-like object, with no visible coma, remain impractically short even with LSST (one or two months at best). There are suggestions to launch missions to chase after departing ISOs, catching up to them at very large distance from the Sun (100 au or more) decades after their perihelion, but these are limited by both available spacecraft technology (propulsion, power, communications) and the navigation and operational practicalities of even locating such a small object so far from the Sun while approaching at tens of km s$^{-1}$ \citep{Hein2019,Hein2021}. Such a mission does not appear feasible in the near term.


\subsection{Technology for Future Comet Missions}
\label{sec:future-mission-types}

There still remains much to be learned about the more accessible JFCs, and new missions to these, or even returns to previously visited comets, are well justified. In this section we  review  current and near-future technology that could enable the next comet missions to JFCs and beyond (see also \citet{Ahearn_after_Rosetta} and \citet{Thomas2019SSRv}).

\subsubsection{Landing on the Nucleus}

As (cryogenic) sample return remains the ultimate goal of comet exploration, a key technology requirement is the ability to land on the nucleus safely: either in a brief `touch and go' maneuver, as used by the asteroid missions \mission{Hayabusa}, \mission{Hayabusa 2} and \mission{OSIRIS-REx}, and proposed at a comet by the \mission{Triple-F} mission \citep{TripleF}, or for a longer period of measurement/drilling.
Landers also have a lot of potential to advance our knowledge of cometary activity, independent of any sample return attempt, through local measurements of the surface (micro-)physics, as was planned for \Philae. Sampling multiple locations on the nucleus would be particularly valuable for this, whether using one mobile lander or a network of static ones (the latter also having the advantage of redundancy and being able to perform seismology or \mission{CONSERT}-like measurements between probes; \citealt{Thomas2019SSRv}).
Future mission architectures to consider include very low speed landers, rovers, and multi-point landers or hoppers, {like the proposed implementation of NASA's Phase A \mission{Comet Hopper} mission concept \citep{CHopper}, and a great deal of autonomous operations.} Taking into account the micro-gravity environment at the surface of a comet and balancing it with the desire to minimize surface alteration upon arrival, various spacecraft designs coupled with landing and holding strategies have been evaluated in concept studies \citep{CHopper,Ulamec2011} or executed without success to date (section \ref{sec:rosetta-lander}). 
With our knowledge of the surface strength and regolith properties still ranging an order of magnitude or more
{after \Philae's inconclusive results \citep{Groussin2019}},
there remains a need for replicating and modeling porous regolith in micro-gravity conditions on Earth for landing tests, and to assess the particle movement near the output of spacecraft thrusters, and the thermal and electrostatic response of the regolith to an approaching spacecraft. Low speed landers will reduce the potential for bouncing on the surface and numerous small thrusters oriented at an angle from the sub-spacecraft point will undoubtedly prove to be beneficial. 

Once in the vicinity of the comet nucleus or safely landed, interacting with the comet's surface is also technologically challenging, whether to rove or hop around to explore, to gather a surface sample, or to penetrate and sample the sub-surface. There are still many unknowns in each individual comet's properties and orders of magnitude in possible strength regimes and surface topographies to design for. A significant difference between classes of comet (e.g., possibly asteroid-like MBCs vs `pristine' LPCs) could also be expected.
Mission designs employing smaller articulating solar panels or nuclear power sources will enable agile traversal of the diverse terrains observed on the comets visited thus far. Additionally, missions powered by nuclear sources will enable operation at larger heliocentric distances with fewer limitations on shared spacecraft resources as well as in non-illuminated regions of a comet, the most likely location to find surface ice \citep{Sunshine_DI_surfaceice,DeSanctis_67PVIRTISH2O,Filacchione_67PVIRTISCO2}. 

\subsubsection{Sample return}

The \Stardust{} mission showed the immense value of laboratory analysis of comet samples. The next step in this field should be to collect a sample from the nucleus, although a more advanced coma sample return mission could also be imagined. \Stardust{} and \Rosetta{} showed that coma dust includes an easily fragmented and fragile component, that would have to be collected very gently (e.g., at near zero relative velocity by a rendezvous mission) to obtain a sample that retains structural information. Measuring the refractory-to-ice abundance in the coma would also be an important, but challenging, goal for such a sample.

For nucleus sample return, there is a need for community guidelines and agreement on the depth to which a sample should be collected to have the greatest science return, from the exposed regolith \citep{noncryo_CSSR} to reaching the least-altered high-volatility sub-surface ice reservoirs. As the surface composition of comets remains uncertain even after \Rosetta{}, there is value in the former approach, which formed the basis of the \mission{CAESAR} proposal that was the runner up in the 2019 NASA New Frontiers call \citep{CAESAR}. A surface regolith sample, which could be returned to Earth in non-cryogenic conditions, would allow for detailed investigation of {thermally processed material and} surface chemistry at a level beyond what \Philae{} could have hoped to achieve, even if it had operated perfectly, and could be compared to similar asteroid samples from \mission{Hayabusa 2} and \mission{OSIRIS-REx}. However,  a {surface sample return} mission {that scrapes, sucks, or brushes up  regolith} makes detailed chemistry measurements at the cost of knowledge of the {physical} structure of the surface material. A more ambitious sample return approach would be to collect a core that preserves both composition and its stratigraphy; information on the size and shape of grains, the bulk strength, porosity, and relative locations of different components \citep[see][for a more detailed discussion]{Thomas2019SSRv}. 
The depth to which a core must reach to sample various ices is not well determined, but is expected to be of order meters, or possibly considerably more to reach `pristine' layers (see discussion in chapter by Guilbert-Lepoutre et al. in this volume).
Comet missions can look to technological advances from ground penetrating radar designed to characterize the subsurface structure 3-10 m deep at Mars \citep{WISDOM_radar} to first locate the true ice depths. This would be a useful measurement for future rendezvous missions whether or not they also attempt to land on or sample the nucleus (e.g., the  \mission{CORE} mission concept, an orbiter whose primary payload is a radar mapping experiment to study nucleus interior structure; \citealt{CORE}).
Drilling into the ice, and extracting an unaltered core, would be technically challenging.  The \mission{SD2} drill on \Philae{} was designed to reach a depth of 23 cm and retrieve material (but not an intact core) for composition analysis \citep{RosIns-SD2}, while \mission{Triple-F} proposed to extract 50 cm cores in touch-and-go landings \citep{TripleF}. Neither of these has been successfully demonstrated at a comet; obtaining meters-long ice cores will require significant advances of these technologies, but may benefit from  current  developments for ocean world applications \citep{OceanWorld_probes}. The largest drill being developed for a near-future flight opportunity is the 1~m long \mission{TRIDENT} drill \citep{Zacny2021}, expected to launch to the Moon in 2022, on a large lander or rover (450 - 2000~kg; a very different class of mission than the 100~kg \Philae).

Cryogenic sample return missions face the challenge not only of acquiring a sample with context and retaining stratigraphy, but also preserving it during the return flight, Earth atmospheric entry, descent, landing, recovery, and curation. {The latter is a significant challenge in its own right; we do not currently have suitable facilities, but progress is being made \citep{McCubbin2021}. The sample would need to be protected} to minimize and ideally prevent alteration that would affect interpretation (this includes shock, vibration, and heat). Space-based retrieval (i.e., docking with a crewed space station, and subsequent `gentle' return of the sample with the astronauts) has even been considered to minimize some of these effects. There have been recent technological advances in cryogenic flight systems and instrumentation, as well as in cryogenic sample curation, handling, and analysis \citep{Westphal_cryo}. \citet{TripleF} proposed a passive system that would keep the sample below 133 K during cruise and 163 K during reentry, which would return a valuable sample at relatively low cost, even if some ices were lost.
\citet{Ambition} suggest that temperatures must be kept below 90 K, and pressure at 1 bar, to preserve most volatile ices (H$_2$O, CO$_2$, HCN). Even lower temperatures or higher pressures would be necessary to keep CO stable. \citet{Westphal_cryo} argue that sample collection and transport at 60 K is necessary to prevent an amorphous-amorphous phase transition in water ice that occurs at 80 K, and that this is considered achievable with current flight system technologies. While challenging, it appears that cryogenic sample preservation is technically possible in the near future. In the end, a small advantage that comet missions have once a sample is collected and contained, is that operating in low gravity environments will allow for easier departure.

\subsubsection{Sub-surface penetrators}
\label{sec:penetrators}

Penetrator probes, which reach sub-surface layers of a planetary body by burying themselves in a high speed impact, rather than a gentle landing and then drilling, have been proposed for some time \citep{Lorenz2011}. Although typically suggested as a solution for larger bodies such as the Moon or icy moons of the outer planets \citep[e.g.,][]{Gowen2011}, where gravity makes soft-landing far more challenging, they were studied in detail for comets as part of the NASA-led \textit{Comet Rendezvous Asteroid Flyby} mission, which was  cancelled in the early 1990s \citep{Swenson1987,Boynton1995}. Penetrator probes rely on sufficiently robust instrumentation that  can survive the sudden deceleration as the probe hits the ground, typically at $\sim$100 m s$^{-1}$, but can reveal physical structure of the surface by measuring this deceleration. They can measure sub-surface composition,  obtain (microscopic) images of the surrounding material, and measure physical properties such as temperatures, conductivity, etc., \textit{in situ}. Although a penetrator  has yet to be {successfully} flown on a planetary mission, they have been studied in detail, including high speed impact tests of instrumentation on Earth, and similar technology has been widely developed for military purposes; penetrators are therefore a realistic and mature technology that would be of value on a suitable rendezvous mission. One may be included on the \ZhengHe{} mission to 311P \citep{ZhengHe-penetrator}, and, if it is, probably represents the best chance for this mission to detect any sub-surface ice.

\subsubsection{Small satellites and multi-point measurements}

A relatively new mission architecture is proving to be beneficial in planetary sciences and has application for comets: the development of small, standardized spacecraft (e.g., CubeSats).
The advantage of small, fast, and relatively cheap missions should not be overlooked. Small-sats can play a role in reconnaissance, rapid response, and could be employed as a fleet. They can also be used to address focused questions and may be the only way in the near future to explore an ISO or LPC up close, either as released probes as part of a larger mission (e.g., \CI) or even as independent spacecraft \citep{PrOVE1}. A \CONTOUR-like tour \citep{NEAT} {and \CI-like `waiting in space' LPC mission \citep{PrOVE2} have}  been proposed within the NASA `SIMPLEx' call for small missions. 

The idea of having a large number of independent spacecraft at a comet has many advantages, particularly for coma sampling and measurements of the plasma environment. 
\CI{} will perform the first multi-point sampling, but the only instrument  common to all three spacecraft is a magnetometer, and it is a fast flyby mission. A plasma-science dedicated rendezvous mission with at least four identical spacecraft measuring the interaction between a comet and the solar wind has been proposed as the next big step forward for this field \citep{Goetz2021}. While this concept assumes four small-to-medium sized independent spacecraft, it is also interesting to consider the possibilities presented by advances in technology for very small platforms. The \CI{} released probes, while small, will still be bespoke spacecraft with rigorous design and build standards. The use of real CubeSats (i.e., standardized units built with {relatively} cheap  off-the-shelf parts) in deep space has been demonstrated to be feasible by the \mission{MarCO} mission to Mars \citep{MarCO}. Even smaller `ChipSats' have been demonstrated to be capable of making simple single measurements at many separate locations in Earth orbit, and have been proposed for small body missions \citep{ChipSats}. Many ($N >> 1$) point measurements with a `coma swarm' would provide fascinating insights into the temporal and spatial variations within a coma, whether simply by tracing dynamics and variations in coma structure (dust or gas density, temperature, or magnetic fields), or even making multiple \textit{in situ} composition measurements (e.g., with CubeSat-sized mass spectrometers based on \Philae{} \mission{Ptolemy} technology; \citealt{RosIns-Ptolemy}). Having a large number of small probes also enables a paradigm shift in build and operation costs that is useful in a cometary environment -- with sufficiently large $N$, failure \emph{is} an option, and results are assured even if some of the probes fail or are lost soon after deployment.

\subsubsection{Rapid reaction}
\label{sec:rapid}

One mission type that is at a true disadvantage in the current program architecture for most, if not all, space agencies is a rapid response mission \citep{Moore_WP_rapid}.  These missions would undoubtedly result in high reward science and could be in response to any number of phenomena or target type, e.g., a major splitting event of an JFC on its inbound perihelion leg, a newly discovered inbound LPC, or an ISO quickly passing through our Solar System. Implementations of rapid response missions may have inherent schedule risk to mission success, or risk failing to achieve their full science potential due to launch vehicle, trajectory, and orbital constraints. Rapid response missions could be executed via a very short and rushed design, build, test, and launch phase, with the major disadvantages of high likelihood of schedule slip, ballooning costs, and potential lack of dedicated launch vehicle {or timely ride-share launch}. In contrast, rapid response missions could take advantage of ground- or space-based storage and thus allow for a more typical and lengthy design and build phase. Having a pre-built spacecraft in storage or in a parking orbit in space waiting for a target has drawbacks, manifesting as having been designed with a target agnostic approach, requiring a minimum intact skeleton team for a long period of low to no mission activity, utilizing technology that is not state-of-the-art by the time of target discovery, degrading parts in orbit, and having the potential need for a dedicated launch vehicle. The key to justifying such a mission will be some reasonably reliable statistical knowledge of how long the wait may have to be, and that the (presumably rare) opportunity  requiring rapid reaction is worth the associated cost.


\section{Summary}
There has been fantastic progress in spacecraft exploration of comets since the beginning of this field around the time of comet Halley's last return. As this comet approaches aphelion in 2024, it is instructive to look at what we have achieved, and what we may look forward to in the next half of its orbit, before its 2061 return. {Reviewing the three major themes we described in the introduction (formation, composition, and activity), we see significant progress with each past mission and clear priorities for future missions and technology. Our understanding of nucleus formation and structure has primarily advanced by the increasing resolution of surface imaging with each mission to date, {revealing an apparent preference for bilobed shapes and a wide variety of surface morphologies.} The next step needs to be probing {microphysical details of the nucleus} and its interior structure. Composition measurements \textit{in situ} have examined gas, dust, and ions, {including isotopic abundances,} and \Stardust{} enabled incredibly detailed laboratory analysis of solid grains -- but increasing the diversity of comets visited and sample return from the nucleus are promising next steps to take, {especially to investigate how the volatile, {organic,} and refractory components are associated}. Finally, processes linked to activity have been localized but fundamentally remain a puzzle that will likely require lander missions, capable of temporally studying the physical, thermal, and chemical properties of the nucleus (sub-)surface at microscopic scales, to solve.} There are still many knowledge gaps that require \textit{in situ} measurement to explore: There are strong cases for missions to comets to better understand nucleus surfaces at small scales, their activity, and the diversity between comets from different source regions or at different evolutionary stages. The return of frozen ices from a comet nucleus remains the highest priority mission objective, yet experiences from flyby, impactor, coma sample return, rendezvous, and landed missions in the past decades have shown that this will be highly challenging. Returning to 1P, and/or cryogenic sample return, in the distant future necessitates long-term planning and investment in technology now. In the meantime, there are still many questions that can, and should, be addressed with a range of smaller missions, taking advantage of current advances in small, fast, and cheap spacecraft. No matter what the mission design and target, the science objectives should be compelling and concise, and drive the spacecraft design and technological advances needed for future cometary exploration.

\vskip .4in
\noindent \textbf{Acknowledgments.} \\

The authors thank the referees, Jessica Sunshine and Horst Uwe Keller, for their helpful and constructive reviews of this chapter. We thank Abbie Donaldson for providing Fig.~\ref{fig:67P_shape}.
C.S. and G.H.J. acknowledge support from the UK Science and Technology Research Council and UK Space Agency.
For the purpose of open access, the author has applied a Creative Commons Attribution (CC BY) licence to any Author Accepted Manuscript version arising from this submission.

\bibliographystyle{sss-three.bst}
\bibliography{refs.bib}

\begin{thebibliography}{298}
\providecommand{\natexlab}[1]{#1}
\parskip=0pt \itemsep=0pt \small \baselineskip=11pt

\bibitem[{\emph{{Accomazzo} et~al.}(2016)\emph{{Accomazzo}, {Lodiot}, and
  {Companys}}}]{Accomazzo_landing}
{Accomazzo} A., {Lodiot} S., and {Companys} V. (2016) \emph{{Rosetta mission
  operations for landing}}, \emph{Acta Astronautica}, \emph{125}, 30--40.

\bibitem[{\emph{{A'Hearn}}(2017)}]{Ahearn_after_Rosetta}
{A'Hearn} M.~F. (2017) \emph{{Comets: looking ahead}}, \emph{Philosophical
  Transactions of the Royal Society of London Series A}, \emph{375}, 20160261.

\bibitem[{\emph{{A'Hearn} et~al.}(2005{\natexlab{a}})\emph{{A'Hearn}, {Belton},
  {Delamere}, and {Blume}}}]{Ahearn_DI_SSR}
{A'Hearn} M.~F., {Belton} M. J.~S., {Delamere} A. et~al. (2005{\natexlab{a}})
  \emph{{Deep Impact: A Large-Scale Active Experiment on a Cometary Nucleus}},
  \emph{\ssr}, \emph{117}, 1--21.

\bibitem[{\emph{{A'Hearn} et~al.}(2011)\emph{{A'Hearn}, {Belton}, {Delamere},
  {Feaga}, {Hampton}, {Kissel}, {Klaasen}, {McFadden}, {Meech}, {Melosh},
  {Schultz}, {Sunshine}, {Thomas}, {Veverka}, {Wellnitz}, {Yeomans}, {Besse},
  {Bodewits}, {Bowling}, {Carcich}, {Collins}, {Farnham}, {Groussin},
  {Hermalyn}, {Kelley}, {Kelley}, {Li}, {Lindler}, {Lisse}, {McLaughlin},
  {Merlin}, {Protopapa}, {Richardson}, and
  {Williams}}}]{Ahearn_Hartley_Science}
{A'Hearn} M.~F., {Belton} M. J.~S., {Delamere} W.~A. et~al. (2011) \emph{{EPOXI
  at Comet Hartley 2}}, \emph{Science}, \emph{332}, 1396.

\bibitem[{\emph{{A'Hearn} et~al.}(2005{\natexlab{b}})\emph{{A'Hearn}, {Belton},
  {Delamere}, {Kissel}, {Klaasen}, {McFadden}, {Meech}, {Melosh}, {Schultz},
  {Sunshine}, {Thomas}, {Veverka}, {Yeomans}, {Baca}, {Busko}, {Crockett},
  {Collins}, {Desnoyer}, {Eberhardy}, {Ernst}, {Farnham}, {Feaga}, {Groussin},
  {Hampton}, {Ipatov}, {Li}, {Lindler}, {Lisse}, {Mastrodemos}, {Owen},
  {Richardson}, {Wellnitz}, and {White}}}]{Ahearn_DI_Science}
{A'Hearn} M.~F., {Belton} M.~J.~S., {Delamere} W.~A. et~al.
  (2005{\natexlab{b}}) \emph{{Deep Impact: Excavating Comet Tempel 1}},
  \emph{Science}, \emph{310}, 258--264.

\bibitem[{\emph{{A'Hearn} and {Johnson}}(2015)}]{AHearn+Johnson}
{A'Hearn} M.~F. and {Johnson} L.~N. (2015) in \emph{Handbook of Cosmic Hazards
  and Planetary Defense}, pp. 513--534.

\bibitem[{\emph{{A'Hearn} et~al.}(1995)\emph{{A'Hearn}, {Millis}, {Schleicher},
  {Osip}, and {Birch}}}]{AHearn1995}
{A'Hearn} M.~F., {Millis} R.~C., {Schleicher} D.~O. et~al. (1995) \emph{{The
  ensemble properties of comets: Results from narrowband photometry of 85
  comets, 1976-1992.}}, \emph{\icarus}, \emph{118}, 223--270.

\bibitem[{\emph{{A'Hearn} et~al.}(1984)\emph{{A'Hearn}, {Schleicher}, {Millis},
  {Feldman}, and {Thompson}}}]{Ahearn1984}
{A'Hearn} M.~F., {Schleicher} D.~G., {Millis} R.~L. et~al. (1984) \emph{{Comet
  Bowell 1980b}}, \emph{\aj}, \emph{89}, 579--591.

\bibitem[{\emph{{Alfv\'en}}(1957)}]{alfven57}
{Alfv\'en} H. (1957) \emph{{On the theory of comet tails.}}, \emph{Tellus},
  \emph{9}.

\bibitem[{\emph{{Altwegg} et~al.}(2015)\emph{{Altwegg}, {Balsiger}, {Bar-Nun},
  {Berthelier}, {Bieler}, {Bochsler}, {Briois}, {Calmonte}, {Combi}, {De
  Keyser}, {Eberhardt}, {Fiethe}, {Fuselier}, {Gasc}, {Gombosi}, {Hansen},
  {H{\"a}ssig}, {J{\"a}ckel}, {Kopp}, {Korth}, {LeRoy}, {Mall}, {Marty},
  {Mousis}, {Neefs}, {Owen}, {R{\`e}me}, {Rubin}, {S{\'e}mon}, {Tzou}, {Waite},
  and {Wurz}}}]{Altwegg2015}
{Altwegg} K., {Balsiger} H., {Bar-Nun} A. et~al. (2015)
  \emph{{67P/Churyumov-Gerasimenko, a Jupiter family comet with a high D/H
  ratio}}, \emph{Science}, \emph{347}, 1261952.

\bibitem[{\emph{{Altwegg} et~al.}(2016)\emph{{Altwegg}, {Balsiger}, {Bar-Nun},
  {Berthelier}, {Bieler}, {Bochsler}, {Briois}, {Calmonte}, {Combi}, {Cottin},
  {De Keyser}, {Dhooghe}, {Fiethe}, {Fuselier}, {Gasc}, {Gombosi}, {Hansen},
  {Haessig}, {Ja ckel}, {Kopp}, {Korth}, {Le Roy}, {Mall}, {Marty}, {Mousis},
  {Owen}, {Reme}, {Rubin}, {Semon}, {Tzou}, {Waite}, and {Wurz}}}]{Altwegg2016}
{Altwegg} K., {Balsiger} H., {Bar-Nun} A. et~al. (2016) \emph{{Prebiotic
  chemicals--amino acid and phosphorus--in the coma of comet
  67P/Churyumov-Gerasimenko}}, \emph{Science Advances}, \emph{2},
  e1600285--e1600285.

\bibitem[{\emph{{Altwegg} et~al.}(2020)\emph{{Altwegg}, {Balsiger},
  {H{\"a}nni}, {Rubin}, {Schuhmann}, {Schroeder}, {S{\'e}mon}, {Wampfler},
  {Berthelier}, {Briois}, {Combi}, {Gombosi}, {Cottin}, {De Keyser}, {Dhooghe},
  {Fiethe}, and {Fuselier}}}]{Altwegg2020}
{Altwegg} K., {Balsiger} H., {H{\"a}nni} N. et~al. (2020) \emph{{Evidence of
  ammonium salts in comet 67P as explanation for the nitrogen depletion in
  cometary comae}}, \emph{Nature Astronomy}, \emph{4}, 533--540.

\bibitem[{\emph{{Ashman} et~al.}(2016)\emph{{Ashman}, {Barth{\'e}l{\'e}my},
  {O`Rourke}, {Almeida}, {Altobelli}, {Costa Sitj{\`a}}, {Garc{\'\i}a Beteta},
  {Geiger}, {Grieger}, {Heather}, {Hoofs}, {K{\"u}ppers}, {Martin}, {Moissl},
  {M{\'u}{\~n}oz Crego}, {P{\'e}rez-Ay{\'u}car}, {Sanchez Suarez}, {Taylor},
  and {Vallat}}}]{Ashman_landing}
{Ashman} M., {Barth{\'e}l{\'e}my} M., {O`Rourke} L. et~al. (2016)
  \emph{{Rosetta science operations in support of the Philae mission}},
  \emph{Acta Astronautica}, \emph{125}, 41--64.

\bibitem[{\emph{{Asphaug}}(2015)}]{CORE}
{Asphaug} E. (2015) in \emph{Spacecraft Reconnaissance of Asteroid and Comet
  Interiors}, vol. 1829 of \emph{LPI Contributions}, p. 6044.

\bibitem[{\emph{{Attree} et~al.}(2019)\emph{{Attree}, {Jorda}, {Groussin},
  {Mottola}, {Thomas}, {Brouet}, {K{\"u}hrt}, {Knapmeyer}, {Preusker},
  {Scholten}, {Knollenberg}, {Hviid}, {Hartogh}, and {Rodrigo}}}]{Attree2019}
{Attree} N., {Jorda} L., {Groussin} O. et~al. (2019) \emph{{Constraining models
  of activity on comet 67P/Churyumov-Gerasimenko with Rosetta trajectory,
  rotation, and water production measurements}}, \emph{\aap}, \emph{630}, A18.

\bibitem[{\emph{{Auster} et~al.}(2007)\emph{{Auster}, {Apathy}, {Berghofer},
  {Remizov}, {Roll}, {Fornacon}, {Glassmeier}, {Haerendel}, {Hejja},
  {K{\"u}hrt}, {Magnes}, {Moehlmann}, {Motschmann}, {Richter}, {Rosenbauer},
  {Russell}, {Rustenbach}, {Sauer}, {Schwingenschuh}, {Szemerey}, and
  {Waesch}}}]{RosIns-ROMAP}
{Auster} H.~U., {Apathy} I., {Berghofer} G. et~al. (2007) \emph{{ROMAP: Rosetta
  Magnetometer and Plasma Monitor}}, \emph{\ssr}, \emph{128}, 221--240.

\bibitem[{\emph{{Baker} et~al.}(2019)\emph{{Baker}, {Colley}, {Essmiller},
  {Klesh}, {Krajewski}, and {Sternberg}}}]{MarCO}
{Baker} J., {Colley} C.~N., {Essmiller} J.~C. et~al. (2019) in \emph{EPSC-DPS
  Joint Meeting 2019}, vol. 2019, pp. EPSC--DPS2019--2009.

\bibitem[{\emph{{Ballard} et~al.}(2011)\emph{{Ballard}, {Christiansen},
  {Charbonneau}, {Deming}, {Holman}, {A'Hearn}, {Wellnitz}, {Barry}, {Kuchner},
  {Livengood}, {Hewagama}, {Sunshine}, {Hampton}, {Lisse}, {Seager}, and
  {Veverka}}}]{EPOCh_Ballard}
{Ballard} S., {Christiansen} J.~L., {Charbonneau} D. et~al. (2011) \emph{{A
  Search for Additional Planets in Five of the Exoplanetary Systems Studied by
  the NASA EPOXI Mission}}, \emph{\apj}, \emph{732}, 41.

\bibitem[{\emph{{Balsiger} et~al.}(2007)\emph{{Balsiger}, {Altwegg},
  {Bochsler}, {Eberhardt}, {Fischer}, {Graf}, {J{\"a}ckel}, {Kopp}, {Langer},
  {Mildner}, {M{\"u}ller}, {Riesen}, {Rubin}, {Scherer}, {Wurz},
  {W{\"u}thrich}, {Arijs}, {Delanoye}, {de Keyser}, {Neefs}, {Nevejans},
  {R{\`e}me}, {Aoustin}, {Mazelle}, {M{\'e}dale}, {Sauvaud}, {Berthelier},
  {Bertaux}, {Duvet}, {Illiano}, {Fuselier}, {Ghielmetti}, {Magoncelli},
  {Shelley}, {Korth}, {Heerlein}, {Lauche}, {Livi}, {Loose}, {Mall}, {Wilken},
  {Gliem}, {Fiethe}, {Gombosi}, {Block}, {Carignan}, {Fisk}, {Waite}, {Young},
  and {Wollnik}}}]{RosIns-ROSINA}
{Balsiger} H., {Altwegg} K., {Bochsler} P. et~al. (2007) \emph{{Rosina Rosetta
  Orbiter Spectrometer for Ion and Neutral Analysis}}, \emph{\ssr}, \emph{128},
  745--801.

\bibitem[{\emph{{Barucci} et~al.}(2015)\emph{{Barucci}, {Fulchignoni}, {Ji},
  {Marchin}, and {Thomas}}}]{ast-flybys}
{Barucci} M.~A., {Fulchignoni} M., {Ji} J. et~al. (2015) \emph{{The Flybys of
  Asteroids 2867 Steins, 21 Lutetia, and 4179 Toutatis}}, pp. 433--450.

\bibitem[{\emph{{Belton} and {A'Hearn}}(1999)}]{Belton_DI_concept}
{Belton} M.~J.~S. and {A'Hearn} M.~F. (1999) \emph{{Deep sub-surface
  exploration of cometary nuclei}}, \emph{Advances in Space Research},
  \emph{24}, 1167--1173.

\bibitem[{\emph{{Belton} et~al.}(2008)\emph{{Belton}, {Feldman}, {A'Hearn}, and
  {Carcich}}}]{Belton_DI_outbursts}
{Belton} M. J.~S., {Feldman} P.~D., {A'Hearn} M.~F. et~al. (2008)
  \emph{{Cometary cryo-volcanism: Source regions and a model for the UT 2005
  June 14 and other mini-outbursts on Comet 9P/Tempel 1}}, \emph{\icarus},
  \emph{198}, 189--207.

\bibitem[{\emph{{Belton} et~al.}(2011)\emph{{Belton}, {Meech}, {Chesley},
  {Pittichov{\'a}}, {Carcich}, {Drahus}, {Harris}, {Gillam}, {Veverka},
  {Mastrodemos}, {Owen}, {A'Hearn}, {Bagnulo}, {Bai}, {Barrera}, {Bastien},
  {Bauer}, {Bedient}, {Bhatt}, {Boehnhardt}, {Brosch}, {Buie}, {Candia},
  {Chen}, {Chiang}, {Choi}, {Cochran}, {Crockett}, {Duddy}, {Farnham},
  {Fern{\'a}ndez}, {Guti{\'e}rrez}, {Hainaut}, {Hampton}, {Herrmann}, {Hsieh},
  {Kadooka}, {Kaluna}, {Keane}, {Kim}, {Klaasen}, {Kleyna}, {Krisciunas},
  {Lara}, {Lauer}, {Li}, {Licandro}, {Lisse}, {Lowry}, {McFadden}, {Moskovitz},
  {Mueller}, {Polishook}, {Raja}, {Riesen}, {Sahu}, {Samarasinha}, {Sarid},
  {Sekiguchi}, {Sonnett}, {Suntzeff}, {Taylor}, {Thomas}, {Tozzi},
  {Vasundhara}, {Vincent}, {Wasserman}, {Webster-Schultz}, {Yang}, {Zenn}, and
  {Zhao}}}]{Belton2011}
{Belton} M. J.~S., {Meech} K.~J., {Chesley} S. et~al. (2011)
  \emph{{Stardust-NExT, Deep Impact, and the accelerating spin of 9P/Tempel
  1}}, \emph{\icarus}, \emph{213}, 345--368.

\bibitem[{\emph{{Belton} and {Melosh}}(2009)}]{Belton_Melosh_outbursts}
{Belton} M. J.~S. and {Melosh} J. (2009) \emph{{Fluidization and multiphase
  transport of particulate cometary material as an explanation of the smooth
  terrains and repetitive outbursts on 9P/Tempel 1}}, \emph{\icarus},
  \emph{200}, 280--291.

\bibitem[{\emph{{Belton} et~al.}(2013{\natexlab{a}})\emph{{Belton}, {Thomas},
  {Carcich}, {Quick}, {Veverka}, {Jay Melosh}, {A'Hearn}, {Li}, {Brownlee},
  {Schultz}, {Klaasen}, and {Sarid}}}]{Belton2013}
{Belton} M. J.~S., {Thomas} P., {Carcich} B. et~al. (2013{\natexlab{a}})
  \emph{{The origin of pits on 9P/Tempel 1 and the geologic signature of
  outbursts in Stardust-NExT images}}, \emph{\icarus}, \emph{222}, 477--486.

\bibitem[{\emph{{Belton} et~al.}(2013{\natexlab{b}})\emph{{Belton}, {Thomas},
  {Li}, {Williams}, {Carcich}, {A'Hearn}, {McLaughlin}, {Farnham}, {McFadden},
  {Lisse}, {Collins}, {Besse}, {Klaasen}, {Sunshine}, {Meech}, and
  {Lindler}}}]{Belton_2013_H2spin}
{Belton} M. J.~S., {Thomas} P., {Li} J.-Y. et~al. (2013{\natexlab{b}})
  \emph{{The complex spin state of 103P/Hartley 2: Kinematics and orientation
  in space}}, \emph{\icarus}, \emph{222}, 595--609.

\bibitem[{\emph{{Benna} et~al.}(2015)\emph{{Benna}, {Mahaffy}, {Grebowsky},
  {Plane}, {Yelle}, and {Jakosky}}}]{Benna2015}
{Benna} M., {Mahaffy} P.~R., {Grebowsky} J.~M. et~al. (2015) \emph{{Metallic
  ions in the upper atmosphere of Mars from the passage of comet C/2013 A1
  (Siding Spring)}}, \emph{\grl}, \emph{42}, 4670--4675.

\bibitem[{\emph{{Bibring} et~al.}(2007{\natexlab{a}})\emph{{Bibring}, {Lamy},
  {Langevin}, {Soufflot}, {Berth{\'e}}, {Borg}, {Poulet}, and
  {Mottola}}}]{RosIns-CIVA}
{Bibring} J.~P., {Lamy} P., {Langevin} Y. et~al. (2007{\natexlab{a}})
  \emph{{Civa}}, \emph{\ssr}, \emph{128}, 397--412.

\bibitem[{\emph{{Bibring} et~al.}(2007{\natexlab{b}})\emph{{Bibring},
  {Rosenbauer}, {Boehnhardt}, {Ulamec}, {Biele}, {Espinasse}, {Feuerbacher},
  {Gaudon}, {Hemmerich}, {Kletzkine}, {Moura}, {Mugnuolo}, {Nietner},
  {P{\"a}tz}, {Roll}, {Scheuerle}, {Szeg{\"o}}, and
  {Wittmann}}}]{Philae-summary}
{Bibring} J.~P., {Rosenbauer} H., {Boehnhardt} H. et~al. (2007{\natexlab{b}})
  \emph{{The Rosetta Lander (``Philae'') Investigations}}, \emph{\ssr},
  \emph{128}, 205--220.

\bibitem[{\emph{{Biele} et~al.}(2015)\emph{{Biele}, {Ulamec}, {Maibaum},
  {Roll}, {Witte}, {Jurado}, {Mu{\~n}oz}, {Arnold}, {Auster}, {Casas}, {Faber},
  {Fantinati}, {Finke}, {Fischer}, {Geurts}, {G{\"u}ttler}, {Heinisch},
  {Herique}, {Hviid}, {Kargl}, {Knapmeyer}, {Knollenberg}, {Kofman},
  {K{\"o}mle}, {K{\"u}hrt}, {Lommatsch}, {Mottola}, {Pardo de Santayana},
  {Remetean}, {Scholten}, {Seidensticker}, {Sierks}, and {Spohn}}}]{Biele2015}
{Biele} J., {Ulamec} S., {Maibaum} M. et~al. (2015) \emph{{The landing(s) of
  Philae and inferences about comet surface mechanical properties}},
  \emph{Science}, \emph{349}, 1.9816.

\bibitem[{\emph{{Bieler} et~al.}(2015)\emph{{Bieler}, {Altwegg}, {Balsiger},
  {Bar-Nun}, {Berthelier}, {Bochsler}, {Briois}, {Calmonte}, {Combi}, {de
  Keyser}, {van Dishoeck}, {Fiethe}, {Fuselier}, {Gasc}, {Gombosi}, {Hansen},
  {H{\"a}ssig}, {J{\"a}ckel}, {Kopp}, {Korth}, {Le Roy}, {Mall}, {Maggiolo},
  {Marty}, {Mousis}, {Owen}, {R{\`e}me}, {Rubin}, {S{\'e}mon}, {Tzou}, {Waite},
  {Walsh}, and {Wurz}}}]{Bieler2015}
{Bieler} A., {Altwegg} K., {Balsiger} H. et~al. (2015) \emph{{Abundant
  molecular oxygen in the coma of comet 67P/Churyumov-Gerasimenko}},
  \emph{\nat}, \emph{526}, 678--681.

\bibitem[{\emph{{Birch} et~al.}(2020)\emph{{Birch}, {Hayes}, {Milam},
  {Agarwal}, {Bodewits}, {Englander}, {Kelley}, {Nakamura}, {Protopapa},
  {Singer}, {Soderblom}, {Sonnett}, {Umurhan}, and {Vincent}}}]{NEAT}
{Birch} S., {Hayes} A., {Milam} S. et~al. (2020) in \emph{AAS/Division for
  Planetary Sciences Meeting Abstracts}, vol.~52 of \emph{AAS/Division for
  Planetary Sciences Meeting Abstracts}, p. 001.04.

\bibitem[{\emph{{Blum} et~al.}(2017)\emph{{Blum}, {Gundlach}, {Krause},
  {Fulle}, {Johansen}, {Agarwal}, {von Borstel}, {Shi}, {Hu}, {Bentley},
  {Capaccioni}, {Colangeli}, {Della Corte}, {Fougere}, {Green}, {Ivanovski},
  {Mannel}, {Merouane}, {Migliorini}, {Rotundi}, {Schmied}, and
  {Snodgrass}}}]{Blum-model}
{Blum} J., {Gundlach} B., {Krause} M. et~al. (2017) \emph{{Evidence for the
  formation of comet 67P/Churyumov-Gerasimenko through gravitational collapse
  of a bound clump of pebbles}}, \emph{\mnras}, \emph{469}, S755--S773.

\bibitem[{\emph{{Blume}}(2005)}]{Blume_DI_SSR}
{Blume} W.~H. (2005) \emph{{Deep Impact Mission Design}}, \emph{\ssr},
  \emph{117}, 23--42.

\bibitem[{\emph{{Bockel{\'e}e-Morvan} et~al.}(2021)\emph{{Bockel{\'e}e-Morvan},
  {Filacchione}, {Altwegg}, {Bianchi}, {Bizzarro}, {Blum}, {Bonal},
  {Capaccioni}, {Choukroun}, {Codella}, {Cottin}, {Davidsson}, {De Sanctis},
  {Drozdovskaya}, {Engrand}, {Galand}, {G{\"u}ttler}, {Henri}, {Herique},
  {Ivanovski}, {Kokotanekova}, {Levasseur-Regourd}, {Miller}, {Rotundi},
  {Sch{\"o}nb{\"a}chler}, {Snodgrass}, {Thomas}, {Tubiana}, {Ulamec}, and
  {Vincent}}}]{Ambition}
{Bockel{\'e}e-Morvan} D., {Filacchione} G., {Altwegg} K. et~al. (2021)
  \emph{{AMBITION - comet nucleus cryogenic sample return}}, \emph{Experimental
  Astronomy}.

\bibitem[{\emph{{Boehnhardt} et~al.}(2016)\emph{{Boehnhardt}, {Riffeser},
  {Kluge}, {Ries}, {Schmidt}, and {Hopp}}}]{Boehnhardt2016}
{Boehnhardt} H., {Riffeser} A., {Kluge} M. et~al. (2016) \emph{{Mt. Wendelstein
  imaging of the post-perihelion dust coma of 67P/Churyumov-Gerasimenko in
  2015/2016}}, \emph{\mnras}, \emph{462}, S376--S393.

\bibitem[{\emph{{Boynton} and {Reinert}}(1995)}]{Boynton1995}
{Boynton} W.~V. and {Reinert} R.~P. (1995) \emph{{The cryo-penetrator: An
  approach to exploration of icy bodies in the solar system}}, \emph{Acta
  Astronautica}, \emph{35}, 59--68.

\bibitem[{\emph{{Britt} et~al.}(2004)\emph{{Britt}, {Boice}, {Buratti},
  {Campins}, {Nelson}, {Oberst}, {Sandel}, {Stern}, {Soderblom}, and
  {Thomas}}}]{Britt2004}
{Britt} D.~T., {Boice} D.~C., {Buratti} B.~J. et~al. (2004) \emph{{The
  morphology and surface processes of Comet 19/P Borrelly}}, \emph{\icarus},
  \emph{167}, 45--53.

\bibitem[{\emph{{Brownlee} et~al.}(2006)\emph{{Brownlee}, {Tsou}, {Al{\'e}on},
  {Alexander}, {Araki}, {Bajt}, {Baratta}, {Bastien}, {Bland}, {Bleuet},
  {Borg}, {Bradley}, {Brearley}, {Brenker}, {Brennan}, {Bridges}, {Browning},
  {Brucato}, {Bullock}, {Burchell}, {Busemann}, {Butterworth}, {Chaussidon},
  {Cheuvront}, {Chi}, {Cintala}, {Clark}, {Clemett}, {Cody}, {Colangeli},
  {Cooper}, {Cordier}, {Daghlian}, {Dai}, {D'Hendecourt}, {Djouadi},
  {Dominguez}, {Duxbury}, {Dworkin}, {Ebel}, {Economou}, {Fakra}, {Fairey},
  {Fallon}, {Ferrini}, {Ferroir}, {Fleckenstein}, {Floss}, {Flynn}, {Franchi},
  {Fries}, {Gainsforth}, {Gallien}, {Genge}, {Gilles}, {Gillet}, {Gilmour},
  {Glavin}, {Gounelle}, {Grady}, {Graham}, {Grant}, {Green}, {Grossemy},
  {Grossman}, {Grossman}, {Guan}, {Hagiya}, {Harvey}, {Heck}, {Herzog},
  {Hoppe}, {H{\"o}rz}, {Huth}, {Hutcheon}, {Ignatyev}, {Ishii}, {Ito}, {Jacob},
  {Jacobsen}, {Jacobsen}, {Jones}, {Joswiak}, {Jurewicz}, {Kearsley}, {Keller},
  {Khodja}, {Kilcoyne}, {Kissel}, {Krot}, {Langenhorst}, {Lanzirotti}, {Le},
  {Leshin}, {Leitner}, {Lemelle}, {Leroux}, {Liu}, {Luening}, {Lyon},
  {MacPherson}, {Marcus}, {Marhas}, {Marty}, {Matrajt}, {McKeegan}, {Meibom},
  {Mennella}, {Messenger}, {Messenger}, {Mikouchi}, {Mostefaoui}, {Nakamura},
  {Nakano}, {Newville}, {Nittler}, {Ohnishi}, {Ohsumi}, {Okudaira},
  {Papanastassiou}, {Palma}, {Palumbo}, {Pepin}, {Perkins}, {Perronnet},
  {Pianetta}, {Rao}, {Rietmeijer}, {Robert}, {Rost}, {Rotundi}, {Ryan},
  {Sandford}, {Schwandt}, {See}, {Schlutter}, {Sheffield-Parker},
  {Simionovici}, {Simon}, {Sitnitsky}, {Snead}, {Spencer}, {Stadermann},
  {Steele}, {Stephan}, {Stroud}, {Susini}, {Sutton}, {Suzuki}, {Taheri},
  {Taylor}, {Teslich}, {Tomeoka}, {Tomioka}, {Toppani}, {Trigo-Rodr{\'\i}guez},
  {Troadec}, {Tsuchiyama}, {Tuzzolino}, {Tyliszczak}, {Uesugi}, {Velbel},
  {Vellenga}, {Vicenzi}, {Vincze}, {Warren}, {Weber}, {Weisberg}, {Westphal},
  {Wirick}, {Wooden}, {Wopenka}, {Wozniakiewicz}, {Wright}, {Yabuta}, {Yano},
  {Young}, {Zare}, {Zega}, {Ziegler}, {Zimmerman}, {Zinner}, and
  {Zolensky}}}]{Brownlee2006}
{Brownlee} D., {Tsou} P., {Al{\'e}on} J. et~al. (2006) \emph{{Comet 81P/Wild 2
  Under a Microscope}}, \emph{Science}, \emph{314}, 1711.

\bibitem[{\emph{{Brownlee} et~al.}(2004)\emph{{Brownlee}, {Horz}, {Newburn},
  {Zolensky}, {Duxbury}, {Sandford}, {Sekanina}, {Tsou}, {Hanner}, {Clark},
  {Green}, and {Kissel}}}]{Brownlee2004}
{Brownlee} D.~E., {Horz} F., {Newburn} R.~L. et~al. (2004) \emph{{Surface of
  Young Jupiter Family Comet 81 P/Wild 2: View from the Stardust Spacecraft}},
  \emph{Science}, \emph{304}, 1764--1769.

\bibitem[{\emph{{Brownlee} et~al.}(2003)\emph{{Brownlee}, {Tsou}, {Anderson},
  {Hanner}, {Newburn}, {Sekanina}, {Clark}, {H{\"o}rz}, {Zolensky}, {Kissel},
  {McDonnell}, {Sandford}, and {Tuzzolino}}}]{Stardust_Brownlee_JGR}
{Brownlee} D.~E., {Tsou} P., {Anderson} J.~D. et~al. (2003) \emph{{Stardust:
  Comet and interstellar dust sample return mission}}, \emph{Journal of
  Geophysical Research (Planets)}, \emph{108}, 8111.

\bibitem[{\emph{{Burnett}}(2006)}]{Burnett2006}
{Burnett} D.~S. (2006) \emph{{NASA Returns Rocks from a Comet}},
  \emph{Science}, \emph{314}, 1709.

\bibitem[{\emph{{Capaccioni} et~al.}(2015)\emph{{Capaccioni}, {Coradini},
  {Filacchione}, {Erard}, {Arnold}, {Drossart}, {De Sanctis},
  {Bockelee-Morvan}, {Capria}, {Tosi}, {Leyrat}, {Schmitt}, {Quirico},
  {Cerroni}, {Mennella}, {Raponi}, {Ciarniello}, {McCord}, {Moroz}, {Palomba},
  {Ammannito}, {Barucci}, {Bellucci}, {Benkhoff}, {Bibring}, {Blanco},
  {Blecka}, {Carlson}, {Carsenty}, {Colangeli}, {Combes}, {Combi}, {Crovisier},
  {Encrenaz}, {Federico}, {Fink}, {Fonti}, {Ip}, {Irwin}, {Jaumann}, {Kuehrt},
  {Langevin}, {Magni}, {Mottola}, {Orofino}, {Palumbo}, {Piccioni}, {Schade},
  {Taylor}, {Tiphene}, {Tozzi}, {Beck}, {Biver}, {Bonal}, {Combe}, {Despan},
  {Flamini}, {Fornasier}, {Frigeri}, {Grassi}, {Gudipati}, {Longobardo},
  {Markus}, {Merlin}, {Orosei}, {Rinaldi}, {Stephan}, {Cartacci}, {Cicchetti},
  {Giuppi}, {Hello}, {Henry}, {Jacquinod}, {Noschese}, {Peter}, {Politi},
  {Reess}, and {Semery}}}]{Capaccioni2015}
{Capaccioni} F., {Coradini} A., {Filacchione} G. et~al. (2015) \emph{{The
  organic-rich surface of comet 67P/Churyumov-Gerasimenko as seen by
  VIRTIS/Rosetta}}, \emph{Science}, \emph{347}, aaa0628.

\bibitem[{\emph{{Carr} et~al.}(2007)\emph{{Carr}, {Cupido}, {Lee}, {Balogh},
  {Beek}, {Burch}, {Dunford}, {Eriksson}, {Gill}, {Glassmeier}, {Goldstein},
  {Lagoutte}, {Lundin}, {Lundin}, {Lybekk}, {Michau}, {Musmann}, {Nilsson},
  {Pollock}, {Richter}, and {Trotignon}}}]{RosIns-RPC}
{Carr} C., {Cupido} E., {Lee} C.~G.~Y. et~al. (2007) \emph{{RPC: The Rosetta
  Plasma Consortium}}, \emph{\ssr}, \emph{128}, 629--647.

\bibitem[{\emph{{Chesley} et~al.}(2013)\emph{{Chesley}, {Belton}, {Carcich},
  {Thomas}, {Pittichov{\'a}}, {Klaasen}, {Li}, {Farnham}, {Gillam}, {Harris},
  and {Veverka}}}]{Chesley2013}
{Chesley} S.~R., {Belton} M.~J.~S., {Carcich} B. et~al. (2013) \emph{{An
  updated rotation model for Comet 9P/Tempel 1}}, \emph{\icarus}, \emph{222},
  516--525.

\bibitem[{\emph{{Choukroun} et~al.}(2020)\emph{{Choukroun}, {Altwegg},
  {K{\"u}hrt}, {Biver}, {Bockel{\'e}e-Morvan}, {Dr{\k{a}}{\.z}kowska},
  {H{\'e}rique}, {Hilchenbach}, {Marschall}, {P{\"a}tzold}, {Taylor}, and
  {Thomas}}}]{Choukroun2020}
{Choukroun} M., {Altwegg} K., {K{\"u}hrt} E. et~al. (2020) \emph{{Dust-to-Gas
  and Refractory-to-Ice Mass Ratios of Comet 67P/Churyumov-Gerasimenko from
  Rosetta Observations}}, \emph{\ssr}, \emph{216}, 44.

\bibitem[{\emph{{Ciarletti} et~al.}(2017)\emph{{Ciarletti}, {Clifford},
  {Plettemeier}, {Le Gall}, {Herv{\'e}}, {Dorizon}, {Quantin-Nataf}, {Benedix},
  {Schwenzer}, {Pettinelli}, {Heggy}, {Herique}, {Berthelier}, {Kofman},
  {Vago}, {Hamran}, and {WISDOM Team}}}]{WISDOM_radar}
{Ciarletti} V., {Clifford} S., {Plettemeier} D. et~al. (2017) \emph{{The WISDOM
  Radar: Unveiling the Subsurface Beneath the ExoMars Rover and Identifying the
  Best Locations for Drilling}}, \emph{Astrobiology}, \emph{17}, 565--584.

\bibitem[{\emph{{Clark} et~al.}(2008)\emph{{Clark}, {Sunshine}, {A'Hearn},
  {Cochran}, {Farnham}, {Harris}, {McCoy}, and {Veverka}}}]{CHopper}
{Clark} B.~C., {Sunshine} J.~M., {A'Hearn} M.~F. et~al. (2008) in
  \emph{Asteroids, Comets, Meteors 2008}, vol. 1405, p. 8131.

\bibitem[{\emph{{Clark} et~al.}(2018)\emph{{Clark}, {Hewagama}, {Aslam},
  {Bauer}, {Daly}, {Feaga}, {Folta}, {Gorius}, {Hughes}, {Hurford}, {Jennings},
  {Livengood}, {Mumma}, {Nixon}, {Sunshine}, {Villanueva}, {Brown}, {Malphrus},
  and {Zucherman}}}]{PrOVE1}
{Clark} P., {Hewagama} T., {Aslam} S. et~al. (2018) in \emph{CubeSats and
  NanoSats for Remote Sensing II}, vol. 10769 of \emph{Society of Photo-Optical
  Instrumentation Engineers (SPIE) Conference Series}, p. 107690J.

\bibitem[{\emph{{Coates} and {Jones}}(2009)}]{CoatesJones2009}
{Coates} A.~J. and {Jones} G.~H. (2009) \emph{{Plasma environment of Jupiter
  family comets}}, \emph{\planss}, \emph{57}, 1175--1191.

\bibitem[{\emph{{Cochran} et~al.}(2002)\emph{{Cochran}, {Veverka}, {Bell},
  {Belton}, {Benkhoff}, {Benkhoff}, {Clark}, {Feldman}, {Kissel}, {Mahaffy},
  {Malin}, {Murchie}, {Neimann}, {Owen}, {Robinson}, {Schwehm}, {Squyres},
  {Thomas}, {Whipple}, and {Yeomans}}}]{Cochran2002}
{Cochran} A., {Veverka} J., {Bell} J. et~al. (2002) \emph{{The Comet Nucleus
  Tour (Contour); A NASA Discovery Mission}}, \emph{Earth Moon and Planets},
  \emph{89}, 289--300.

\bibitem[{\emph{{Colangeli} et~al.}(2007)\emph{{Colangeli}, {Lopez-Moreno},
  {Palumbo}, {Rodriguez}, {Cosi}, {Della Corte}, {Esposito}, {Fulle},
  {Herranz}, {Jeronimo}, {Lopez-Jimenez}, {Epifani}, {Morales}, {Moreno},
  {Palomba}, and {Rotundi}}}]{RosIns-GIADA}
{Colangeli} L., {Lopez-Moreno} J.~J., {Palumbo} P. et~al. (2007) \emph{{The
  Grain Impact Analyser and Dust Accumulator (GIADA) Experiment for the Rosetta
  Mission: Design, Performances and First Results}}, \emph{\ssr}, \emph{128},
  803--821.

\bibitem[{\emph{{Combi} et~al.}(2008)\emph{{Combi}, {M{\"a}kinen}, {Henry},
  {Bertaux}, and {Quem{\'e}rais}}}]{Combi2008}
{Combi} M.~R., {M{\"a}kinen} J.~T.~T., {Henry} N.~J. et~al. (2008) \emph{{Solar
  and Heliospheric Observatory/solar Wind Anisotropies Observations of Five
  Moderately Bright Comets: 1999-2002}}, \emph{\aj}, \emph{135}, 1533--1550.

\bibitem[{\emph{{Combi} et~al.}(2019)\emph{{Combi}, {M{\"a}kinen}, {Bertaux},
  {Qu{\'e}merais}, and {Ferron}}}]{Combi2019}
{Combi} M.~R., {M{\"a}kinen} T.~T., {Bertaux} J.~L. et~al. (2019) \emph{{A
  survey of water production in 61 comets from SOHO/SWAN observations of
  hydrogen Lyman-alpha: Twenty-one years 1996-2016}}, \emph{\icarus},
  \emph{317}, 610--620.

\bibitem[{\emph{{Combi} et~al.}(2021)\emph{{Combi}, {Shou}, {M{\"a}kinen},
  {Bertaux}, {Qu{\'e}merais}, {Ferron}, and {Coronel}}}]{Combi2021}
{Combi} M.~R., {Shou} Y., {M{\"a}kinen} T. et~al. (2021) \emph{{Water
  production rates from SOHO/SWAN observations of six comets: 2017-2020}},
  \emph{\icarus}, \emph{365}, 114509.

\bibitem[{\emph{{Coradini} et~al.}(2007)\emph{{Coradini}, {Capaccioni},
  {Drossart}, {Arnold}, {Ammannito}, {Angrilli}, {Barucci}, {Bellucci},
  {Benkhoff}, {Bianchini}, {Bibring}, {Blecka}, {Bockelee-Morvan}, {Capria},
  {Carlson}, {Carsenty}, {Cerroni}, {Colangeli}, {Combes}, {Combi},
  {Crovisier}, {De Sanctis}, {Encrenaz}, {Erard}, {Federico}, {Filacchione},
  {Fink}, {Fonti}, {Formisano}, {Ip}, {Jaumann}, {Kuehrt}, {Langevin}, {Magni},
  {McCord}, {Mennella}, {Mottola}, {Neukum}, {Palumbo}, {Piccioni}, {Rauer},
  {Saggin}, {Schmitt}, {Tiphene}, and {Tozzi}}}]{RosIns-VIRTIS}
{Coradini} A., {Capaccioni} F., {Drossart} P. et~al. (2007) \emph{{Virtis: An
  Imaging Spectrometer for the Rosetta Mission}}, \emph{\ssr}, \emph{128},
  529--559.

\bibitem[{\emph{{Crow} et~al.}(2011)\emph{{Crow}, {McFadden}, {Robinson},
  {Meadows}, {Livengood}, {Hewagama}, {Barry}, {Deming}, {Lisse}, and
  {Wellnitz}}}]{EPOCh_Crow}
{Crow} C.~A., {McFadden} L.~A., {Robinson} T. et~al. (2011) \emph{{Views from
  EPOXI: Colors in Our Solar System as an Analog for Extrasolar Planets}},
  \emph{\apj}, \emph{729}, 130.

\bibitem[{\emph{{Dachwald} et~al.}(2020)\emph{{Dachwald}, {Ulamec}, {Postberg},
  {Sohl}, {de Vera}, {Waldmann}, {Lorenz}, {Zacny}, {Hellard}, {Biele}, and
  {Rettberg}}}]{OceanWorld_probes}
{Dachwald} B., {Ulamec} S., {Postberg} F. et~al. (2020) \emph{{Key Technologies
  and Instrumentation for Subsurface Exploration of Ocean Worlds}},
  \emph{\ssr}, \emph{216}, 83.

\bibitem[{\emph{{Davidsson} and {Guti{\'e}rrez}}(2005)}]{67Ppole1}
{Davidsson} B. J.~R. and {Guti{\'e}rrez} P.~J. (2005) \emph{{Nucleus properties
  of Comet 67P/Churyumov Gerasimenko estimated from non-gravitational force
  modeling}}, \emph{\icarus}, \emph{176}, 453--477.

\bibitem[{\emph{{De Sanctis} et~al.}(2015)\emph{{De Sanctis}, {Capaccioni},
  {Ciarniello}, {Filacchione}, {Formisano}, {Mottola}, {Raponi}, {Tosi},
  {Bockel{\'e}e-Morvan}, {Erard}, {Leyrat}, {Schmitt}, {Ammannito}, {Arnold},
  {Barucci}, {Combi}, {Capria}, {Cerroni}, {Ip}, {Kuehrt}, {McCord}, {Palomba},
  {Beck}, {Quirico}, {VIRTIS Team}, {Piccioni}, {Bellucci}, {Fulchignoni},
  {Jaumann}, {Stephan}, {Longobardo}, {Mennella}, {Migliorini}, {Benkhoff},
  {Bibring}, {Blanco}, {Blecka}, {Carlson}, {Carsenty}, {Colangeli}, {Combes},
  {Crovisier}, {Drossart}, {Encrenaz}, {Federico}, {Fink}, {Fonti}, {Irwin},
  {Langevin}, {Magni}, {Moroz}, {Orofino}, {Schade}, {Taylor}, {Tiphene},
  {Tozzi}, {Biver}, {Bonal}, {Combe}, {Despan}, {Flamini}, {Fornasier},
  {Frigeri}, {Grassi}, {Gudipati}, {Mancarella}, {Markus}, {Merlin}, {Orosei},
  {Rinaldi}, {Cartacci}, {Cicchetti}, {Giuppi}, {Hello}, {Henry}, {Jacquinod},
  {Rees}, {Noschese}, {Politi}, and {Peter}}}]{DeSanctis_67PVIRTISH2O}
{De Sanctis} M.~C., {Capaccioni} F., {Ciarniello} M. et~al. (2015) \emph{{The
  diurnal cycle of water ice on comet 67P/Churyumov-Gerasimenko}}, \emph{\nat},
  \emph{525}, 500--503.

\bibitem[{\emph{{Dello Russo} et~al.}(2016)\emph{{Dello Russo}, {Kawakita},
  {Vervack}, and {Weaver}}}]{DelloRusso2016}
{Dello Russo} N., {Kawakita} H., {Vervack} R.~J. et~al. (2016) \emph{{Emerging
  trends and a comet taxonomy based on the volatile chemistry measured in
  thirty comets with high-resolution infrared spectroscopy between 1997 and
  2013}}, \emph{\icarus}, \emph{278}, 301--332.

\bibitem[{\emph{{Dello Russo} et~al.}(2011)\emph{{Dello Russo}, {Vervack},
  {Lisse}, {Weaver}, {Kawakita}, {Kobayashi}, {Cochran}, {Harris}, {McKay},
  {Biver}, {Bockel{\'e}e-Morvan}, and {Crovisier}}}]{dellorusso2011}
{Dello Russo} N., {Vervack} R.~J. J., {Lisse} C.~M. et~al. (2011) \emph{{The
  Volatile Composition and Activity of Comet 103P/Hartley 2 During the EPOXI
  Closest Approach}}, \emph{\apj}, \emph{734}, L8.

\bibitem[{\emph{{Devog{\`e}le} et~al.}(2015)\emph{{Devog{\`e}le}, {Rivet},
  {Tanga}, {Bendjoya}, {Surdej}, {Bartczak}, and {Hanus}}}]{Devogele2015}
{Devog{\`e}le} M., {Rivet} J.~P., {Tanga} P. et~al. (2015) \emph{{A method to
  search for large-scale concavities in asteroid shape models}}, \emph{\mnras},
  \emph{453}, 2232--2240.

\bibitem[{\emph{{Dunham} et~al.}(2015)\emph{{Dunham}, {Farquhar}, {Loucks},
  {Roberts}, {Wingo}, {Cowing}, {Garcia}, {Craychee}, {Nickel}, {Ford},
  {Colleluori}, {Folta}, {Giorgini}, {Nace}, {Spohr}, {Dove}, {Mogk},
  {Furfaro}, and {Martin}}}]{Dunham2015}
{Dunham} D.~W., {Farquhar} R.~W., {Loucks} M. et~al. (2015) \emph{{The 2014
  Earth return of the ISEE-3/ICE spacecraft}}, \emph{Acta Astronautica},
  \emph{110}, 29--42.

\bibitem[{\emph{{Duxbury} et~al.}(2004)\emph{{Duxbury}, {Newburn}, and
  {Brownlee}}}]{Duxbury2004}
{Duxbury} T.~C., {Newburn} R.~L., and {Brownlee} D.~E. (2004) \emph{{Comet
  81P/Wild 2 size, shape, and orientation}}, \emph{Journal of Geophysical
  Research (Planets)}, \emph{109}, E12S02.

\bibitem[{\emph{{Economou} et~al.}(2013)\emph{{Economou}, {Green}, {Brownlee},
  and {Clark}}}]{Economou2013}
{Economou} T.~E., {Green} S.~F., {Brownlee} D.~E. et~al. (2013) \emph{{Dust
  Flux Monitor Instrument measurements during Stardust-NExT Flyby of Comet
  9P/Tempel 1}}, \emph{\icarus}, \emph{222}, 526--539.

\bibitem[{\emph{{Elsila} et~al.}(2009)\emph{{Elsila}, {Glavin}, and
  {Dworkin}}}]{Elsilaetal2009}
{Elsila} J.~E., {Glavin} D.~P., and {Dworkin} J.~P. (2009) \emph{{Cometary
  glycine detected in samples returned by Stardust}}, \emph{\maps}, \emph{44},
  1323--1330.

\bibitem[{\emph{{Ernst} and {Schultz}}(2007)}]{Ernst_DI_flash}
{Ernst} C.~M. and {Schultz} P.~H. (2007) \emph{{Evolution of the Deep Impact
  flash: Implications for the nucleus surface based on laboratory
  experiments}}, \emph{\icarus}, \emph{191}, 123--133.

\bibitem[{\emph{{Espley} et~al.}(2015)\emph{{Espley}, {DiBraccio}, {Connerney},
  {Brain}, {Gruesbeck}, {Soobiah}, {Halekas}, {Combi}, {Luhmann}, {Ma}, {Jia},
  and {Jakosky}}}]{Espley2015}
{Espley} J.~R., {DiBraccio} G.~A., {Connerney} J. E.~P. et~al. (2015) \emph{{A
  comet engulfs Mars: MAVEN observations of comet Siding Spring's influence on
  the Martian magnetosphere}}, \emph{\grl}, \emph{42}, 8810--8818.

\bibitem[{\emph{{Farnham} et~al.}(2013)\emph{{Farnham}, {Bodewits}, {Li},
  {Veverka}, {Thomas}, and {Belton}}}]{Farnham_SDN}
{Farnham} T.~L., {Bodewits} D., {Li} J.~Y. et~al. (2013) \emph{{Connections
  between the jet activity and surface features on Comet 9P/Tempel 1}},
  \emph{\icarus}, \emph{222}, 540--549.

\bibitem[{\emph{{Farnham} et~al.}(2015)\emph{{Farnham}, {Delamere}, {Kelley},
  {Heyd}, and {Li}}}]{Farnham2015}
{Farnham} T.~L., {Delamere} W.~A., {Kelley} M. S.~P. et~al. (2015) in
  \emph{AAS/Division for Planetary Sciences Meeting Abstracts \#47}, vol.~47 of
  \emph{AAS/Division for Planetary Sciences Meeting Abstracts}, p. 415.04.

\bibitem[{\emph{{Farnham} et~al.}(2017)\emph{{Farnham}, {Kelley}, {A'Hearn},
  {Feaga}, {Bodewits}, {Sunshine}, {Wellnitz}, and {Wissler}}}]{Farnham_ISON}
{Farnham} T.~L., {Kelley} M.~S.~P., {A'Hearn} M.~F. et~al. (2017) \emph{{Comet
  C/2012 S1 (ISON): Final observations from the Deep Impact spacecraft}},
  \emph{\icarus}, \emph{284}, 106--113.

\bibitem[{\emph{{Farnham} et~al.}(2007)\emph{{Farnham}, {Wellnitz}, {Hampton},
  {Li}, {Sunshine}, {Groussin}, {McFadden}, {Crockett}, {A'Hearn}, {Belton},
  {Schultz}, and {Lisse}}}]{Farnham_DI_outbursts}
{Farnham} T.~L., {Wellnitz} D.~D., {Hampton} D.~L. et~al. (2007) \emph{{Dust
  coma morphology in the Deep Impact images of Comet 9P/Tempel 1}},
  \emph{\icarus}, \emph{191}, 146--160.

\bibitem[{\emph{{Farnocchia} et~al.}(2016)\emph{{Farnocchia}, {Chesley},
  {Micheli}, {Delamere}, {Heyd}, {Tholen}, {Giorgini}, {Owen}, and
  {Tamppari}}}]{Farnocchia2016}
{Farnocchia} D., {Chesley} S.~R., {Micheli} M. et~al. (2016) \emph{{High
  precision comet trajectory estimates: The Mars flyby of C/2013 A1 (Siding
  Spring)}}, \emph{\icarus}, \emph{266}, 279--287.

\bibitem[{\emph{{Feaga} et~al.}(2014)\emph{{Feaga}, {A'Hearn}, {Farnham},
  {Bodewits}, {Sunshine}, {Gersch}, {Protopapa}, {Yang}, {Drahus}, and
  {Schleicher}}}]{Feaga_Garradd}
{Feaga} L.~M., {A'Hearn} M.~F., {Farnham} T.~L. et~al. (2014)
  \emph{{Uncorrelated Volatile Behavior during the 2011 Apparition of Comet
  C/2009 P1 Garradd}}, \emph{\aj}, \emph{147}, 24.

\bibitem[{\emph{{Feaga} et~al.}(2007)\emph{{Feaga}, {A'Hearn}, {Sunshine},
  {Groussin}, and {Farnham}}}]{Feaga_DI_coma}
{Feaga} L.~M., {A'Hearn} M.~F., {Sunshine} J.~M. et~al. (2007)
  \emph{{Asymmetries in the distribution of H$_{2}$O and CO$_{2}$ in the inner
  coma of Comet 9P/Tempel 1 as observed by Deep Impact}}, \emph{\icarus},
  \emph{191}, 134--145.

\bibitem[{\emph{{Feldman} et~al.}(2015)\emph{{Feldman}, {A'Hearn}, {Bertaux},
  {Feaga}, {Parker}, {Schindhelm}, {Steffl}, {Stern}, {Weaver}, {Sierks}, and
  {Vincent}}}]{Feldman2015}
{Feldman} P.~D., {A'Hearn} M.~F., {Bertaux} J.-L. et~al. (2015)
  \emph{{Measurements of the near-nucleus coma of comet
  67P/Churyumov-Gerasimenko with the Alice far-ultraviolet spectrograph on
  Rosetta}}, \emph{\aap}, \emph{583}, A8.

\bibitem[{\emph{{Fern{\'a}ndez} et~al.}(2016)\emph{{Fern{\'a}ndez}, {Gallardo},
  and {Young}}}]{Fernandez2016}
{Fern{\'a}ndez} J.~A., {Gallardo} T., and {Young} J.~D. (2016) \emph{{The end
  states of long-period comets and the origin of Halley-type comets}},
  \emph{\mnras}, \emph{461}, 3075--3088.

\bibitem[{\emph{{Festou} et~al.}(2004)\emph{{Festou}, {Keller}, and
  {Weaver}}}]{CometsIIbook}
{Festou} M.~C., {Keller} H.-U., and {Weaver} H.~A. (Eds.) (2004) \emph{{Comets
  II}}, Univ. of Arizona Press.

\bibitem[{\emph{{Filacchione} et~al.}(2016)\emph{{Filacchione}, {Raponi},
  {Capaccioni}, {Ciarniello}, {Tosi}, {Capria}, {De Sanctis}, {Migliorini},
  {Piccioni}, {Cerroni}, {Barucci}, {Fornasier}, {Schmitt}, {Quirico}, {Erard},
  {Bockelee-Morvan}, {Leyrat}, {Arnold}, {Mennella}, {Ammannito}, {Bellucci},
  {Benkhoff}, {Bibring}, {Blanco}, {Blecka}, {Carlson}, {Carsenty},
  {Colangeli}, {Combes}, {Combi}, {Crovisier}, {Drossart}, {Encrenaz},
  {Federico}, {Fink}, {Fonti}, {Fulchignoni}, {Ip}, {Irwin}, {Jaumann},
  {Kuehrt}, {Langevin}, {Magni}, {McCord}, {Moroz}, {Mottola}, {Palomba},
  {Schade}, {Stephan}, {Taylor}, {Tiphene}, {Tozzi}, {Beck}, {Biver}, {Bonal},
  {Combe}, {Despan}, {Flamini}, {Formisano}, {Frigeri}, {Grassi}, {Gudipati},
  {Kappel}, {Longobardo}, {Mancarella}, {Markus}, {Merlin}, {Orosei},
  {Rinaldi}, {Cartacci}, {Cicchetti}, {Hello}, {Henry}, {Jacquinod}, {Reess},
  {Noschese}, {Politi}, and {Peter}}}]{Filacchione_67PVIRTISCO2}
{Filacchione} G., {Raponi} A., {Capaccioni} F. et~al. (2016) \emph{{Seasonal
  exposure of carbon dioxide ice on the nucleus of comet
  67P/Churyumov-Gerasimenko}}, \emph{Science}, \emph{354}, 1563--1566.

\bibitem[{\emph{{Fink} et~al.}(1999)\emph{{Fink}, {Hicks}, and
  {Fevig}}}]{fink1999}
{Fink} U., {Hicks} M.~P., and {Fevig} R.~A. (1999) \emph{{Production Rates for
  the Stardust Mission Target: 81P/Wild 2}}, \emph{\icarus}, \emph{141},
  331--340.

\bibitem[{\emph{{Finzi} et~al.}(2007)\emph{{Finzi}, {Zazzera}, {Dainese},
  {Malnati}, {Magnani}, {Re}, {Bologna}, {Espinasse}, and
  {Olivieri}}}]{RosIns-SD2}
{Finzi} A.~E., {Zazzera} F.~B., {Dainese} C. et~al. (2007) \emph{{SD2 How To
  Sample A Comet}}, \emph{\ssr}, \emph{128}, 281--299.

\bibitem[{\emph{{Fornasier} et~al.}(2016)\emph{{Fornasier}, {Mottola},
  {Keller}, {Barucci}, {Davidsson}, {Feller}, {Deshapriya}, {Sierks},
  {Barbieri}, {Lamy}, {Rodrigo}, {Koschny}, {Rickman}, {A'Hearn}, {Agarwal},
  {Bertaux}, {Bertini}, {Besse}, {Cremonese}, {Da Deppo}, {Debei}, {De Cecco},
  {Deller}, {El-Maarry}, {Fulle}, {Groussin}, {Gutierrez}, {G{\"u}ttler},
  {Hofmann}, {Hviid}, {Ip}, {Jorda}, {Knollenberg}, {Kovacs}, {Kramm},
  {K{\"u}hrt}, {K{\"u}ppers}, {Lara}, {Lazzarin}, {Moreno}, {Marzari},
  {Massironi}, {Naletto}, {Oklay}, {Pajola}, {Pommerol}, {Preusker},
  {Scholten}, {Shi}, {Thomas}, {Toth}, {Tubiana}, and
  {Vincent}}}]{Fornasier2016}
{Fornasier} S., {Mottola} S., {Keller} H.~U. et~al. (2016)
  \emph{{Rosetta{\textquoteright}s comet 67P/Churyumov-Gerasimenko sheds its
  dusty mantle to reveal its icy nature}}, \emph{Science}, \emph{354},
  1566--1570.

\bibitem[{\emph{{Friedman}}(1976)}]{friedman1976}
{Friedman} L.~D. (1976) in \emph{Shuttle-Based Cometary Science Workshop}, pp.
  251--256.

\bibitem[{\emph{{Galvin} and {Simunac}}(2015)}]{galvin2015}
{Galvin} A.~B. and {Simunac} K. (2015) in \emph{IAU General Assembly}, vol.~29,
  p. 2257907.

\bibitem[{\emph{{Ge} et~al.}(2019)\emph{{Ge}, {Cui}, and {Zhu}}}]{auto-landing}
{Ge} D., {Cui} P., and {Zhu} S. (2019) \emph{{Recent development of autonomous
  GNC technologies for small celestial body descent and landing}},
  \emph{Progress in Aerospace Sciences}, \emph{110}, 100551.

\bibitem[{\emph{{Gilbert} et~al.}(2015)\emph{{Gilbert}, {Lepri}, {Rubin},
  {Combi}, and {Zurbuchen}}}]{gilbert2015}
{Gilbert} J.~A., {Lepri} S.~T., {Rubin} M. et~al. (2015) \emph{{In Situ Plasma
  Measurements of Fragmented Comet 73P Schwassmann-Wachmann 3}}, \emph{\apj},
  \emph{815}, 12.

\bibitem[{\emph{{Glavin} et~al.}(2008)\emph{{Glavin}, {Dworkin}, and
  {Sandford}}}]{Glavinetal2008}
{Glavin} D.~P., {Dworkin} J.~P., and {Sandford} S.~A. (2008) \emph{{Detection
  of cometary amines in samples returned by Stardust}}, \emph{\maps},
  \emph{43}, 399--413.

\bibitem[{\emph{{Gloeckler} et~al.}(2004)\emph{{Gloeckler}, {Allegrini},
  {Elliott}, {McComas}, {Schwadron}, {Geiss}, {von Steiger}, and
  {Jones}}}]{gloeckler2004}
{Gloeckler} G., {Allegrini} F., {Elliott} H.~A. et~al. (2004) \emph{{Cometary
  Ions Trapped in a Coronal Mass Ejection}}, \emph{\apjl}, \emph{604},
  L121--L124.

\bibitem[{\emph{{Gloeckler} et~al.}(2000)\emph{{Gloeckler}, {Geiss},
  {Schwadron}, {Fisk}, {Zurbuchen}, {Ipavich}, {von Steiger}, {Balsiger}, and
  {Wilken}}}]{gloeckler00}
{Gloeckler} G., {Geiss} J., {Schwadron} N.~A. et~al. (2000) \emph{{Interception
  of comet Hyakutake's ion tail at a distance of 500 million kilometres}},
  \emph{\nat}, \emph{404}, 576--578.

\bibitem[{\emph{{Goesmann} et~al.}(2007)\emph{{Goesmann}, {Rosenbauer}, {Roll},
  {Szopa}, {Raulin}, {Sternberg}, {Israel}, {Meierhenrich}, {Thiemann}, and
  {Munoz-Caro}}}]{RosIns-COSAC}
{Goesmann} F., {Rosenbauer} H., {Roll} R. et~al. (2007) \emph{{Cosac, The
  Cometary Sampling and Composition Experiment on Philae}}, \emph{\ssr},
  \emph{128}, 257--280.

\bibitem[{\emph{{Goetz} et~al.}(2021)\emph{{Goetz}, {Gunell}, {Volwerk},
  {Beth}, {Eriksson}, {Galand}, {Henri}, {Nilsson}, {Wedlund}, {Alho},
  {Andersson}, {Andre}, {De Keyser}, {Deca}, {Ge}, {Glassmeier}, {Hajra},
  {Karlsson}, {Kasahara}, {Kolmasova}, {LLera}, {Madanian}, {Mann}, {Mazelle},
  {Odelstad}, {Plaschke}, {Rubin}, {Sanchez-Cano}, {Snodgrass}, and
  {Vigren}}}]{Goetz2021}
{Goetz} C., {Gunell} H., {Volwerk} M. et~al. (2021) \emph{{Cometary plasma
  science}}, \emph{Experimental Astronomy}.

\bibitem[{\emph{{Gowen} et~al.}(2011)\emph{{Gowen}, {Smith}, {Fortes},
  {Barber}, {Brown}, {Church}, {Collinson}, {Coates}, {Collins}, {Crawford},
  {Dehant}, {Chela-Flores}, {Griffiths}, {Grindrod}, {Gurvits}, {Hagermann},
  {Hussmann}, {Jaumann}, {Jones}, {Joy}, {Karatekin}, {Miljkovic}, {Palomba},
  {Pike}, {Prieto-Ballesteros}, {Raulin}, {Sephton}, {Sheridan}, {Sims},
  {Storrie-Lombardi}, {Ambrosi}, {Fielding}, {Fraser}, {Gao}, {Jones}, {Kargl},
  {Karl}, {Macagnano}, {Mukherjee}, {Muller}, {Phipps}, {Pullan}, {Richter},
  {Sohl}, {Snape}, {Sykes}, and {Wells}}}]{Gowen2011}
{Gowen} R.~A., {Smith} A., {Fortes} A.~D. et~al. (2011) \emph{{Penetrators for
  in situ subsurface investigations of Europa}}, \emph{Advances in Space
  Research}, \emph{48}, 725--742.

\bibitem[{\emph{{Graham} et~al.}(2006)\emph{{Graham}, {Teslich}, {Dai},
  {Bradley}, {Kearsley}, and {H{\"o}rz}}}]{Grahametal2006}
{Graham} G.~A., {Teslich} N., {Dai} Z.~R. et~al. (2006) \emph{{Focused ion beam
  recovery of hypervelocity impact residue in experimental craters on metallic
  foils}}, \emph{\maps}, \emph{41}, 159--165.

\bibitem[{\emph{{Green} et~al.}(2004)\emph{{Green}, {McDonnell}, {McBride},
  {Colwell}, {Tuzzolino}, {Economou}, {Tsou}, {Clark}, and
  {Brownlee}}}]{Green2004}
{Green} S.~F., {McDonnell} J.~A.~M., {McBride} N. et~al. (2004) \emph{{The dust
  mass distribution of comet 81P/Wild 2}}, \emph{Journal of Geophysical
  Research (Planets)}, \emph{109}, E12S04.

\bibitem[{\emph{{Grensemann} and {Schwehm}}(1993)}]{GrensSch1993}
{Grensemann} M.~G. and {Schwehm} G. (1993) \emph{{Giotto's second encounter:
  The mission to comet P/Grigg-Skjellerup}}, \emph{\jgr}, \emph{98},
  20907--20910.

\bibitem[{\emph{{Groussin} et~al.}(2007)\emph{{Groussin}, {A'Hearn}, {Li},
  {Thomas}, {Sunshine}, {Lisse}, {Meech}, {Farnham}, {Feaga}, and
  {Delamere}}}]{Groussin_DI_thermal}
{Groussin} O., {A'Hearn} M.~F., {Li} J.~Y. et~al. (2007) \emph{{Surface
  temperature of the nucleus of Comet 9P/Tempel 1}}, \emph{\icarus},
  \emph{191}, 63--72.

\bibitem[{\emph{{Groussin} et~al.}(2019)\emph{{Groussin}, {Attree}, {Brouet},
  {Ciarletti}, {Davidsson}, {Filacchione}, {Fischer}, {Gundlach}, {Knapmeyer},
  {Knollenberg}, {Kokotanekova}, {K{\"u}hrt}, {Leyrat}, {Marshall}, {Pelivan},
  {Skorov}, {Snodgrass}, {Spohn}, and {Tosi}}}]{Groussin2019}
{Groussin} O., {Attree} N., {Brouet} Y. et~al. (2019) \emph{{The Thermal,
  Mechanical, Structural, and Dielectric Properties of Cometary Nuclei After
  Rosetta}}, \emph{\ssr}, \emph{215}, 29.

\bibitem[{\emph{{Groussin} et~al.}(2004)\emph{{Groussin}, {Lamy}, {Jorda}, and
  {Toth}}}]{Groussin_103P_hyper}
{Groussin} O., {Lamy} P., {Jorda} L. et~al. (2004) \emph{{The nuclei of comets
  126P/IRAS and 103P/Hartley 2}}, \emph{\aap}, \emph{419}, 375--383.

\bibitem[{\emph{{Gr{\"u}n} et~al.}(2016)\emph{{Gr{\"u}n}, {Agarwal},
  {Altobelli}, {Altwegg}, {Bentley}, {Biver}, {Della Corte}, {Edberg},
  {Feldman}, {Galand}, {Geiger}, {G{\"o}tz}, {Grieger}, {G{\"u}ttler}, {Henri},
  {Hofstadter}, {Horanyi}, {Jehin}, {Kr{\"u}ger}, {Lee}, {Mannel}, {Morales},
  {Mousis}, {M{\"u}ller}, {Opitom}, {Rotundi}, {Schmied}, {Schmidt}, {Sierks},
  {Snodgrass}, {Soja}, {Sommer}, {Srama}, {Tzou}, {Vincent},
  {Yanamandra-Fisher}, {A'Hearn}, {Erikson}, {Barbieri}, {Barucci}, {Bertaux},
  {Bertini}, {Burch}, {Colangeli}, {Cremonese}, {Da Deppo}, {Davidsson},
  {Debei}, {De Cecco}, {Deller}, {Feaga}, {Ferrari}, {Fornasier}, {Fulle},
  {Gicquel}, {Gillon}, {Green}, {Groussin}, {Guti{\'e}rrez}, {Hofmann},
  {Hviid}, {Ip}, {Ivanovski}, {Jorda}, {Keller}, {Knight}, {Knollenberg},
  {Koschny}, {Kramm}, {K{\"u}hrt}, {K{\"u}ppers}, {Lamy}, {Lara}, {Lazzarin},
  {L{\`o}pez-Moreno}, {Manfroid}, {Epifani}, {Marzari}, {Naletto}, {Oklay},
  {Palumbo}, {Parker}, {Rickman}, {Rodrigo}, {Rodr{\`\i}guez}, {Schindhelm},
  {Shi}, {Sordini}, {Steffl}, {Stern}, {Thomas}, {Tubiana}, {Weaver},
  {Weissman}, {Zakharov}, and {Taylor}}}]{Gruen-outburst}
{Gr{\"u}n} E., {Agarwal} J., {Altobelli} N. et~al. (2016) \emph{{The 2016 Feb
  19 outburst of comet 67P/CG: an ESA Rosetta multi-instrument study}},
  \emph{\mnras}, \emph{462}, S220--S234.

\bibitem[{\emph{{Gulkis} et~al.}(2007)\emph{{Gulkis}, {Frerking}, {Crovisier},
  {Beaudin}, {Hartogh}, {Encrenaz}, {Koch}, {Kahn}, {Salinas}, {Nowicki},
  {Irigoyen}, {Janssen}, {Stek}, {Hofstadter}, {Allen}, {Backus}, {Kamp},
  {Jarchow}, {Steinmetz}, {Deschamps}, {Krieg}, {Gheudin},
  {Bockel{\'e}e-Morvan}, {Biver}, {Encrenaz}, {Despois}, {Ip}, {Lellouch},
  {Mann}, {Muhleman}, {Rauer}, {Schloerb}, and {Spilker}}}]{RosIns-MIRO}
{Gulkis} S., {Frerking} M., {Crovisier} J. et~al. (2007) \emph{{MIRO: Microwave
  Instrument for Rosetta Orbiter}}, \emph{\ssr}, \emph{128}, 561--597.

\bibitem[{\emph{{Gurnett} et~al.}(2015)\emph{{Gurnett}, {Morgan}, {Persoon},
  {Granroth}, {Kopf}, {Plaut}, and {Green}}}]{Gurnett2015}
{Gurnett} D.~A., {Morgan} D.~D., {Persoon} A.~M. et~al. (2015) \emph{{An
  ionized layer in the upper atmosphere of Mars caused by dust impacts from
  comet Siding Spring}}, \emph{\grl}, \emph{42}, 4745--4751.

\bibitem[{\emph{{G{\"u}ttler} et~al.}(2019)\emph{{G{\"u}ttler}, {Mannel},
  {Rotundi}, {Merouane}, {Fulle}, {Bockel{\'e}e-Morvan}, {Lasue},
  {Levasseur-Regourd}, {Blum}, {Naletto}, {Sierks}, {Hilchenbach}, {Tubiana},
  {Capaccioni}, {Paquette}, {Flandes}, {Moreno}, {Agarwal}, {Bodewits},
  {Bertini}, {Tozzi}, {Hornung}, {Langevin}, {Kr{\"u}ger}, {Longobardo}, {Della
  Corte}, {T{\'o}th}, {Filacchione}, {Ivanovski}, {Mottola}, and
  {Rinaldi}}}]{Guettler-dust}
{G{\"u}ttler} C., {Mannel} T., {Rotundi} A. et~al. (2019) \emph{{Synthesis of
  the morphological description of cometary dust at comet
  67P/Churyumov-Gerasimenko}}, \emph{\aap}, \emph{630}, A24.

\bibitem[{\emph{{Hampton} et~al.}(2005)\emph{{Hampton}, {Baer}, {Huisjen},
  {Varner}, {Delamere}, {Wellnitz}, {A'Hearn}, and {Klaasen}}}]{Hampton_DI_SSR}
{Hampton} D.~L., {Baer} J.~W., {Huisjen} M.~A. et~al. (2005) \emph{{An Overview
  of the Instrument Suite for the Deep Impact Mission}}, \emph{\ssr},
  \emph{117}, 43--93.

\bibitem[{\emph{{Hansen} et~al.}(2016)\emph{{Hansen}, {Altwegg}, {Berthelier},
  {Bieler}, {Biver}, {Bockel{\'e}e-Morvan}, {Calmonte}, {Capaccioni}, {Combi},
  {de Keyser}, {Fiethe}, {Fougere}, {Fuselier}, {Gasc}, {Gombosi}, {Huang}, {Le
  Roy}, {Lee}, {Nilsson}, {Rubin}, {Shou}, {Snodgrass}, {Tenishev}, {Toth},
  {Tzou}, {Simon Wedlund}, and {Rosina Team}}}]{KC-waterproduction}
{Hansen} K.~C., {Altwegg} K., {Berthelier} J.~J. et~al. (2016) \emph{{Evolution
  of water production of 67P/Churyumov-Gerasimenko: An empirical model and a
  multi-instrument study}}, \emph{\mnras}, \emph{462}, S491--S506.

\bibitem[{\emph{{Harris} et~al.}(2019)\emph{{Harris}, {Woodney}, and
  {Villanueva}}}]{chimera}
{Harris} W., {Woodney} L., and {Villanueva} G. (2019) in \emph{EPSC-DPS Joint
  Meeting 2019}, vol. 2019, pp. EPSC--DPS2019--1094.

\bibitem[{\emph{{He} et~al.}(2021)\emph{{He}, {Cui}, {Yang}, {Hou}, {Zhang},
  {Ip}, {Jia}, {Dong}, {Duan}, {Zong}, {Bale}, {Pulupa}, {Bonnell}, {Dudok De
  Wit}, {Goetz}, {Harvey}, {MacDowall}, and {Malaspina}}}]{Jiansen2021}
{He} J., {Cui} B., {Yang} L. et~al. (2021) \emph{{The Encounter of the Parker
  Solar Probe and a Comet-like Object Near the Sun: Model Predictions and
  Measurements}}, \emph{\apj}, \emph{910}, 7.

\bibitem[{\emph{{Hein} et~al.}(2022)\emph{{Hein}, {Eubanks}, {Lingam},
  {Hibberd}, {Fries}, {Schneider}, {Kervella}, {Kennedy}, {Perakis}, and
  {Dachwald}}}]{Hein2021}
{Hein} A.~M., {Eubanks} T.~M., {Lingam} M. et~al. (2022) \emph{{Interstellar
  Now! Missions to Explore Nearby Interstellar Objects}}, \emph{Advances in
  Space Research}, \emph{69}, 402--414.

\bibitem[{\emph{{Hein} et~al.}(2019)\emph{{Hein}, {Perakis}, {Eubanks},
  {Hibberd}, {Crowl}, {Hayward}, {Kennedy}, and {Osborne}}}]{Hein2019}
{Hein} A.~M., {Perakis} N., {Eubanks} T.~M. et~al. (2019) \emph{{Project Lyra:
  Sending a spacecraft to 1I/'Oumuamua (former A/2017 U1), the interstellar
  asteroid}}, \emph{Acta Astronautica}, \emph{161}, 552--561.

\bibitem[{\emph{{Heinisch} et~al.}(2019)\emph{{Heinisch}, {Auster}, {Gundlach},
  {Blum}, {G{\"u}ttler}, {Tubiana}, {Sierks}, {Hilchenbach}, {Biele},
  {Richter}, and {Glassmeier}}}]{Heinisch2019}
{Heinisch} P., {Auster} H.~U., {Gundlach} B. et~al. (2019) \emph{{Compressive
  strength of comet 67P/Churyumov-Gerasimenko derived from Philae surface
  contacts}}, \emph{\aap}, \emph{630}, A2.

\bibitem[{\emph{{Herique} et~al.}(2018)\emph{{Herique}, {Agnus}, {Asphaug},
  {Barucci}, {Beck}, {Bellerose}, {Biele}, {Bonal}, {Bousquet}, {Bruzzone},
  {Buck}, {Carnelli}, {Cheng}, {Ciarletti}, {Delbo}, {Du}, {Du}, {Eyraud},
  {Fa}, {Gil Fernandez}, {Gassot}, {Granados-Alfaro}, {Green}, {Grieger},
  {Grundmann}, {Grygorczuk}, {Hahnel}, {Heggy}, {Ho}, {Karatekin}, {Kasaba},
  {Kobayashi}, {Kofman}, {Krause}, {Kumamoto}, {K{\"u}ppers}, {Laabs}, {Lange},
  {Lasue}, {Levasseur-Regourd}, {Mallet}, {Michel}, {Mottola}, {Murdoch},
  {M{\"u}tze}, {Oberst}, {Orosei}, {Plettemeier}, {Rochat}, {RodriguezSuquet},
  {Rogez}, {Schaffer}, {Snodgrass}, {Souyris}, {Tokarz}, {Ulamec}, {Wahlund},
  and {Zine}}}]{Herique2018}
{Herique} A., {Agnus} B., {Asphaug} E. et~al. (2018) \emph{{Direct observations
  of asteroid interior and regolith structure: Science measurement
  requirements}}, \emph{Advances in Space Research}, \emph{62}, 2141--2162.

\bibitem[{\emph{{Herique} et~al.}(2015)\emph{{Herique}, {Rogez}, {Pasquero},
  {Zine}, {Puget}, and {Kofman}}}]{Herique2015}
{Herique} A., {Rogez} Y., {Pasquero} O.~P. et~al. (2015) \emph{{Philae
  localization from CONSERT/Rosetta measurement}}, \emph{\planss}, \emph{117},
  475--484.

\bibitem[{\emph{{Heritier} et~al.}(2017)\emph{{Heritier}, {Altwegg},
  {Balsiger}, {Berthelier}, {Beth}, {Bieler}, {Biver}, {Calmonte}, {Combi}, {De
  Keyser}, {Eriksson}, {Fiethe}, {Fougere}, {Fuselier}, {Galand}, {Gasc},
  {Gombosi}, {Hansen}, {Hassig}, {Kopp}, {Odelstad}, {Rubin}, {Tzou}, {Vigren},
  and {Vuitton}}}]{Heritier2017}
{Heritier} K.~L., {Altwegg} K., {Balsiger} H. et~al. (2017) \emph{{Ion
  composition at comet 67P near perihelion: Rosetta observations and
  model-based interpretation}}, \emph{\mnras}, \emph{469}, S427--S442.

\bibitem[{\emph{{Hermalyn} et~al.}(2013)\emph{{Hermalyn}, {Farnham}, {Collins},
  {Kelley}, {A'Hearn}, {Bodewits}, {Carcich}, {Lindler}, {Lisse}, {Meech},
  {Schultz}, and {Thomas}}}]{Hermalyn2013}
{Hermalyn} B., {Farnham} T.~L., {Collins} S.~M. et~al. (2013) \emph{{The
  detection, localization, and dynamics of large icy particles surrounding
  Comet 103P/Hartley 2}}, \emph{\icarus}, \emph{222}, 625--633.

\bibitem[{\emph{{Hewagama} et~al.}(2018)\emph{{Hewagama}, {Aslam}, {Clark},
  {Daly}, {Feaga}, {Folta}, {Gorius}, {Hurford}, {Livengood}, {Malphrus},
  {Mumma}, {Nixon}, {Sunshine}, {Villanueva}, and {Zucherman}}}]{PrOVE2}
{Hewagama} T., {Aslam} S., {Clark} P. et~al. (2018) in \emph{49th Annual Lunar
  and Planetary Science Conference}, Lunar and Planetary Science Conference, p.
  2800.

\bibitem[{\emph{{Hilchenbach} et~al.}(2016)\emph{{Hilchenbach}, {Kissel},
  {Langevin}, {Briois}, {von Hoerner}, {Koch}, {Schulz}, {Sil{\'e}n},
  {Altwegg}, {Colangeli}, {Cottin}, {Engrand}, {Fischer}, {Glasmachers},
  {Gr{\"u}n}, {Haerendel}, {Henkel}, {H{\"o}fner}, {Hornung}, {Jessberger},
  {Lehto}, {Lehto}, {Raulin}, {Le Roy}, {Ryn{\"o}}, {Steiger}, {Stephan},
  {Thirkell}, {Thomas}, {Torkar}, {Varmuza}, {Wanczek}, {Altobelli},
  {Baklouti}, {Bardyn}, {Fray}, {Kr{\"u}ger}, {Ligier}, {Lin}, {Martin},
  {Merouane}, {Orthous-Daunay}, {Paquette}, {Revillet}, {Siljestr{\"o}m},
  {Stenzel}, and {Zaprudin}}}]{Hilchenbach2016}
{Hilchenbach} M., {Kissel} J., {Langevin} Y. et~al. (2016) \emph{{Comet
  67P/Churyumov-Gerasimenko: Close-up on Dust Particle Fragments}},
  \emph{\apjl}, \emph{816}, L32.

\bibitem[{\emph{{Hoover} et~al.}(2022)\emph{{Hoover}, {Seligman}, and
  {Payne}}}]{Hoover2021}
{Hoover} D.~J., {Seligman} D.~Z., and {Payne} M.~J. (2022) \emph{{The
  Population of Interstellar Objects Detectable with the LSST and Accessible
  for In Situ Rendezvous with Various Mission Designs}}, \emph{\psj}, \emph{3},
  71.

\bibitem[{\emph{{Hoppe} et~al.}(2018)\emph{{Hoppe}, {Rubin}, and
  {Altwegg}}}]{Hoppe2018}
{Hoppe} P., {Rubin} M., and {Altwegg} K. (2018) \emph{{Presolar Isotopic
  Signatures in Meteorites and Comets: New Insights from the Rosetta Mission to
  Comet 67P/Churyumov-Gerasimenko}}, \emph{\ssr}, \emph{214}, 106.

\bibitem[{\emph{{Howell} et~al.}(2018)\emph{{Howell}, {Chou}, {Thompson},
  {Bouchard}, {Cusson}, {Marcus}, {Smith}, {Bhattaru}, {Blalock}, {Brueshaber},
  {Eggl}, {Jawin}, {Miller}, {Rizzo}, {Steakley}, {Thomas}, {Trent}, {Ugelow},
  {Budney}, {Mitchell}, and {Lowes}}}]{chariklo-mission}
{Howell} S.~M., {Chou} L., {Thompson} M. et~al. (2018) \emph{{Camilla: A
  centaur reconnaissance and impact mission concept}}, \emph{\planss},
  \emph{164}, 184--193.

\bibitem[{\emph{{Hu} et~al.}(2017)\emph{{Hu}, {Shi}, {Sierks}, {Fulle}, {Blum},
  {Keller}, {K{\"u}hrt}, {Davidsson}, {G{\"u}ttler}, {Gundlach}, {Pajola},
  {Bodewits}, {Vincent}, {Oklay}, {Massironi}, {Fornasier}, {Tubiana},
  {Groussin}, {Boudreault}, {H{\"o}fner}, {Mottola}, {Barbieri}, {Lamy},
  {Rodrigo}, {Koschny}, {Rickman}, {A'Hearn}, {Agarwal}, {Barucci}, {Bertaux},
  {Bertini}, {Cremonese}, {Da Deppo}, {Debei}, {De Cecco}, {Deller},
  {El-Maarry}, {Gicquel}, {Gutierrez-Marques}, {Guti{\'e}rrez}, {Hofmann},
  {Hviid}, {Ip}, {Jorda}, {Knollenberg}, {Kovacs}, {Kramm}, {K{\"u}ppers},
  {Lara}, {Lazzarin}, {Lopez-Moreno}, {Marzari}, {Naletto}, and
  {Thomas}}}]{Hu2017-fallback}
{Hu} X., {Shi} X., {Sierks} H. et~al. (2017) \emph{{Seasonal erosion and
  restoration of the dust cover on comet 67P/Churyumov-Gerasimenko as observed
  by OSIRIS onboard Rosetta}}, \emph{\aap}, \emph{604}, A114.

\bibitem[{\emph{{Ishibashi} et~al.}(2020)\emph{{Ishibashi}, {Hong}, {Okamoto},
  {Ishimaru}, {Sato}, {Yamada}, {Okudaira}, {Arai}, {Yoshida}, {Kameda},
  {Kagitani}, {Iwata}, {Okada}, and {Takashima}}}]{Ishibashi2020}
{Ishibashi} K., {Hong} P., {Okamoto} T. et~al. (2020) in \emph{51st Annual
  Lunar and Planetary Science Conference}, Lunar and Planetary Science
  Conference, p. 1698.

\bibitem[{\emph{{Jewitt} and {Li}}(2010)}]{Jewitt2010-Phaethon}
{Jewitt} D. and {Li} J. (2010) \emph{{Activity in Geminid Parent (3200)
  Phaethon}}, \emph{\aj}, \emph{140}, 1519--1527.

\bibitem[{\emph{{Jewitt} et~al.}(2013)\emph{{Jewitt}, {Li}, and
  {Agarwal}}}]{Jewitt2013-Phaethon}
{Jewitt} D., {Li} J., and {Agarwal} J. (2013) \emph{{The Dust Tail of Asteroid
  (3200) Phaethon}}, \emph{\apjl}, \emph{771}, L36.

\bibitem[{\emph{{Jewitt} et~al.}(2018)\emph{{Jewitt}, {Weaver}, {Mutchler},
  {Li}, {Agarwal}, and {Larson}}}]{Jewitt2018}
{Jewitt} D., {Weaver} H., {Mutchler} M. et~al. (2018) \emph{{The Nucleus of
  Active Asteroid 311P/(2013 P5) PANSTARRS}}, \emph{\aj}, \emph{155}, 231.

\bibitem[{\emph{{Johnstone} et~al.}(1993)\emph{{Johnstone}, {Coates},
  {Huddleston}, {Jockers}, {Wilken}, {Borg}, {Gurgiolo}, {Winningham}, and
  {Amata}}}]{johnstone1993}
{Johnstone} A.~D., {Coates} A.~J., {Huddleston} D.~E. et~al. (1993)
  \emph{{Observations of the Solar Wind and Cometary Ions during the Encounter
  Between Giotto and Comet Grigg-Skjellerup}}, \emph{\aap}, \emph{273}, L1--L4.

\bibitem[{\emph{Jones et~al.}(2022)\emph{Jones, Elliott, McComas, Hill,
  Vandegriff, Smith, Crary, and Waite}}]{JONES2022115199}
Jones G., Elliott H., McComas D. et~al. (2022) \emph{Cometary ions detected by
  the cassini spacecraft 6.5 au downstream of comet 153p/ikeya-zhang},
  \emph{Icarus}, p. 115199.

\bibitem[{\emph{{Jones} et~al.}(2020)\emph{{Jones}, {Afghan}, and
  {Price}}}]{jones2020}
{Jones} G.~H., {Afghan} Q., and {Price} O. (2020) \emph{{Prospects for the In
  Situ detection of Comet C/2019 Y4 ATLAS by Solar Orbiter}}, \emph{Research
  Notes of the American Astronomical Society}, \emph{4}, 62.

\bibitem[{\emph{{Jones} et~al.}(2018)\emph{{Jones}, {Agarwal}, {Bowles},
  {Burchell}, {Coates}, {Fitzsimmons}, {Graps}, {Hsieh}, {Lisse}, {Lowry},
  {Masters}, {Snodgrass}, and {Tubiana}}}]{Caroline}
{Jones} G.~H., {Agarwal} J., {Bowles} N. et~al. (2018) \emph{{The proposed
  Caroline ESA M3 mission to a Main Belt Comet}}, \emph{Advances in Space
  Research}, \emph{62}, 1921--1946.

\bibitem[{\emph{{Jones} et~al.}(2000)\emph{{Jones}, {Balogh}, and
  {Horbury}}}]{jones00}
{Jones} G.~H., {Balogh} A., and {Horbury} T.~S. (2000) \emph{{Identification of
  comet Hyakutake's extremely long ion tail from magnetic field signatures}},
  \emph{\nat}, \emph{404}, 574--576.

\bibitem[{\emph{{Jones} et~al.}(2003{\natexlab{a}})\emph{{Jones}, {Balogh},
  {McComas}, and {MacDowall}}}]{jones2003}
{Jones} G.~H., {Balogh} A., {McComas} D.~J. et~al. (2003{\natexlab{a}})
  \emph{{Strong interplanetary field enhancements at
  Ulysses{\textemdash}evidence of dust trails' interaction with the solar
  wind?}}, \emph{\icarus}, \emph{166}, 297--310.

\bibitem[{\emph{{Jones} et~al.}(2003{\natexlab{b}})\emph{{Jones}, {Balogh},
  {Russell}, and {Dougherty}}}]{jones2003a}
{Jones} G.~H., {Balogh} A., {Russell} C.~T. et~al. (2003{\natexlab{b}})
  \emph{{Possible Distortion of the Interplanetary Magnetic Field by the Dust
  Trail of Comet 122P/de Vico}}, \emph{\apjl}, \emph{597}, L61--L64.

\bibitem[{\emph{{Jorda} et~al.}(2016)\emph{{Jorda}, {Gaskell}, {Capanna},
  {Hviid}, {Lamy}, {{\v{D}}urech}, {Faury}, {Groussin}, {Guti{\'e}rrez},
  {Jackman}, {Keihm}, {Keller}, {Knollenberg}, {K{\"u}hrt}, {Marchi},
  {Mottola}, {Palmer}, {Schloerb}, {Sierks}, {Vincent}, {A'Hearn}, {Barbieri},
  {Rodrigo}, {Koschny}, {Rickman}, {Barucci}, {Bertaux}, {Bertini},
  {Cremonese}, {Da Deppo}, {Davidsson}, {Debei}, {De Cecco}, {Fornasier},
  {Fulle}, {G{\"u}ttler}, {Ip}, {Kramm}, {K{\"u}ppers}, {Lara}, {Lazzarin},
  {Lopez Moreno}, {Marzari}, {Naletto}, {Oklay}, {Thomas}, {Tubiana}, and
  {Wenzel}}}]{Jorda2016}
{Jorda} L., {Gaskell} R., {Capanna} C. et~al. (2016) \emph{{The global shape,
  density and rotation of Comet 67P/Churyumov-Gerasimenko from preperihelion
  Rosetta/OSIRIS observations}}, \emph{\icarus}, \emph{277}, 257--278.

\bibitem[{\emph{{Kearsley} et~al.}(2012)\emph{{Kearsley}, {Burchell}, {Price},
  {Cole}, {Wozniakiewicz}, {Ishii}, {Bradley}, {Fries}, and
  {Foster}}}]{Kearsley2012}
{Kearsley} A.~T., {Burchell} M.~J., {Price} M.~C. et~al. (2012)
  \emph{{Experimental impact features in Stardust aerogel: How track morphology
  reflects particle structure, composition, and density}}, \emph{\maps},
  \emph{47}, 737--762.

\bibitem[{\emph{{Keeney} et~al.}(2019)\emph{{Keeney}, {Stern}, {Feldman},
  {A'Hearn}, {Bertaux}, {Feaga}, {Knight}, {Medina}, {Noonan}, {Parker},
  {Pineau}, {Schindhelm}, {Steffl}, {Versteeg}, {Vervack}, and
  {Weaver}}}]{Keeney_O2}
{Keeney} B.~A., {Stern} S.~A., {Feldman} P.~D. et~al. (2019) \emph{{Stellar
  Occultation by Comet 67P/Churyumov-Gerasimenko Observed with Rosetta's Alice
  Far-ultraviolet Spectrograph}}, \emph{\aj}, \emph{157}, 173.

\bibitem[{\emph{{Keller}}(2006)}]{Keller2006-OSIRIS-comets}
{Keller} H.~U. (2006) in \emph{IAU Joint Discussion}, vol.~26 of \emph{IAU
  Joint Discussion}, p.~29.

\bibitem[{\emph{{Keller} et~al.}(1986)\emph{{Keller}, {Arpigny}, {Barbieri},
  {Bonnet}, {Cazes}, {Coradini}, {Cosmovici}, {Delamere}, {Huebner}, {Hughes},
  {Jamar}, {Malaise}, {Reitsema}, {Schmidt}, {Schmidt}, {Seige}, {Whipple}, and
  {Wilhelm}}}]{Keller1986}
{Keller} H.~U., {Arpigny} C., {Barbieri} C. et~al. (1986) \emph{{First Halley
  Multicolour Camera imaging results from Giotto}}, \emph{\nat}, \emph{321},
  320--326.

\bibitem[{\emph{{Keller} et~al.}(2007{\natexlab{a}})\emph{{Keller}, {Barbieri},
  {Lamy}, {Rickman}, {Rodrigo}, {Wenzel}, {Sierks}, {A'Hearn}, {Angrilli},
  {Angulo}, {Bailey}, {Barthol}, {Barucci}, {Bertaux}, {Bianchini}, {Boit},
  {Brown}, {Burns}, {B{\"u}ttner}, {Castro}, {Cremonese}, {Curdt}, {da Deppo},
  {Debei}, {de Cecco}, {Dohlen}, {Fornasier}, {Fulle}, {Germerott}, {Gliem},
  {Guizzo}, {Hviid}, {Ip}, {Jorda}, {Koschny}, {Kramm}, {K{\"u}hrt},
  {K{\"u}ppers}, {Lara}, {Llebaria}, {L{\'o}pez}, {L{\'o}pez-Jimenez},
  {L{\'o}pez-Moreno}, {Meller}, {Michalik}, {Michelena}, {M{\"u}ller},
  {Naletto}, {Orign{\'e}}, {Parzianello}, {Pertile}, {Quintana}, {Ragazzoni},
  {Ramous}, {Reiche}, {Reina}, {Rodr{\'\i}guez}, {Rousset}, {Sabau}, {Sanz},
  {Sivan}, {St{\"o}ckner}, {Tabero}, {Telljohann}, {Thomas}, {Timon},
  {Tomasch}, {Wittrock}, and {Zaccariotto}}}]{RosIns-OSIRIS}
{Keller} H.~U., {Barbieri} C., {Lamy} P. et~al. (2007{\natexlab{a}})
  \emph{{OSIRIS The Scientific Camera System Onboard Rosetta}}, \emph{\ssr},
  \emph{128}, 433--506.

\bibitem[{\emph{{Keller} et~al.}(2004)\emph{{Keller}, {Britt}, {Buratti}, and
  {Thomas}}}]{Keller-cometsII}
{Keller} H.~U., {Britt} D., {Buratti} B.~J. et~al. (2004) \emph{{In situ
  observations of cometary nuclei}}, p. 211.

\bibitem[{\emph{{Keller} et~al.}(2005)\emph{{Keller}, {Jorda}, {K{\"u}ppers},
  {Gutierrez}, {Hviid}, {Knollenberg}, {Lara}, {Sierks}, {Barbieri}, {Lamy},
  {Rickman}, and {Rodrigo}}}]{Keller2005}
{Keller} H.~U., {Jorda} L., {K{\"u}ppers} M. et~al. (2005) \emph{{Deep Impact
  Observations by OSIRIS Onboard the Rosetta Spacecraft}}, \emph{Science},
  \emph{310}, 281--283.

\bibitem[{\emph{{Keller} and {K{\"u}hrt}}(2020)}]{grumpy-Keller}
{Keller} H.~U. and {K{\"u}hrt} E. (2020) \emph{{Cometary
  Nuclei{\textemdash}From Giotto to Rosetta}}, \emph{\ssr}, \emph{216}, 14.

\bibitem[{\emph{{Keller} et~al.}(2007{\natexlab{b}})\emph{{Keller},
  {K{\"u}ppers}, {Fornasier}, {Guti{\'e}rrez}, {Hviid}, {Jorda}, {Knollenberg},
  {Lowry}, {Rengel}, {Bertini}, {Cremonese}, {Ip}, {Koschny}, {Kramm},
  {K{\"u}hrt}, {Lara}, {Sierks}, {Thomas}, {Barbieri}, {Lamy}, {Rickman},
  {Rodrigo}, {A'Hearn}, {Angrilli}, {Barucci}, {Bertaux}, {da Deppo},
  {Davidsson}, {de Cecco}, {Debei}, {Fulle}, {Gliem}, {Groussin}, {Lopez
  Moreno}, {Marzari}, {Naletto}, {Sabau}, {Sanz Andr{\'e}s}, and
  {Wenzel}}}]{Keller2007}
{Keller} H.~U., {K{\"u}ppers} M., {Fornasier} S. et~al. (2007{\natexlab{b}})
  \emph{{Observations of Comet 9P/Tempel 1 around the Deep Impact event by the
  OSIRIS cameras onboard Rosetta}}, \emph{\icarus}, \emph{187}, 87--103.

\bibitem[{\emph{{Keller} et~al.}(2017)\emph{{Keller}, {Mottola}, {Hviid},
  {Agarwal}, {K{\"u}hrt}, {Skorov}, {Otto}, {Vincent}, {Oklay}, {Schr{\"o}der},
  {Davidsson}, {Pajola}, {Shi}, {Bodewits}, {Toth}, {Preusker}, {Scholten},
  {Sierks}, {Barbieri}, {Lamy}, {Rodrigo}, {Koschny}, {Rickman}, {A'Hearn},
  {Barucci}, {Bertaux}, {Bertini}, {Cremonese}, {Da Deppo}, {Debei}, {De
  Cecco}, {Deller}, {Fornasier}, {Fulle}, {Groussin}, {Guti{\'e}rrez},
  {G{\"u}ttler}, {Hofmann}, {Ip}, {Jorda}, {Knollenberg}, {Kramm},
  {K{\"u}ppers}, {Lara}, {Lazzarin}, {Lopez-Moreno}, {Marzari}, {Naletto},
  {Tubiana}, and {Thomas}}}]{Keller2017}
{Keller} H.~U., {Mottola} S., {Hviid} S.~F. et~al. (2017) \emph{{Seasonal mass
  transfer on the nucleus of comet 67P/Chuyumov-Gerasimenko}}, \emph{\mnras},
  \emph{469}, S357--S371.

\bibitem[{\emph{{Keller} et~al.}(2015)\emph{{Keller}, {Mottola}, {Skorov}, and
  {Jorda}}}]{Keller2015}
{Keller} H.~U., {Mottola} S., {Skorov} Y. et~al. (2015) \emph{{The changing
  rotation period of comet 67P/Churyumov-Gerasimenko controlled by its
  activity}}, \emph{\aap}, \emph{579}, L5.

\bibitem[{\emph{{Keller} et~al.}(2006)\emph{{Keller}, {Bajt}, {Baratta},
  {Borg}, {Bradley}, {Brownlee}, {Busemann}, {Brucato}, {Burchell},
  {Colangeli}, {D'Hendecourt}, {Djouadi}, {Ferrini}, {Flynn}, {Franchi},
  {Fries}, {Grady}, {Graham}, {Grossemy}, {Kearsley}, {Matrajt},
  {Nakamura-Messenger}, {Mennella}, {Nittler}, {Palumbo}, {Stadermann}, {Tsou},
  {Rotundi}, {Sandford}, {Snead}, {Steele}, {Wooden}, and
  {Zolensky}}}]{Kelleretal2006}
{Keller} L.~P., {Bajt} S., {Baratta} G.~A. et~al. (2006) \emph{{Infrared
  Spectroscopy of Comet 81P/Wild 2 Samples Returned by Stardust}},
  \emph{Science}, \emph{314}, 1728.

\bibitem[{\emph{{Kelley} et~al.}(2013)\emph{{Kelley}, {Lindler}, {Bodewits},
  {A'Hearn}, {Lisse}, {Kolokolova}, {Kissel}, and
  {Hermalyn}}}]{Kelley_H2_particles}
{Kelley} M.~S., {Lindler} D.~J., {Bodewits} D. et~al. (2013) \emph{{A
  distribution of large particles in the coma of Comet 103P/Hartley 2}},
  \emph{\icarus}, \emph{222}, 634--652.

\bibitem[{\emph{{Kissel} et~al.}(2007)\emph{{Kissel}, {Altwegg}, {Clark},
  {Colangeli}, {Cottin}, {Czempiel}, {Eibl}, {Engrand}, {Fehringer},
  {Feuerbacher}, {Fomenkova}, {Glasmachers}, {Greenberg}, {Gr{\"u}n},
  {Haerendel}, {Henkel}, {Hilchenbach}, {von Hoerner}, {H{\"o}fner}, {Hornung},
  {Jessberger}, {Koch}, {Kr{\"u}ger}, {Langevin}, {Parigger}, {Raulin},
  {R{\"u}denauer}, {Ryn{\"o}}, {Schmid}, {Schulz}, {Sil{\'e}n}, {Steiger},
  {Stephan}, {Thirkell}, {Thomas}, {Torkar}, {Utterback}, {Varmuza}, {Wanczek},
  {Werther}, and {Zscheeg}}}]{RosIns-COSIMA}
{Kissel} J., {Altwegg} K., {Clark} B.~C. et~al. (2007) \emph{{Cosima High
  Resolution Time-of-Flight Secondary Ion Mass Spectrometer for the Analysis of
  Cometary Dust Particles onboard Rosetta}}, \emph{\ssr}, \emph{128}, 823--867.

\bibitem[{\emph{{Kissel} et~al.}(2003)\emph{{Kissel}, {Glasmachers},
  {Gr{\"u}n}, {Henkel}, {H{\"o}fner}, {Haerendel}, {von Hoerner}, {Hornung},
  {Jessberger}, {Krueger}, {M{\"o}hlmann}, {Greenberg}, {Langevin},
  {Sil{\'e}n}, {Brownlee}, {Clark}, {Hanner}, {Hoerz}, {Sandford}, {Sekanina},
  {Tsou}, {Utterback}, {Zolensky}, and {Heiss}}}]{Kissel2003}
{Kissel} J., {Glasmachers} A., {Gr{\"u}n} E. et~al. (2003) \emph{{Cometary and
  Interstellar Dust Analyzer for comet Wild 2}}, \emph{Journal of Geophysical
  Research (Planets)}, \emph{108}, 8114.

\bibitem[{\emph{{Kissel} et~al.}(2004)\emph{{Kissel}, {Krueger}, {Sil{\'e}n},
  and {Clark}}}]{Kissel2004}
{Kissel} J., {Krueger} F.~R., {Sil{\'e}n} J. et~al. (2004) \emph{{The Cometary
  and Interstellar Dust Analyzer at Comet 81P/Wild 2}}, \emph{Science},
  \emph{304}, 1774--1776.

\bibitem[{\emph{{Klaasen} et~al.}(2013{\natexlab{a}})\emph{{Klaasen},
  {A'Hearn}, {Besse}, {Bodewits}, {Carcich}, {Farnham}, {Feaga}, {Groussin},
  {Hampton}, {Huisjen}, {Kelley}, {McLaughlin}, {Merlin}, {Protopapa},
  {Sunshine}, {Thomas}, and {Wellnitz}}}]{Klaasen_DIXI_cal_paper}
{Klaasen} K.~P., {A'Hearn} M., {Besse} S. et~al. (2013{\natexlab{a}})
  \emph{{EPOXI instrument calibration}}, \emph{\icarus}, \emph{225}, 643--680.

\bibitem[{\emph{{Klaasen} et~al.}(2008)\emph{{Klaasen}, {A'Hearn}, {Baca},
  {Delamere}, {Desnoyer}, {Farnham}, {Groussin}, {Hampton}, {Ipatov}, {Li},
  {Lisse}, {Mastrodemos}, {McLaughlin}, {Sunshine}, {Thomas}, and
  {Wellnitz}}}]{Klaasen_DI_cal}
{Klaasen} K.~P., {A'Hearn} M.~F., {Baca} M. et~al. (2008) \emph{{Invited
  Article: Deep Impact instrument calibration}}, \emph{Review of Scientific
  Instruments}, \emph{79}, 091301--091301.

\bibitem[{\emph{{Klaasen} et~al.}(2013{\natexlab{b}})\emph{{Klaasen}, {Brown},
  {Carcich}, {Farnham}, {Owen}, and {Thomas}}}]{Klaasen2013}
{Klaasen} K.~P., {Brown} D., {Carcich} B. et~al. (2013{\natexlab{b}})
  \emph{{Stardust-NExT NAVCAM calibration and performance}}, \emph{\icarus},
  \emph{222}, 436--452.

\bibitem[{\emph{{Klingelh{\"o}fer} et~al.}(2007)\emph{{Klingelh{\"o}fer},
  {Br{\"u}ckner}, {D'Uston}, {Gellert}, and {Rieder}}}]{RosIns-APXS}
{Klingelh{\"o}fer} G., {Br{\"u}ckner} J., {D'Uston} C. et~al. (2007) \emph{{The
  Rosetta Alpha Particle X-Ray Spectrometer (APXS)}}, \emph{\ssr}, \emph{128},
  383--396.

\bibitem[{\emph{{Knight} et~al.}(2016)\emph{{Knight}, {Fitzsimmons}, {Kelley},
  and {Snodgrass}}}]{Knight2016}
{Knight} M.~M., {Fitzsimmons} A., {Kelley} M. S.~P. et~al. (2016) \emph{{Comet
  322P/SOHO 1: An Asteroid with the Smallest Perihelion Distance?}},
  \emph{\apjl}, \emph{823}, L6.

\bibitem[{\emph{{Knight} et~al.}(2015)\emph{{Knight}, {Mueller}, {Samarasinha},
  and {Schleicher}}}]{103P-rotation}
{Knight} M.~M., {Mueller} B. E.~A., {Samarasinha} N.~H. et~al. (2015) \emph{{A
  Further Investigation of Apparent Periodicities and the Rotational State of
  Comet 103P/Hartley 2 from Combined Coma Morphology and Light Curve Data
  Sets}}, \emph{\aj}, \emph{150}, 22.

\bibitem[{\emph{{Kofman} et~al.}(2007)\emph{{Kofman}, {Herique}, {Goutail},
  {Hagfors}, {Williams}, {Nielsen}, {Barriot}, {Barbin}, {Elachi}, {Edenhofer},
  {Levasseur-Regourd}, {Plettemeier}, {Picardi}, {Seu}, and
  {Svedhem}}}]{RosIns-CONSERT}
{Kofman} W., {Herique} A., {Goutail} J.~P. et~al. (2007) \emph{{The Comet
  Nucleus Sounding Experiment by Radiowave Transmission (CONSERT): A Short
  Description of the Instrument and of the Commissioning Stages}}, \emph{\ssr},
  \emph{128}, 413--432.

\bibitem[{\emph{{Kofman} et~al.}(2020)\emph{{Kofman}, {Zine}, {Herique},
  {Rogez}, {Jorda}, and {Levasseur-Regourd}}}]{Kofman2020}
{Kofman} W., {Zine} S., {Herique} A. et~al. (2020) \emph{{The interior of Comet
  67P/C-G; revisiting CONSERT results with the exact position of the Philae
  lander}}, \emph{\mnras}, \emph{497}, 2616--2622.

\bibitem[{\emph{{Kramer} and {L{\"a}uter}}(2019)}]{Kramer2019}
{Kramer} T. and {L{\"a}uter} M. (2019) \emph{{Outgassing-induced acceleration
  of comet 67P/Churyumov-Gerasimenko}}, \emph{\aap}, \emph{630}, A4.

\bibitem[{\emph{{Krankowsky} et~al.}(1986)\emph{{Krankowsky}, {Lammerzahl},
  {Herrwerth}, {Woweries}, {Eberhardt}, {Dolder}, {Herrmann}, {Schulte},
  {Berthelier}, {Illiano}, {Hodges}, and {Hoffman}}}]{krankowsky1986}
{Krankowsky} D., {Lammerzahl} P., {Herrwerth} I. et~al. (1986) \emph{{In situ
  gas and ion measurements at comet Halley}}, \emph{\nat}, \emph{321},
  326--329.

\bibitem[{\emph{{K\"uppers}}(2017)}]{Kueppers2017}
{K\"uppers} M. (2017) in \emph{{From Giotto to Rosetta: 30 Years of Cometary
  Science from Space and Ground}} (C.~{Barbieri} and C.~G. {Someda}, eds.), pp.
  167--181, Accademia Galileiana di Sicienze, Lettere ed Arti, Padova.

\bibitem[{\emph{{K{\"u}ppers} et~al.}(2005)\emph{{K{\"u}ppers}, {Bertini},
  {Fornasier}, {Gutierrez}, {Hviid}, {Jorda}, {Keller}, {Knollenberg},
  {Koschny}, {Kramm}, {Lara}, {Sierks}, {Thomas}, {Barbieri}, {Lamy},
  {Rickman}, {Rodrigo}, {A'Hearn}, {Angrilli}, {Bailey}, {Barthol}, {Barucci},
  {Bertaux}, {Burns}, {Cremonese}, {Curdt}, {De Cecco}, {Debei}, {Fulle},
  {Gliem}, {Ip}, {Huhrt}, {Llebaria}, {Lopez Moreno}, {Marzari}, {Naletto},
  {Sabau}, {Sanz Andres}, {Sivan}, {Tondello}, and {Wenzel}}}]{Kueppers2005}
{K{\"u}ppers} M., {Bertini} I., {Fornasier} S. et~al. (2005) \emph{{A large
  dust/ice ratio in the nucleus of comet 9P/Tempel 1}}, \emph{\nat},
  \emph{437}, 987--990.

\bibitem[{\emph{{K{\"u}ppers} et~al.}(2009)\emph{{K{\"u}ppers}, {Keller},
  {K{\"u}hrt}, {A'Hearn}, {Altwegg}, {Bertrand}, {Busemann}, {Capria},
  {Colangeli}, {Davidsson}, {Ehrenfreund}, {Knollenberg}, {Mottola}, {Rathke},
  {Weiss}, {Zolensky}, {Akim}, {Basilevsky}, {Galimov}, {Gerasimov},
  {Korablev}, {Lomakin}, {Marov}, {Martynov}, {Nazarov}, {Zakharov}, {Zelenyi},
  {Aronica}, {Ball}, {Barbieri}, {Bar-Nun}, {Benkhoff}, {Biele}, {Biver},
  {Blum}, {Bockel{\'e}e-Morvan}, {Botta}, {Bredeh{\"o}ft}, {Capaccioni},
  {Charnley}, {Cloutis}, {Cottin}, {Cremonese}, {Crovisier}, {Crowther},
  {Epifani}, {Esposito}, {Ferrari}, {Ferri}, {Fulle}, {Gilmour}, {Goesmann},
  {Gortsas}, {Green}, {Groussin}, {Gr{\"u}n}, {Guti{\'e}rrez}, {Hartogh},
  {Henkel}, {Hilchenbach}, {Ho}, {Horneck}, {Hviid}, {Ip}, {J{\"a}ckel},
  {Jessberger}, {Kallenbach}, {Kargl}, {K{\"o}mle}, {Korth}, {Kossacki},
  {Krause}, {Kr{\"u}ger}, {Li}, {Licandro}, {Lopez-Moreno}, {Lowry}, {Lyon},
  {Magni}, {Mall}, {Mann}, {Markiewicz}, {Martins}, {Maurette}, {Meierhenrich},
  {Mennella}, {Ng}, {Nittler}, {Palumbo}, {P{\"a}tzold}, {Prialnik}, {Rengel},
  {Rickman}, {Rodriguez}, {Roll}, {Rost}, {Rotundi}, {Sandford},
  {Sch{\"o}nb{\"a}chler}, {Sierks}, {Srama}, {Stroud}, {Szutowicz}, {Tornow},
  {Ulamec}, {Wallis}, {Waniak}, {Weissman}, {Wieler}, {Wurz}, {Yung}, and
  {Zarnecki}}}]{TripleF}
{K{\"u}ppers} M., {Keller} H.~U., {K{\"u}hrt} E. et~al. (2009) \emph{{Triple
  F{\textemdash}a comet nucleus sample return mission}}, \emph{Experimental
  Astronomy}, \emph{23}, 809--847.

\bibitem[{\emph{{Lai} et~al.}(2015)\emph{{Lai}, {Russell}, {Jia}, {Wei}, and
  {Angelopoulos}}}]{lai2015}
{Lai} H.~R., {Russell} C.~T., {Jia} Y.~D. et~al. (2015) \emph{{Momentum
  transfer from solar wind to interplanetary field enhancements inferred from
  magnetic field draping signatures}}, \emph{\grl}, \emph{42}, 1640--1645.

\bibitem[{\emph{{Lamy} et~al.}(2006)\emph{{Lamy}, {Toth}, {Weaver}, {Jorda},
  {Kaasalainen}, and {Guti{\'e}rrez}}}]{Lamy2006}
{Lamy} P.~L., {Toth} I., {Weaver} H.~A. et~al. (2006) \emph{{Hubble Space
  Telescope observations of the nucleus and inner coma of comet
  67P/Churyumov-Gerasimenko}}, \emph{\aap}, \emph{458}, 669--678.

\bibitem[{\emph{{Lara} et~al.}(2006)\emph{{Lara}, {Boehnhardt}, {Gredel},
  {Guti{\'e}rrez}, {Ortiz}, {Rodrigo}, and {Vidal-Nu{\~n}ez}}}]{Lara2006}
{Lara} L.~M., {Boehnhardt} H., {Gredel} R. et~al. (2006) \emph{{Pre-impact
  monitoring of Comet 9P/Tempel 1, the Deep Impact target}}, \emph{\aap},
  \emph{445}, 1151--1157.

\bibitem[{\emph{{Ledbetter} et~al.}(2018)\emph{{Ledbetter}, {Sood}, and
  {Keane}}}]{ChipSats}
{Ledbetter} W.~G., {Sood} R., and {Keane} J.~T. (2018) in \emph{49th Annual
  Lunar and Planetary Science Conference}, Lunar and Planetary Science
  Conference, p. 2136.

\bibitem[{\emph{{Levasseur-Regourd} et~al.}(1993)\emph{{Levasseur-Regourd},
  {Goidet}, {Le Duin}, {Malique}, {Renard}, and {Bertaux}}}]{ACLR1993}
{Levasseur-Regourd} A.~C., {Goidet} B., {Le Duin} T. et~al. (1993) \emph{{Short
  Communication: Optical probing of dust in comet Grigg-Skjellerup from the
  Giotto spacecraft}}, \emph{\planss}, \emph{41}, 167--169.

\bibitem[{\emph{{Levison} et~al.}(2006)\emph{{Levison}, {Duncan}, {Dones}, and
  {Gladman}}}]{Levison2006}
{Levison} H.~F., {Duncan} M.~J., {Dones} L. et~al. (2006) \emph{{The scattered
  disk as a source of Halley-type comets}}, \emph{\icarus}, \emph{184},
  619--633.

\bibitem[{\emph{{Li} and {Jewitt}}(2013)}]{Li2013-Phaethon}
{Li} J. and {Jewitt} D. (2013) \emph{{Recurrent Perihelion Activity in (3200)
  Phaethon}}, \emph{\aj}, \emph{145}, 154.

\bibitem[{\emph{{Li} et~al.}(2007)\emph{{Li}, {A'Hearn}, {Belton}, {Crockett},
  {Farnham}, {Lisse}, {McFadden}, {Meech}, {Sunshine}, {Thomas}, and
  {Veverka}}}]{Li_DI_photometry}
{Li} J.-Y., {A'Hearn} M.~F., {Belton} M. J.~S. et~al. (2007) \emph{{Deep Impact
  photometry of Comet 9P/Tempel 1}}, \emph{\icarus}, \emph{191}, 161--175.

\bibitem[{\emph{{Lindler} et~al.}(2007)\emph{{Lindler}, {Busko}, {A'Hearn}, and
  {White}}}]{Lindler_deconvolve}
{Lindler} D., {Busko} I., {A'Hearn} M.~F. et~al. (2007) \emph{{Restoration of
  Images of Comet 9P/Tempel 1 Taken with the Deep Impact High Resolution
  Instrument}}, \emph{\pasp}, \emph{119}, 427--436.

\bibitem[{\emph{{Lindler} et~al.}(2012)\emph{{Lindler}, {A'Hearn}, and
  {McLaughlin}}}]{LindlerPDS}
{Lindler} D.~J., {A'Hearn} M.~F., and {McLaughlin} S.~A. (2012) \emph{{EPOXI
  103P/Hartley 2 Encounter - HRIV Deconvolved Images V1.0}}, \emph{NASA
  Planetary Data System}, DIF-C-HRIV-5-EPOXI-HARTLEY2-DECONV-V1.0.

\bibitem[{\emph{{Lis} et~al.}(2019)\emph{{Lis}, {Bockel{\'e}e-Morvan},
  {G{\"u}sten}, {Biver}, {Stutzki}, {Delorme}, {Dur{\'a}n}, {Wiesemeyer}, and
  {Okada}}}]{Lis2019}
{Lis} D.~C., {Bockel{\'e}e-Morvan} D., {G{\"u}sten} R. et~al. (2019)
  \emph{{Terrestrial deuterium-to-hydrogen ratio in water in hyperactive
  comets}}, \emph{\aap}, \emph{625}, L5.

\bibitem[{\emph{{Lisse}}(2015)}]{Lisse2015}
{Lisse} C. (2015) \emph{{Comet Siding Spring, up close and personal}},
  \emph{Science}, \emph{350}, 277--278.

\bibitem[{\emph{{Lorenz}}(2011)}]{Lorenz2011}
{Lorenz} R.~D. (2011) \emph{{Planetary penetrators: Their origins, history and
  future}}, \emph{Advances in Space Research}, \emph{48}, 403--431.

\bibitem[{\emph{{Lowry} et~al.}(2012)\emph{{Lowry}, {Duddy}, {Rozitis},
  {Green}, {Fitzsimmons}, {Snodgrass}, {Hsieh}, and {Hainaut}}}]{67Ppole2}
{Lowry} S., {Duddy} S.~R., {Rozitis} B. et~al. (2012) \emph{{The nucleus of
  Comet 67P/Churyumov-Gerasimenko. A new shape model and thermophysical
  analysis}}, \emph{\aap}, \emph{548}, A12.

\bibitem[{\emph{{Luspay-Kuti} et~al.}(2022)\emph{{Luspay-Kuti}, {Mousis},
  {Pauzat}, {Ozgurel}, {Ellinger}, {Lunine}, {Fuselier}, {Mandt}, {Trattner},
  and {Petrinec}}}]{Luspaykuti_O2}
{Luspay-Kuti} A., {Mousis} O., {Pauzat} F. et~al. (2022) \emph{{Dual storage
  and release of molecular oxygen in comet 67P/Churyumov-Gerasimenko}},
  \emph{Nature Astronomy}.

\bibitem[{\emph{{Marschall} et~al.}(2020)\emph{{Marschall}, {Skorov},
  {Zakharov}, {Rezac}, {Gerig}, {Christou}, {Dadzie}, {Migliorini}, {Rinaldi},
  {Agarwal}, {Vincent}, and {Kappel}}}]{Marschall2020}
{Marschall} R., {Skorov} Y., {Zakharov} V. et~al. (2020) \emph{{Cometary
  Comae-Surface Links}}, \emph{\ssr}, \emph{216}, 130.

\bibitem[{\emph{{Marty} et~al.}(2017)\emph{{Marty}, {Altwegg}, {Balsiger},
  {Bar-Nun}, {Bekaert}, {Berthelier}, {Bieler}, {Briois}, {Calmonte}, {Combi},
  {De Keyser}, {Fiethe}, {Fuselier}, {Gasc}, {Gombosi}, {Hansen}, {H{\"a}ssig},
  {J{\"a}ckel}, {Kopp}, {Korth}, {Le Roy}, {Mall}, {Mousis}, {Owen},
  {R{\`e}me}, {Rubin}, {S{\'e}mon}, {Tzou}, {Waite}, and {Wurz}}}]{Marty2017}
{Marty} B., {Altwegg} K., {Balsiger} H. et~al. (2017) \emph{{Xenon isotopes in
  67P/Churyumov-Gerasimenko show that comets contributed to Earth's
  atmosphere}}, \emph{Science}, \emph{356}, 1069--1072.

\bibitem[{\emph{{Mastrodemos} et~al.}(2005)\emph{{Mastrodemos}, {Kubitschek},
  and {Synnott}}}]{Mastrodemos_DI_SSR}
{Mastrodemos} N., {Kubitschek} D.~G., and {Synnott} S.~P. (2005)
  \emph{{Autonomous Navigation for the Deep Impact Mission Encounter with Comet
  Tempel 1}}, \emph{\ssr}, \emph{117}, 95--121.

\bibitem[{\emph{{Matousek}}(2007)}]{Juno_spacecraft}
{Matousek} S. (2007) \emph{{The Juno New Frontiers mission}}, \emph{Acta
  Astronautica}, \emph{61}, 932--939.

\bibitem[{\emph{{Matteini} et~al.}(2021)\emph{{Matteini}, {Laker}, {Horbury},
  {Woodham}, {Bale}, {Stawarz}, {Woolley}, {Steinvall}, {Jones}, {Grant},
  {Afghan}, {Galand}, {O'Brien}, {Evans}, {Angelini}, {Maksimovic}, {Chust},
  {Khotyaintsev}, {Krasnoselskikh}, {Kretzschmar}, {Lorf{\`e}vre},
  {Plettemeier}, {Sou{\v{c}}ek}, {Steller}, {{\v{S}}tver{\'a}k},
  {Tr{\'a}vn{\'\i}{\v{c}}ek}, {Vaivads}, {Vecchio}, {Wimmer-Schweingruber},
  {Ho}, {G{\'o}mez-Herrero}, {Rodr{\'\i}guez-Pacheco}, {Louarn}, {Fedorov},
  {Owen}, {Bruno}, {Livi}, {Zouganelis}, and {M{\"u}ller}}}]{Matteini2021}
{Matteini} L., {Laker} R., {Horbury} T. et~al. (2021) \emph{{Solar Orbiter's
  encounter with the tail of comet C/2019 Y4 (ATLAS): Magnetic field draping
  and cometary pick-up ion waves}}, \emph{\aap}, \emph{656}, A39.

\bibitem[{\emph{{Mayyasi} et~al.}(2020)\emph{{Mayyasi}, {Clarke}, {Combi},
  {Fougere}, {Quemerais}, {Katushkina}, {Bhattacharyya}, {Crismani}, {Deighan},
  {Jain}, {Schneider}, and {Jakosky}}}]{Mayyasi2020}
{Mayyasi} M., {Clarke} J., {Combi} M. et~al. (2020)
  \emph{{Ly{\ensuremath{\alpha}} Observations of Comet C/2013 A1 (Siding
  Spring) Using MAVEN IUVS Echelle}}, \emph{\aj}, \emph{160}, 10.

\bibitem[{\emph{{McBride} et~al.}(1997)\emph{{McBride}, {Green}, {Chantal
  Levasseur-Regourd}, {Goidet-Devel}, and {Renard}}}]{McBride1997}
{McBride} N., {Green} S.~F., {Chantal Levasseur-Regourd} A. et~al. (1997)
  \emph{{The inner dust coma of Comet 26P/Grigg-Skjellerup: multiple jets and
  nucleus fragments?}}, \emph{\mnras}, \emph{289}, 535--553.

\bibitem[{\emph{{McCubbin} et~al.}(2021)\emph{{McCubbin}, {Allton}, {Barnes},
  {Calaway}, {Corrigan}, {Filiberto}, {Fries}, {Gross}, {Harrington}, {Herd},
  {Hutzler}, {Ishii}, {McCoy}, {McKeegan}, {Mitchell}, {Nittler}, {Regberg},
  {Righter}, {Snead}, {Stroud}, {Tait}, {Yada}, {Zeigler}, {Zolensky}, and
  {Stansbery}}}]{McCubbin2021}
{McCubbin} F., {Allton} J.~H., {Barnes} J.~J. et~al. (2021) in \emph{Bulletin
  of the American Astronomical Society}, vol.~53, p. 021.

\bibitem[{\emph{{McFadden} et~al.}(2005)\emph{{McFadden}, {Rountree-Brown},
  {Warner}, {Claughlin}, {Behne}, {Ristvey}, {Baird-Wilkerson}, {Duncan},
  {Gillam}, {Walker}, and {Meech}}}]{McFadden2005}
{McFadden} L.~A., {Rountree-Brown} M.~K., {Warner} E.~M. et~al. (2005)
  \emph{{Education and Public Outreach for Nasa's Deep Impact Mission}},
  \emph{\ssr}, \emph{117}, 373--396.

\bibitem[{\emph{{McLaughlin} et~al.}(2014)\emph{{McLaughlin}, {Carcich},
  {Sackett}, {McCarthy}, {Desnoyer}, {Klaasen}, and
  {Wellnitz}}}]{McLaughlinPDS}
{McLaughlin} S.~A., {Carcich} B., {Sackett} S.~E. et~al. (2014) \emph{{Deep
  Impact 9P/Tempel Encounter - Reduced HRIV Images V3.0}}, \emph{NASA Planetary
  Data System}, DIF-C-HRIV-3/4-9P-ENCOUNTER-V3.0.

\bibitem[{\emph{{Meech} et~al.}(2005{\natexlab{a}})\emph{{Meech}, {Ageorges},
  {A'Hearn}, {Arpigny}, {Ates}, {Aycock}, {Bagnulo}, {Bailey}, {Barber},
  {Barrera}, {Barrena}, {Bauer}, {Belton}, {Bensch}, {Bhattacharya}, {Biver},
  {Blake}, {Bockel{\'e}e-Morvan}, {Boehnhardt}, {Bonev}, {Bonev}, {Buie},
  {Burton}, {Butner}, {Cabanac}, {Campbell}, {Campins}, {Capria}, {Carroll},
  {Chaffee}, {Charnley}, {Cleis}, {Coates}, {Cochran}, {Colom}, {Conrad},
  {Coulson}, {Crovisier}, {deBuizer}, {Dekany}, {de L{\'e}on}, {Dello Russo},
  {Delsanti}, {DiSanti}, {Drummond}, {Dundon}, {Etzel}, {Farnham}, {Feldman},
  {Fern{\'a}ndez}, {Filipovic}, {Fisher}, {Fitzsimmons}, {Fong}, {Fugate},
  {Fujiwara}, {Fujiyoshi}, {Furusho}, {Fuse}, {Gibb}, {Groussin}, {Gulkis},
  {Gurwell}, {Hadamcik}, {Hainaut}, {Harker}, {Harrington}, {Harwit},
  {Hasegawa}, {Hergenrother}, {Hirst}, {Hodapp}, {Honda}, {Howell},
  {Hutsem{\'e}kers}, {Iono}, {Ip}, {Jackson}, {Jehin}, {Jiang}, {Jones},
  {Jones}, {Kadono}, {Kamath}, {K{\"a}ufl}, {Kasuga}, {Kawakita}, {Kelley},
  {Kerber}, {Kidger}, {Kinoshita}, {Knight}, {Lara}, {Larson}, {Lederer},
  {Lee}, {Levasseur-Regourd}, {Li}, {Li}, {Licandro}, {Lin}, {Lisse},
  {LoCurto}, {Lovell}, {Lowry}, {Lyke}, {Lynch}, {Ma}, {Magee-Sauer},
  {Maheswar}, {Manfroid}, {Marco}, {Martin}, {Melnick}, {Miller}, {Miyata},
  {Moriarty-Schieven}, {Moskovitz}, {Mueller}, {Mumma}, {Muneer}, {Neufeld},
  {Ootsubo}, {Osip}, {Pandea}, {Pantin}, {Paterno-Mahler}, {Patten},
  {Penprase}, {Peck}, {Petitpas}, {Pinilla-Alonso}, {Pittichova}, {Pompei},
  {Prabhu}, {Qi}, {Rao}, {Rauer}, {Reitsema}, {Rodgers}, {Rodriguez}, {Ruane},
  {Ruch}, {Rujopakarn}, {Sahu}, {Sako}, {Sakon}, {Samarasinha}, {Sarkissian},
  {Saviane}, {Schirmer}, {Schultz}, {Schulz}, {Seitzer}, {Sekiguchi}, {Selman},
  {Serra-Ricart}, {Sharp}, {Snell}, {Snodgrass}, {Stallard}, {Stecklein},
  {Sterken}, {St{\"u}we}, {Sugita}, {Sumner}, {Suntzeff}, {Swaters},
  {Takakuwa}, {Takato}, {Thomas-Osip}, {Thompson}, {Tokunaga}, {Tozzi}, {Tran},
  {Troy}, {Trujillo}, {Van Cleve}, {Vasundhara}, {Vazquez}, {Vilas},
  {Villanueva}, {von Braun}, {Vora}, {Wainscoat}, {Walsh}, {Watanabe},
  {Weaver}, {Weaver}, {Weiler}, {Weissman}, {Welsh}, {Wilner}, {Wolk},
  {Womack}, {Wooden}, {Woodney}, {Woodward}, {Wu}, {Wu}, {Yamashita}, {Yang},
  {Yang}, {Yokogawa}, {Zook}, {Zauderer}, {Zhao}, {Zhou}, and
  {Zucconi}}}]{Meech2005}
{Meech} K.~J., {Ageorges} N., {A'Hearn} M.~F. et~al. (2005{\natexlab{a}})
  \emph{{Deep Impact: Observations from a Worldwide Earth-Based Campaign}},
  \emph{Science}, \emph{310}, 265--269.

\bibitem[{\emph{{Meech} et~al.}(2011{\natexlab{a}})\emph{{Meech}, {A'Hearn},
  {Adams}, {Bacci}, {Bai}, {Barrera}, {Battelino}, {Bauer}, {Becklin}, {Bhatt},
  {Biver}, {Bockel{\'e}e-Morvan}, {Bodewits}, {B{\"o}hnhardt}, {Boissier},
  {Bonev}, {Borghini}, {Brucato}, {Bryssinck}, {Buie}, {Canovas}, {Castellano},
  {Charnley}, {Chen}, {Chiang}, {Choi}, {Christian}, {Chuang}, {Cochran},
  {Colom}, {Combi}, {Coulson}, {Crovisier}, {Dello Russo}, {Dennerl}, {DeWahl},
  {DiSanti}, {Facchini}, {Farnham}, {Fern{\'a}ndez}, {Flor{\'e}n}, {Frisk},
  {Fujiyoshi}, {Furusho}, {Fuse}, {Galli}, {Garc{\'\i}a-Hern{\'a}ndez},
  {Gersch}, {Getu}, {Gibb}, {Gillon}, {Guido}, {Guillermo}, {Hadamcik},
  {Hainaut}, {Hammel}, {Harker}, {Harmon}, {Harris}, {Hartogh}, {Hashimoto},
  {H{\"a}usler}, {Herter}, {Hjalmarson}, {Holland}, {Honda}, {Hosseini},
  {Howell}, {Howes}, {Hsieh}, {Hsiao}, {Hutsem{\'e}kers}, {Immler}, {Jackson},
  {Jeffers}, {Jehin}, {Jones}, {de Juan Ovelar}, {Kaluna}, {Karlsson},
  {Kawakita}, {Keane}, {Keller}, {Kelley}, {Kinoshita}, {Kiselev}, {Kleyna},
  {Knight}, {Kobayashi}, {Kobulnicky}, {Kolokolova}, {Kreiny}, {Kuan},
  {K{\"u}ppers}, {Lacruz}, {Landsman}, {Lara}, {Lecacheux},
  {Levasseur-Regourd}, {Li}, {Licandro}, {Ligustri}, {Lin}, {Lippi}, {Lis},
  {Lisse}, {Lovell}, {Lowry}, {Lu}, {Lundin}, {Magee-Sauer}, {Magain},
  {Manfroid}, {Mazzotta Epifani}, {McKay}, {Melita}, {Mikuz}, {Milam},
  {Milani}, {Min}, {Moreno}, {Mueller}, {Mumma}, {Nicolini}, {Nolan}, {Nordh},
  {Nowajewski}, {Odin Team}, {Ootsubo}, {Paganini}, {Perrella},
  {Pittichov{\'a}}, {Prosperi}, {Radeva}, {Reach}, {Remijan}, {Rengel},
  {Riesen}, {Rodenhuis}, {Rodr{\'\i}guez}, {Russell}, {Sahu}, {Samarasinha},
  {S{\'a}nchez Caso}, {Sandqvist}, {Sarid}, {Sato}, {Schleicher},
  {Schwieterman}, {Sen}, {Shenoy}, {Shi}, {Shinnaka}, {Skvarc}, {Snodgrass},
  {Sitko}, {Sonnett}, {Sosseini}, {Sostero}, {Sugita}, {Swinyard}, {Szutowicz},
  {Takato}, {Tanga}, {Taylor}, {Tozzi}, {Trabatti}, {Trigo-Rodr{\'\i}guez},
  {Tubiana}, {de Val-Borro}, {Vacca}, {Vandenbussche}, {Vaubaillion},
  {Velichko}, {Velichko}, {Vervack}, {Vidal-Nunez}, {Villanueva}, {Vinante},
  {Vincent}, {Wang}, {Wasserman}, {Watanabe}, {Weaver}, {Weissman}, {Wolk},
  {Wooden}, {Woodward}, {Yamaguchi}, {Yamashita}, {Yanamandra-Fischer}, {Yang},
  {Yao}, {Yeomans}, {Zenn}, {Zhao}, and {Ziffer}}}]{Meech-EPOXI}
{Meech} K.~J., {A'Hearn} M.~F., {Adams} J.~A. et~al. (2011{\natexlab{a}})
  \emph{{EPOXI: Comet 103P/Hartley 2 Observations from a Worldwide Campaign}},
  \emph{\apjl}, \emph{734}, L1.

\bibitem[{\emph{{Meech} et~al.}(2005{\natexlab{b}})\emph{{Meech}, {A'Hearn},
  {Fern{\'a}ndez}, {Lisse}, {Weaver}, {Biver}, and {Woodney}}}]{Meech_DI_SSR}
{Meech} K.~J., {A'Hearn} M.~F., {Fern{\'a}ndez} Y.~R. et~al.
  (2005{\natexlab{b}}) \emph{{The Deep Impact Earth-Based Campaign}},
  \emph{\ssr}, \emph{117}, 297--334.

\bibitem[{\emph{{Meech} and {Castillo-Rogez}}(2015)}]{Proteus}
{Meech} K.~J. and {Castillo-Rogez} J.~C. (2015) in \emph{IAU General Assembly},
  vol.~29, p. 2257859.

\bibitem[{\emph{{Meech} et~al.}(2013)\emph{{Meech}, {Kleyna}, {Hainaut},
  {Lowry}, {Fuse}, {A'Hearn}, {Chesley}, {Yeomans}, {Fern{\'a}ndez}, {Lisse},
  {Reach}, {Bauer}, {Mainzer}, {Pittichov{\'a}}, {Christensen}, {Osip},
  {Brink}, {Mateo}, {Motta}, {Challis}, {Holman}, and
  {Ferr{\'\i}n}}}]{Meech_Boethin}
{Meech} K.~J., {Kleyna} J., {Hainaut} O.~R. et~al. (2013) \emph{{The demise of
  Comet 85P/Boethin, the first EPOXI mission target}}, \emph{\icarus},
  \emph{222}, 662--678.

\bibitem[{\emph{{Meech} et~al.}(2017)\emph{{Meech}, {Kleyna}, {Hainaut},
  {Micheli}, {Bauer}, {Denneau}, {Keane}, {Stephens}, {Jedicke}, {Wainscoat},
  {Weryk}, {Flewelling}, {Schunov{\'a}-Lilly}, {Magnier}, and
  {Chambers}}}]{MeechK2}
{Meech} K.~J., {Kleyna} J.~T., {Hainaut} O. et~al. (2017) \emph{{CO-driven
  Activity in Comet C/2017 K2 (PANSTARRS)}}, \emph{\apjl}, \emph{849}, L8.

\bibitem[{\emph{{Meech} et~al.}(2011{\natexlab{b}})\emph{{Meech},
  {Pittichov{\'a}}, {Yang}, {Zenn}, {Belton}, {A'Hearn}, {Bagnulo}, {Bai},
  {Barrera}, {Bauer}, {Bedient}, {Bhatt}, {Boehnhardt}, {Brosch}, {Buie},
  {Candia}, {Chen}, {Chesley}, {Chiang}, {Choi}, {Cochran}, {Duddy}, {Farnham},
  {Fern{\'a}ndez}, {Guti{\'e}rrez}, {Hainaut}, {Hampton}, {Herrmann}, {Hsieh},
  {Kadooka}, {Kaluna}, {Keane}, {Kim}, {Kleyna}, {Krisciunas}, {Lauer}, {Lara},
  {Licandro}, {Lowry}, {McFadden}, {Moskovitz}, {Mueller}, {Polishook}, {Raja},
  {Riesen}, {Sahu}, {Samarasinha}, {Sarid}, {Sekiguchi}, {Sonnett}, {Suntzeff},
  {Taylor}, {Tozzi}, {Vasundhara}, {Vincent}, {Wasserman}, {Webster-Schultz},
  and {Zhao}}}]{Meech2011}
{Meech} K.~J., {Pittichov{\'a}} J., {Yang} B. et~al. (2011{\natexlab{b}})
  \emph{{Deep Impact, Stardust-NExT and the behavior of Comet 9P/Tempel 1 from
  1997 to 2010}}, \emph{\icarus}, \emph{213}, 323--344.

\bibitem[{\emph{{Meech} and {Svoren}}(2004)}]{Meech-cometsII}
{Meech} K.~J. and {Svoren} J. (2004) \emph{{Using cometary activity to trace
  the physical and chemical evolution of cometary nuclei}}, p. 317.

\bibitem[{\emph{{Merouane} et~al.}(2016)\emph{{Merouane}, {Zaprudin},
  {Stenzel}, {Langevin}, {Altobelli}, {Della Corte}, {Fischer}, {Fulle},
  {Hornung}, {Sil{\'e}n}, {Ligier}, {Rotundi}, {Ryno}, {Schulz}, {Hilchenbach},
  {Kissel}, and {Cosima Team}}}]{Merouane2016}
{Merouane} S., {Zaprudin} B., {Stenzel} O. et~al. (2016) \emph{{Dust particle
  flux and size distribution in the coma of 67P/Churyumov-Gerasimenko measured
  in situ by the COSIMA instrument on board Rosetta}}, \emph{\aap}, \emph{596},
  A87.

\bibitem[{\emph{{M{\"o}hlmann} et~al.}(2018)\emph{{M{\"o}hlmann},
  {Seidensticker}, {Fischer}, {Faber}, {Flandes}, {Knapmeyer}, {Kr{\"u}ger},
  {Roll}, {Scholten}, {Thiel}, and {Arnold}}}]{Moehlmann2018}
{M{\"o}hlmann} D., {Seidensticker} K.~J., {Fischer} H.-H. et~al. (2018)
  \emph{{Compressive strength and elastic modulus at Agilkia on comet
  67P/Churyumov-Gerasimenko derived from the SESAME/CASSE touchdown signals}},
  \emph{\icarus}, \emph{303}, 251--264.

\bibitem[{\emph{Moore et~al.}(2021)\emph{Moore, Castillo-Rogez, Meech,
  Courville, Donitz, Ferguson, Llera, and French}}]{Moore_WP_rapid}
Moore K., Castillo-Rogez J., Meech K.~J. et~al. (2021) \emph{Rapid response
  missions to explore fast, high-value targets such as interstellar objects and
  long period comets}, \emph{Bulletin of the AAS}, \emph{53},
  https://baas.aas.org/pub/2021n4i481.

\bibitem[{\emph{{Moore} et~al.}(2021)\emph{{Moore}, {Courville}, {Ferguson},
  {Schoenfeld}, {Llera}, {Agrawal}, {Brack}, {Buhler}, {Connour}, {Czaplinski},
  {DeLuca}, {Deutsch}, {Hammond}, {Kuettel}, {Marusiak}, {Nerozzi}, {Stuart},
  {Tarnas}, {Thelen}, {Castillo-Rogez}, {Smythe}, {Landau}, {Mitchell}, and
  {Budney}}}]{Moore2021}
{Moore} K., {Courville} S., {Ferguson} S. et~al. (2021) \emph{{Bridge to the
  stars: A mission concept to an interstellar object}}, \emph{\planss},
  \emph{197}, 105137.

\bibitem[{\emph{{Mottola} et~al.}(2007)\emph{{Mottola}, {Arnold}, {Grothues},
  {Jaumann}, {Michaelis}, {Neukum}, and {Bibring}}}]{RosIns-ROLIS}
{Mottola} S., {Arnold} G., {Grothues} H.-G. et~al. (2007) \emph{{The Rolis
  Experiment on the Rosetta Lander}}, \emph{\ssr}, \emph{128}, 241--255.

\bibitem[{\emph{{Mottola} et~al.}(2014)\emph{{Mottola}, {Lowry}, {Snodgrass},
  {Lamy}, {Toth}, {Ro{\.z}ek}, {Sierks}, {A'Hearn}, {Angrilli}, {Barbieri},
  {Barucci}, {Bertaux}, {Cremonese}, {Da Deppo}, {Davidsson}, {De Cecco},
  {Debei}, {Fornasier}, {Fulle}, {Groussin}, {Guti{\'e}rrez}, {Hviid}, {Ip},
  {Jorda}, {Keller}, {Knollenberg}, {Koschny}, {Kramm}, {K{\"u}hrt},
  {K{\"u}ppers}, {Lara}, {Lazzarin}, {Lopez Moreno}, {Marzari}, {Michalik},
  {Naletto}, {Rickman}, {Rodrigo}, {Sabau}, {Thomas}, {Wenzel}, {Agarwal},
  {Bertini}, {Ferri}, {G{\"u}ttler}, {Magrin}, {Oklay}, {Tubiana}, and
  {Vincent}}}]{Mottola67Pspinup}
{Mottola} S., {Lowry} S., {Snodgrass} C. et~al. (2014) \emph{{The rotation
  state of 67P/Churyumov-Gerasimenko from approach observations with the OSIRIS
  cameras on Rosetta}}, \emph{\aap}, \emph{569}, L2.

\bibitem[{\emph{Nakamura-Messenger et~al.}(2021)\emph{Nakamura-Messenger,
  Hayes, Sandford, Raymond, Squyres, Nittler, Birch, Bodewits, Chabot, Wadhwa,
  Choukroun, Clemett, Bose, Dello~Russo, Dworkin, Elsila, Fisher, Gerakines,
  Glavin, Mitchell, Mumma, Nguyen, Pace, Soderblom, and
  Sunshine}}]{noncryo_CSSR}
Nakamura-Messenger K., Hayes A.~G., Sandford S. et~al. (2021) \emph{The case
  for non-cryogenic comet nucleus sample return}, \emph{Bulletin of the AAS},
  \emph{53}, https://baas.aas.org/pub/2021n4i168.

\bibitem[{\emph{NASA}(2003)}]{CONTOUR-report}
NASA (2003) \emph{{Comet Nucleus Tour Mishap Investigation Board Report}},
  \emph{technical report}.

\bibitem[{\emph{{Neugebauer} et~al.}(2007)\emph{{Neugebauer}, {Gloeckler},
  {Gosling}, {Rees}, {Skoug}, {Goldstein}, {Armstrong}, {Combi}, {M{\"a}kinen},
  {McComas}, {von Steiger}, {Zurbuchen}, {Smith}, {Geiss}, and
  {Lanzerotti}}}]{neugebauer07}
{Neugebauer} M., {Gloeckler} G., {Gosling} J.~T. et~al. (2007) \emph{{Encounter
  of the Ulysses Spacecraft with the Ion Tail of Comet MCNaught}}, \emph{\apj},
  \emph{667}, 1262--1266.

\bibitem[{\emph{{Newburn} et~al.}(2003)\emph{{Newburn}, {Bhaskaran}, {Duxbury},
  {Fraschetti}, {Radey}, and {Schwochert}}}]{Newburn2003}
{Newburn} R.~L., {Bhaskaran} S., {Duxbury} T.~C. et~al. (2003) \emph{{Stardust
  Imaging Camera}}, \emph{Journal of Geophysical Research (Planets)},
  \emph{108}, 8116.

\bibitem[{\emph{{Olkin} et~al.}(2021)\emph{{Olkin}, {Levison}, {Vincent},
  {Noll}, {Andrews}, {Gray}, {Good}, {Marchi}, {Christensen}, {Reuter},
  {Weaver}, {P{\"a}tzold}, {Bell}, {Hamilton}, {Dello Russo}, {Simon},
  {Beasley}, {Grundy}, {Howett}, {Spencer}, {Ravine}, and
  {Caplinger}}}]{Lucy_spacecraft}
{Olkin} C.~B., {Levison} H.~F., {Vincent} M. et~al. (2021) \emph{{Lucy Mission
  to the Trojan Asteroids: Instrumentation and Encounter Concept of
  Operations}}, \emph{\psj}, \emph{2}, 172.

\bibitem[{\emph{{Opitom} et~al.}(2017)\emph{{Opitom}, {Snodgrass},
  {Fitzsimmons}, {Jehin}, {Manfroid}, {Tozzi}, {Faggi}, and
  {Gillon}}}]{Opitom67Pspec}
{Opitom} C., {Snodgrass} C., {Fitzsimmons} A. et~al. (2017) \emph{{Ground-based
  monitoring of comet 67P/Churyumov-Gerasimenko gas activity throughout the
  Rosetta mission}}, \emph{\mnras}, \emph{469}, S222--S229.

\bibitem[{\emph{{O'Rourke} et~al.}(2019)\emph{{O'Rourke}, {Tubiana},
  {G{\"u}ttler}, {Lodiot}, {Mu{\~n}oz}, {Herique}, {Rogez}, {Durand},
  {Charpentier}, {Sierks}, {Gutierrez-Marques}, {Deller}, {Grieger}, {Andres},
  {Geiger}, {Geurts}, {Ulamec}, {K{\"o}mle}, {Lommatsch}, {Maibaum}, {Pellon},
  {Bielsa}, {Garmier}, {Taylor}, {Martin}, {K{\"u}ppers}, {Accomazzo},
  {Companys}, {Bibring}, {Kofman}, {Mckenna Lawlor}, {Salatti}, and
  {Gaudon}}}]{2019ORourke}
{O'Rourke} L., {Tubiana} C., {G{\"u}ttler} C. et~al. (2019) \emph{{The search
  campaign to identify and image the Philae Lander on the surface of comet
  67P/Churyumov-Gerasimenko}}, \emph{Acta Astronautica}, \emph{157}, 199--214.

\bibitem[{\emph{{Ozaki} et~al.}(2022)\emph{{Ozaki}, {Yamamoto},
  {Gonzalez-Franquesa}, {Gutierrez-Ramon}, {Pushparaj}, {Chikazawa}, {Dei Tos},
  {{\c{C}}elik}, {Marmo}, {Kawakatsu}, {Arai}, {Nishiyama}, and
  {Takashima}}}]{Ozaki2022}
{Ozaki} N., {Yamamoto} T., {Gonzalez-Franquesa} F. et~al. (2022) \emph{{Mission
  Design of DESTINY+: Toward Active Asteroid (3200) Phaethon and Multiple Small
  Bodies}}, \emph{Acta Astronautica}, \emph{in press}, arXiv:2201.01933.

\bibitem[{\emph{{Ozaki} et~al.}(2021)\emph{{Ozaki}, {Yoshioka}, {Kameda},
  {Murakami}, {Kobayashi}, {Kawakita}, {Shinnaka}, {Funase}, {Dei Tos},
  {Fujimoto}, {Kasahara}, {Hyodo}, {Bonardi}, {Takao}, and
  {Sugahara}}}]{Halley2061}
{Ozaki} N., {Yoshioka} K., {Kameda} S. et~al. (2021) in \emph{43rd COSPAR
  Scientific Assembly. Held 28 January - 4 February}, vol.~43, p. 254.

\bibitem[{\emph{{Pajola} et~al.}(2017)\emph{{Pajola}, {H{\"o}fner}, {Vincent},
  {Oklay}, {Scholten}, {Preusker}, {Mottola}, {Naletto}, {Fornasier}, {Lowry},
  {Feller}, {Hasselmann}, {G{\"u}ttler}, {Tubiana}, {Sierks}, {Barbieri},
  {Lamy}, {Rodrigo}, {Koschny}, {Rickman}, {Keller}, {Agarwal}, {A'Hearn},
  {Barucci}, {Bertaux}, {Bertini}, {Besse}, {Boudreault}, {Cremonese}, {da
  Deppo}, {Davidsson}, {Debei}, {de Cecco}, {Deller}, {Deshapriya},
  {El-Maarry}, {Ferrari}, {Ferri}, {Fulle}, {Groussin}, {Gutierrez}, {Hofmann},
  {Hviid}, {Ip}, {Jorda}, {Knollenberg}, {Kovacs}, {Kramm}, {K{\"u}hrt},
  {K{\"u}ppers}, {Lara}, {Lin}, {Lazzarin}, {Lucchetti}, {Lopez Moreno},
  {Marzari}, {Massironi}, {Michalik}, {Penasa}, {Pommerol}, {Simioni},
  {Thomas}, {Toth}, and {Baratti}}}]{Pajola2017}
{Pajola} M., {H{\"o}fner} S., {Vincent} J.~B. et~al. (2017) \emph{{The pristine
  interior of comet 67P revealed by the combined Aswan outburst and cliff
  collapse}}, \emph{Nature Astronomy}, \emph{1}, 0092.

\bibitem[{\emph{{P{\"a}tzold} et~al.}(2016)\emph{{P{\"a}tzold}, {Andert},
  {Hahn}, {Asmar}, {Barriot}, {Bird}, {H{\"a}usler}, {Peter}, {Tellmann},
  {Gr{\"u}n}, {Weissman}, {Sierks}, {Jorda}, {Gaskell}, {Preusker}, and
  {Scholten}}}]{Paetzold2016}
{P{\"a}tzold} M., {Andert} T., {Hahn} M. et~al. (2016) \emph{{A homogeneous
  nucleus for comet 67P/Churyumov-Gerasimenko from its gravity field}},
  \emph{\nat}, \emph{530}, 63--65.

\bibitem[{\emph{{P{\"a}tzold} et~al.}(2007)\emph{{P{\"a}tzold}, {H{\"a}usler},
  {Aksnes}, {Anderson}, {Asmar}, {Barriot}, {Bird}, {Boehnhardt}, {Eidel},
  {Gr{\"u}n}, {Ip}, {Marouf}, {Morley}, {Neubauer}, {Rickman}, {Thomas},
  {Tsurutani}, {Wallis}, {Wickramasinghe}, {Mysen}, {Olson}, {Remus},
  {Tellmann}, {Andert}, {Carone}, {Fels}, {Stanzel}, {Audenrieth-Kersten},
  {Gahr}, {M{\"u}ller}, {Stupar}, and {Walter}}}]{RosIns-RSI}
{P{\"a}tzold} M., {H{\"a}usler} B., {Aksnes} K. et~al. (2007) \emph{{Rosetta
  Radio Science Investigations (RSI)}}, \emph{\ssr}, \emph{128}, 599--627.

\bibitem[{\emph{{Preusker} et~al.}(2017)\emph{{Preusker}, {Scholten}, {Matz},
  {Roatsch}, {Hviid}, {Mottola}, {Knollenberg}, {K{\"u}hrt}, {Pajola}, {Oklay},
  {Vincent}, {Davidsson}, {A'Hearn}, {Agarwal}, {Barbieri}, {Barucci},
  {Bertaux}, {Bertini}, {Cremonese}, {Da Deppo}, {Debei}, {De Cecco},
  {Fornasier}, {Fulle}, {Groussin}, {Guti{\'e}rrez}, {G{\"u}ttler}, {Ip},
  {Jorda}, {Keller}, {Koschny}, {Kramm}, {K{\"u}ppers}, {Lamy}, {Lara},
  {Lazzarin}, {Lopez Moreno}, {Marzari}, {Massironi}, {Naletto}, {Rickman},
  {Rodrigo}, {Sierks}, {Thomas}, and {Tubiana}}}]{Preusker2017}
{Preusker} F., {Scholten} F., {Matz} K.~D. et~al. (2017) \emph{{The global
  meter-level shape model of comet 67P/Churyumov-Gerasimenko}}, \emph{\aap},
  \emph{607}, L1.

\bibitem[{\emph{{Protopapa} et~al.}(2014)\emph{{Protopapa}, {Sunshine},
  {Feaga}, {Kelley}, {A'Hearn}, {Farnham}, {Groussin}, {Besse}, {Merlin}, and
  {Li}}}]{Protopapa_H2_ice}
{Protopapa} S., {Sunshine} J.~M., {Feaga} L.~M. et~al. (2014) \emph{{Water ice
  and dust in the innermost coma of comet 103P/Hartley 2}}, \emph{\icarus},
  \emph{238}, 191--204.

\bibitem[{\emph{{Rayman}}(2002)}]{Rayman2002}
{Rayman} M.~D. (2002) \emph{{The Deep Space 1 extended mission: Challenges in
  preparing for an encounter with comet Borrelly}}, \emph{Acta Astronautica},
  \emph{51}, 507--516.

\bibitem[{\emph{{Rayman} and {Varghese}}(2001)}]{Rayman2001}
{Rayman} M.~D. and {Varghese} P. (2001) \emph{{The deep space 1 extended
  mission}}, \emph{Acta Astronautica}, \emph{48}, 693--705.

\bibitem[{\emph{{Rayman} et~al.}(2000)\emph{{Rayman}, {Varghese}, {Lehman}, and
  {Livesay}}}]{Rayman2000}
{Rayman} M.~D., {Varghese} P., {Lehman} D.~H. et~al. (2000) \emph{{Results from
  the Deep Space 1 Technology Validation Mission}}, \emph{Acta Astronautica},
  \emph{47}, 475--487.

\bibitem[{\emph{{Rengel} et~al.}(2007)\emph{{Rengel}, {Jones}, {K{\"u}ppers},
  {Keller}, and {Owens}}}]{Rengel2007-OSIRIS-comets}
{Rengel} M., {Jones} G.~H., {K{\"u}ppers} M. et~al. (2007) in \emph{Geophysical
  Research Abstracts}, vol.~9, p. 02744.

\bibitem[{\emph{{Richardson} et~al.}(2005)\emph{{Richardson}, {Melosh},
  {Artemeiva}, and {Pierazzo}}}]{Richardson_DI_SSR}
{Richardson} J.~E., {Melosh} H.~J., {Artemeiva} N.~A. et~al. (2005)
  \emph{{Impact Cratering Theory and Modeling for the Deep Impact Mission: From
  Mission Planning to Data Analysis}}, \emph{\ssr}, \emph{117}, 241--267.

\bibitem[{\emph{{Richardson} et~al.}(2007)\emph{{Richardson}, {Melosh},
  {Lisse}, and {Carcich}}}]{Richardson_DI_ejecta}
{Richardson} J.~E., {Melosh} H.~J., {Lisse} C.~M. et~al. (2007) \emph{{A
  ballistics analysis of the Deep Impact ejecta plume: Determining Comet Tempel
  1's gravity, mass, and density}}, \emph{\icarus}, \emph{191}, 176--209.

\bibitem[{\emph{{Riedler} et~al.}(2007)\emph{{Riedler}, {Torkar}, {Jeszenszky},
  {Romstedt}, {Alleyne}, {Arends}, {Barth}, {Biezen}, {Butler}, {Ehrenfreund},
  {Fehringer}, {Fremuth}, {Gavira}, {Havnes}, {Jessberger}, {Kassing},
  {Kl{\"o}ck}, {Koeberl}, {Levasseur-Regourd}, {Maurette}, {R{\"u}denauer},
  {Schmidt}, {Stangl}, {Steller}, and {Weber}}}]{RosIns-MIDAS}
{Riedler} W., {Torkar} K., {Jeszenszky} H. et~al. (2007) \emph{{MIDAS The
  Micro-Imaging Dust Analysis System for the Rosetta Mission}}, \emph{\ssr},
  \emph{128}, 869--904.

\bibitem[{\emph{{Robinson} et~al.}(2011)\emph{{Robinson}, {Meadows}, {Crisp},
  {Deming}, {A'Hearn}, {Charbonneau}, {Livengood}, {Seager}, {Barry}, {Hearty},
  {Hewagama}, {Lisse}, {McFadden}, and {Wellnitz}}}]{EPOCh_Robinson}
{Robinson} T.~D., {Meadows} V.~S., {Crisp} D. et~al. (2011) \emph{{Earth as an
  Extrasolar Planet: Earth Model Validation Using EPOXI Earth Observations}},
  \emph{Astrobiology}, \emph{11}, 393--408.

\bibitem[{\emph{{Rosenbush} et~al.}(2017)\emph{{Rosenbush}, {Ivanova},
  {Kiselev}, {Kolokolova}, and {Afanasiev}}}]{Rosenbush2017}
{Rosenbush} V.~K., {Ivanova} O.~V., {Kiselev} N.~N. et~al. (2017)
  \emph{{Spatial variations of brightness, colour and polarization of dust in
  comet 67P/Churyumov-Gerasimenko}}, \emph{\mnras}, \emph{469}, S475--S491.

\bibitem[{\emph{{Rubin} et~al.}(2018)\emph{{Rubin}, {Altwegg}, {Balsiger},
  {Bar-Nun}, {Berthelier}, {Briois}, {Calmonte}, {Combi}, {De Keyser},
  {Fiethe}, {Fuselier}, {Gasc}, {Gombosi}, {Hansen}, {Kopp}, {Korth}, {Laufer},
  {Le Roy}, {Mall}, {Marty}, {Mousis}, {Owen}, {R{\`e}me}, {S{\'e}mon}, {Tzou},
  {Waite}, and {Wurz}}}]{Rubin2018}
{Rubin} M., {Altwegg} K., {Balsiger} H. et~al. (2018) \emph{{Krypton isotopes
  and noble gas abundances in the coma of comet 67P/Churyumov-Gerasimenko}},
  \emph{Science Advances}, \emph{4}, eaar6297.

\bibitem[{\emph{{Rubin} et~al.}(2015)\emph{{Rubin}, {Altwegg}, {van Dishoeck},
  and {Schwehm}}}]{Rubin2015}
{Rubin} M., {Altwegg} K., {van Dishoeck} E.~F. et~al. (2015) \emph{{Molecular
  Oxygen in Oort Cloud Comet 1P/Halley}}, \emph{\apjl}, \emph{815}, L11.

\bibitem[{\emph{{Rubin} et~al.}(2020)\emph{{Rubin}, {Engrand}, {Snodgrass},
  {Weissman}, {Altwegg}, {Busemann}, {Morbidelli}, and
  {Mumma}}}]{Rubin2020-SSR}
{Rubin} M., {Engrand} C., {Snodgrass} C. et~al. (2020) \emph{{On the Origin and
  Evolution of the Material in 67P/Churyumov-Gerasimenko}}, \emph{\ssr},
  \emph{216}, 102.

\bibitem[{\emph{{Russell} et~al.}(1984)\emph{{Russell}, {Arghavani}, and
  {Luhmann}}}]{russell1984}
{Russell} C.~T., {Arghavani} M.~R., and {Luhmann} J.~G. (1984)
  \emph{{Interplanetary field enhancements in the solar wind: Statistical
  properties at 0.72 AU}}, \emph{\icarus}, \emph{60}, 332--350.

\bibitem[{\emph{{Russell} et~al.}(1983)\emph{{Russell}, {Luhmann}, {Barnes},
  {Mihalov}, and {Elphic}}}]{russell1983}
{Russell} C.~T., {Luhmann} J.~G., {Barnes} A. et~al. (1983) \emph{{An unusual
  interplanetary event - Encounter with a comet?}}, \emph{\nat}, \emph{305},
  612--615.

\bibitem[{\emph{{S{\'a}nchez} et~al.}(2021)\emph{{S{\'a}nchez}, {Morante},
  {Hermosin}, {Ranuschio}, {Estalella}, {Viera}, {Centuori}, {Jones},
  {Snodgrass}, {Levasseur-Regourd}, and {Tubiana}}}]{Pau-CI}
{S{\'a}nchez} J.~P., {Morante} D., {Hermosin} P. et~al. (2021) \emph{{ESA
  F-Class Comet Interceptor: Trajectory design to intercept a
  yet-to-be-discovered comet}}, \emph{Acta Astronautica}, \emph{188}, 265--277.

\bibitem[{\emph{{S{\'a}nchez-Cano} et~al.}(2020)\emph{{S{\'a}nchez-Cano},
  {Lester}, {Witasse}, {Morgan}, {Opgenoorth}, {Andrews}, {Blelly}, {Cowley},
  {Kopf}, {Leblanc}, {Espley}, and {Cardes{\'\i}n-Moinelo}}}]{SanchezCano2020}
{S{\'a}nchez-Cano} B., {Lester} M., {Witasse} O. et~al. (2020) \emph{{Mars'
  Ionospheric Interaction With Comet C/2013 A1 Siding Spring's Coma at Their
  Closest Approach as Seen by Mars Express}}, \emph{Journal of Geophysical
  Research (Space Physics)}, \emph{125}, e27344.

\bibitem[{\emph{{S{\'a}nchez-Cano} et~al.}(2018)\emph{{S{\'a}nchez-Cano},
  {Witasse}, {Lester}, {Rahmati}, {Ambrosi}, {Lillis}, {Leblanc}, {Blelly},
  {Costa}, {Cowley}, {Espley}, {Milan}, {Plaut}, {Lee}, and
  {Larson}}}]{SanchezCano2018}
{S{\'a}nchez-Cano} B., {Witasse} O., {Lester} M. et~al. (2018) \emph{{Energetic
  Particle Showers Over Mars from Comet C/2013 A1 Siding Spring}},
  \emph{Journal of Geophysical Research (Space Physics)}, \emph{123},
  8778--8796.

\bibitem[{\emph{{Sandford} et~al.}(2006)\emph{{Sandford}, {Al{\'e}on},
  {Alexander}, {Araki}, {Bajt}, {Baratta}, {Borg}, {Bradley}, {Brownlee},
  {Brucato}, {Burchell}, {Busemann}, {Butterworth}, {Clemett}, {Cody},
  {Colangeli}, {Cooper}, {D'Hendecourt}, {Djouadi}, {Dworkin}, {Ferrini},
  {Fleckenstein}, {Flynn}, {Franchi}, {Fries}, {Gilles}, {Glavin}, {Gounelle},
  {Grossemy}, {Jacobsen}, {Keller}, {Kilcoyne}, {Leitner}, {Matrajt}, {Meibom},
  {Mennella}, {Mostefaoui}, {Nittler}, {Palumbo}, {Papanastassiou}, {Robert},
  {Rotundi}, {Snead}, {Spencer}, {Stadermann}, {Steele}, {Stephan}, {Tsou},
  {Tyliszczak}, {Westphal}, {Wirick}, {Wopenka}, {Yabuta}, {Zare}, and
  {Zolensky}}}]{Sandfordetal2006}
{Sandford} S.~A., {Al{\'e}on} J., {Alexander} C. M.~O.~D. et~al. (2006)
  \emph{{Organics Captured from Comet 81P/Wild 2 by the Stardust Spacecraft}},
  \emph{Science}, \emph{314}, 1720.

\bibitem[{\emph{{Sandford} et~al.}(2021)\emph{{Sandford}, {Brownlee}, and
  {Zolensky}}}]{stardust-review}
{Sandford} S.~A., {Brownlee} D.~E., and {Zolensky} M.~E. (2021) in \emph{Sample
  Return Missions: The Last Frontier of Solar System Exploration. Editor:
  Andrea Longobardo. Elsevier} (A.~{Longobardo}, ed.), pp. 79--104.

\bibitem[{\emph{{Sarid} et~al.}(2019)\emph{{Sarid}, {Volk}, {Steckloff},
  {Harris}, {Womack}, and {Woodney}}}]{29P}
{Sarid} G., {Volk} K., {Steckloff} J.~K. et~al. (2019)
  \emph{{29P/Schwassmann-Wachmann 1, A Centaur in the Gateway to the
  Jupiter-family Comets}}, \emph{\apjl}, \emph{883}, L25.

\bibitem[{\emph{{Sarli} et~al.}(2018)\emph{{Sarli}, {Horikawa}, {Yam},
  {Kawakatsu}, and {Yamamoto}}}]{Sarli2018}
{Sarli} B.~V., {Horikawa} M., {Yam} C.~H. et~al. (2018) \emph{{DESTINY+
  Trajectory Design to (3200) Phaethon}}, \emph{Journal of the Astronautical
  Sciences}, \emph{65}, 82--110.

\bibitem[{\emph{{Schleicher} et~al.}(2006)\emph{{Schleicher}, {Barnes}, and
  {Baugh}}}]{schleicher2006}
{Schleicher} D.~G., {Barnes} K.~L., and {Baugh} N.~F. (2006) \emph{{Photometry
  and Imaging Results for Comet 9P/Tempel 1 and Deep Impact: Gas Production
  Rates, Postimpact Light Curves, and Ejecta Plume Morphology}}, \emph{\aj},
  \emph{131}, 1130--1137.

\bibitem[{\emph{{Schleicher} and {Farnham}}(2004)}]{Schleicher+Farnham2004}
{Schleicher} D.~G. and {Farnham} T.~L. (2004) \emph{{Photometry and imaging of
  the coma with narrowband filters}}, p. 449.

\bibitem[{\emph{{Schultz} et~al.}(2007)\emph{{Schultz}, {Eberhardy}, {Ernst},
  {A'Hearn}, {Sunshine}, and {Lisse}}}]{Schultz_DI_ejecta}
{Schultz} P.~H., {Eberhardy} C.~A., {Ernst} C.~M. et~al. (2007) \emph{{The Deep
  Impact oblique impact cratering experiment}}, \emph{\icarus}, \emph{191},
  84--122.

\bibitem[{\emph{{Schultz} et~al.}(2005)\emph{{Schultz}, {Ernst}, and
  {Anderson}}}]{Schultz_DI_SSR}
{Schultz} P.~H., {Ernst} C.~M., and {Anderson} J. L.~B. (2005)
  \emph{{Expectations for Crater Size and Photometric Evolution from the Deep
  Impact Collision}}, \emph{\ssr}, \emph{117}, 207--239.

\bibitem[{\emph{{Schultz} et~al.}(2013)\emph{{Schultz}, {Hermalyn}, and
  {Veverka}}}]{Schultz2013}
{Schultz} P.~H., {Hermalyn} B., and {Veverka} J. (2013) \emph{{The Deep Impact
  crater on 9P/Tempel-1 from Stardust-NExT}}, \emph{\icarus}, \emph{222},
  502--515.

\bibitem[{\emph{{Schwamb} et~al.}(2020)\emph{{Schwamb}, {Knight}, {Jones},
  {Snodgrass}, {Bucci}, {S{\'a}nchez P{\'e}rez}, {Skuppin}, and {Comet
  Interceptor Science Team}}}]{Schwamb-CI}
{Schwamb} M.~E., {Knight} M.~M., {Jones} G.~H. et~al. (2020) \emph{{Potential
  Backup Targets for Comet Interceptor}}, \emph{Research Notes of the American
  Astronomical Society}, \emph{4}, 21.

\bibitem[{\emph{{Seidensticker} et~al.}(2007)\emph{{Seidensticker},
  {M{\"o}hlmann}, {Apathy}, {Schmidt}, {Thiel}, {Arnold}, {Fischer},
  {Kretschmer}, {Madlener}, {P{\'e}ter}, {Trautner}, and
  {Schieke}}}]{RosIns-SESAME}
{Seidensticker} K.~J., {M{\"o}hlmann} D., {Apathy} I. et~al. (2007)
  \emph{{Sesame - An Experiment of the Rosetta Lander Philae: Objectives and
  General Design}}, \emph{\ssr}, \emph{128}, 301--337.

\bibitem[{\emph{{Seligman} and {Laughlin}}(2018)}]{Seligman2018}
{Seligman} D. and {Laughlin} G. (2018) \emph{{The Feasibility and Benefits of
  In Situ Exploration of {\textquoteleft}Oumuamua-like Objects}}, \emph{\aj},
  \emph{155}, 217.

\bibitem[{\emph{{Shinnaka} et~al.}(2017)\emph{{Shinnaka}, {Fougere},
  {Kawakita}, {Kameda}, {Combi}, {Ikezawa}, {Seki}, {Kuwabara}, {Sato},
  {Taguchi}, and {Yoshikawa}}}]{LAICA}
{Shinnaka} Y., {Fougere} N., {Kawakita} H. et~al. (2017) \emph{{Imaging
  Observations of the Hydrogen Coma of Comet 67P/Churyumov-Gerasimenko in 2015
  September by the PROCYON/LAICA}}, \emph{\aj}, \emph{153}, 76.

\bibitem[{\emph{{Singer} et~al.}(2021)\emph{{Singer}, {Stern}, {Elliott},
  {Karimi}, {Stern}, {Chimelewski}, {Fong}, {Andrews}, {Bottke}, {Olkin},
  {Propster}, and {Thurman}}}]{centaurus}
{Singer} K.~N., {Stern} S.~A., {Elliott} J. et~al. (2021) \emph{{A new
  spacecraft mission concept combining the first exploration of the Centaurs
  and an astrophysical space telescope for the outer solar system}},
  \emph{\planss}, \emph{205}, 105290.

\bibitem[{\emph{{Smith} et~al.}(1986)\emph{{Smith}, {Slavin}, {Bame},
  {Thomsen}, {Cowley}, {Richardson}, {Hovestadt}, {Ipavich}, {Ogilive}, and
  {Coplan}}}]{smith1986}
{Smith} E.~J., {Slavin} J.~A., {Bame} S.~J. et~al. (1986) in \emph{ESLAB
  Symposium on the Exploration of Halley's Comet} (B.~{Battrick}, E.~J.
  {Rolfe}, and R.~{Reinhard}, eds.), vol. 250 of \emph{ESA Special
  Publication}.

\bibitem[{\emph{{Snodgrass} et~al.}(2017{\natexlab{a}})\emph{{Snodgrass},
  {Agarwal}, {Combi}, {Fitzsimmons}, {Guilbert-Lepoutre}, {Hsieh}, {Hui},
  {Jehin}, {Kelley}, {Knight}, {Opitom}, {Orosei}, {de Val-Borro}, and
  {Yang}}}]{Snodgrass-MBCs}
{Snodgrass} C., {Agarwal} J., {Combi} M. et~al. (2017{\natexlab{a}}) \emph{{The
  Main Belt Comets and ice in the Solar System}}, \emph{\aapr}, \emph{25}, 5.

\bibitem[{\emph{{Snodgrass} et~al.}(2017{\natexlab{b}})\emph{{Snodgrass},
  {A'Hearn}, {Aceituno}, {Afanasiev}, {Bagnulo}, {Bauer}, {Bergond}, {Besse},
  {Biver}, {Bodewits}, {Boehnhardt}, {Bonev}, {Borisov}, {Carry}, {Casanova},
  {Cochran}, {Conn}, {Davidsson}, {Davies}, {de Le{\'o}n}, {de Mooij}, {de
  Val-Borro}, {Delacruz}, {DiSanti}, {Drew}, {Duffard}, {Edberg}, {Faggi},
  {Feaga}, {Fitzsimmons}, {Fujiwara}, {Gibb}, {Gillon}, {Green}, {Guijarro},
  {Guilbert-Lepoutre}, {Guti{\'e}rrez}, {Hadamcik}, {Hainaut}, {Haque},
  {Hedrosa}, {Hines}, {Hopp}, {Hoyo}, {Hutsem{\'e}kers}, {Hyland}, {Ivanova},
  {Jehin}, {Jones}, {Keane}, {Kelley}, {Kiselev}, {Kleyna}, {Kluge}, {Knight},
  {Kokotanekova}, {Koschny}, {Kramer}, {L{\'o}pez-Moreno}, {Lacerda}, {Lara},
  {Lasue}, {Lehto}, {Levasseur-Regourd}, {Licandro}, {Lin}, {Lister}, {Lowry},
  {Mainzer}, {Manfroid}, {Marchant}, {McKay}, {McNeill}, {Meech}, {Micheli},
  {Mohammed}, {Mongui{\'o}}, {Moreno}, {Mu{\~n}oz}, {Mumma}, {Nikolov},
  {Opitom}, {Ortiz}, {Paganini}, {Pajuelo}, {Pozuelos}, {Protopapa}, {Pursimo},
  {Rajkumar}, {Ramanjooloo}, {Ramos}, {Ries}, {Riffeser}, {Rosenbush},
  {Rousselot}, {Ryan}, {Santos-Sanz}, {Schleicher}, {Schmidt}, {Schulz}, {Sen},
  {Somero}, {Sota}, {Stinson}, {Sunshine}, {Thompson}, {Tozzi}, {Tubiana},
  {Villanueva}, {Wang}, {Wooden}, {Yagi}, {Yang}, {Zaprudin}, and
  {Zegmott}}}]{Rosetta-campaign}
{Snodgrass} C., {A'Hearn} M.~F., {Aceituno} F. et~al. (2017{\natexlab{b}})
  \emph{{The 67P/Churyumov-Gerasimenko observation campaign in support of the
  Rosetta mission}}, \emph{Philosophical Transactions of the Royal Society of
  London Series A}, \emph{375}, 20160249.

\bibitem[{\emph{{Snodgrass} and {Jones}}(2019)}]{Snodgrass+Jones-CI}
{Snodgrass} C. and {Jones} G.~H. (2019) \emph{{The European Space Agency's
  Comet Interceptor lies in wait}}, \emph{Nature Communications}, \emph{10},
  5418.

\bibitem[{\emph{{Snodgrass} et~al.}(2018)\emph{{Snodgrass}, {Jones},
  {Boehnhardt}, {Gibbings}, {Homeister}, {Andre}, {Beck}, {Bentley}, {Bertini},
  {Bowles}, {Capria}, {Carr}, {Ceriotti}, {Coates}, {Della Corte}, {Donaldson
  Hanna}, {Fitzsimmons}, {Guti{\'e}rrez}, {Hainaut}, {Herique}, {Hilchenbach},
  {Hsieh}, {Jehin}, {Karatekin}, {Kofman}, {Lara}, {Laudan}, {Licandro},
  {Lowry}, {Marzari}, {Masters}, {Meech}, {Moreno}, {Morse}, {Orosei}, {Pack},
  {Plettemeier}, {Prialnik}, {Rotundi}, {Rubin}, {S{\'a}nchez}, {Sheridan},
  {Trieloff}, and {Winterboer}}}]{Castalia}
{Snodgrass} C., {Jones} G.~H., {Boehnhardt} H. et~al. (2018) \emph{{The
  Castalia mission to Main Belt Comet 133P/Elst-Pizarro}}, \emph{Advances in
  Space Research}, \emph{62}, 1947--1976.

\bibitem[{\emph{{Snodgrass} et~al.}(2013)\emph{{Snodgrass}, {Tubiana},
  {Bramich}, {Meech}, {Boehnhardt}, and {Barrera}}}]{Snodgrass2013}
{Snodgrass} C., {Tubiana} C., {Bramich} D.~M. et~al. (2013) \emph{{Beginning of
  activity in 67P/Churyumov-Gerasimenko and predictions for 2014-2015}},
  \emph{\aap}, \emph{557}, A33.

\bibitem[{\emph{{Snodgrass} et~al.}(2010)\emph{{Snodgrass}, {Tubiana},
  {Vincent}, {Sierks}, {Hviid}, {Moissi}, {Boehnhardt}, {Barbieri}, {Koschny},
  {Lamy}, {Rickman}, {Rodrigo}, {Carry}, {Lowry}, {Laird}, {Weissman},
  {Fitzsimmons}, {Marchi}, and {OSIRIS Team}}}]{2010a2}
{Snodgrass} C., {Tubiana} C., {Vincent} J.-B. et~al. (2010) \emph{{A collision
  in 2009 as the origin of the debris trail of asteroid P/2010A2}},
  \emph{\nat}, \emph{467}, 814--816.

\bibitem[{\emph{{Spohn} et~al.}(2015)\emph{{Spohn}, {Knollenberg}, {Ball},
  {Banaszkiewicz}, {Benkhoff}, {Grott}, {Grygorczuk}, {H{\"u}ttig},
  {Hagermann}, {Kargl}, {Kaufmann}, {K{\"o}mle}, {K{\"u}hrt}, {Kossacki},
  {Marczewski}, {Pelivan}, {Schr{\"o}dter}, and {Seiferlin}}}]{Spohn2015}
{Spohn} T., {Knollenberg} J., {Ball} A.~J. et~al. (2015) \emph{{Thermal and
  mechanical properties of the near-surface layers of comet
  67P/Churyumov-Gerasimenko}}, \emph{Science}, \emph{349}, 2.464.

\bibitem[{\emph{{Spohn} et~al.}(2007)\emph{{Spohn}, {Seiferlin}, {Hagermann},
  {Knollenberg}, {Ball}, {Banaszkiewicz}, {Benkhoff}, {Gadomski}, {Gregorczyk},
  {Grygorczuk}, {Hlond}, {Kargl}, {K{\"u}hrt}, {K{\"o}mle}, {Krasowski},
  {Marczewski}, and {Zarnecki}}}]{RosIns-MUPUS}
{Spohn} T., {Seiferlin} K., {Hagermann} A. et~al. (2007) \emph{{Mupus A Thermal
  and Mechanical Properties Probe for the Rosetta Lander Philae}}, \emph{\ssr},
  \emph{128}, 339--362.

\bibitem[{\emph{{Squyres} et~al.}(2018)\emph{{Squyres}, {Nakamura-Messenger},
  {Mitchell}, {Moran}, {Houghton}, {Glavin}, {Hayes}, {Lauretta}, and {Caesar
  Project Team}}}]{CAESAR}
{Squyres} S.~W., {Nakamura-Messenger} K., {Mitchell} D.~F. et~al. (2018) in
  \emph{Lunar and Planetary Science Conference}, Lunar and Planetary Science
  Conference, p. 1332.

\bibitem[{\emph{{Srama} et~al.}(2004)\emph{{Srama}, {Ahrens}, {Altobelli},
  {Auer}, {Bradley}, {Burton}, {Dikarev}, {Economou}, {Fechtig}, {G{\"o}rlich},
  {Grande}, {Graps}, {Gr{\"u}n}, {Havnes}, {Helfert}, {Horanyi}, {Igenbergs},
  {Jessberger}, {Johnson}, {Kempf}, {Krivov}, {Kr{\"u}ger}, {Mocker-Ahlreep},
  {Moragas-Klostermeyer}, {Lamy}, {Landgraf}, {Linkert}, {Linkert}, {Lura},
  {McDonnell}, {M{\"o}hlmann}, {Morfill}, {M{\"u}ller}, {Roy}, {Sch{\"a}fer},
  {Schlotzhauer}, {Schwehm}, {Spahn}, {St{\"u}big}, {Svestka}, {Tschernjawski},
  {Tuzzolino}, {W{\"a}sch}, and {Zook}}}]{CDA}
{Srama} R., {Ahrens} T.~J., {Altobelli} N. et~al. (2004) \emph{{The Cassini
  Cosmic Dust Analyzer}}, \emph{\ssr}, \emph{114}, 465--518.

\bibitem[{\emph{{Stern} et~al.}(2007)\emph{{Stern}, {Slater}, {Scherrer},
  {Stone}, {Versteeg}, {A'Hearn}, {Bertaux}, {Feldman}, {Festou}, {Parker}, and
  {Siegmund}}}]{RosIns-Alice}
{Stern} S.~A., {Slater} D.~C., {Scherrer} J. et~al. (2007) \emph{{Alice: The
  rosetta Ultraviolet Imaging Spectrograph}}, \emph{\ssr}, \emph{128},
  507--527.

\bibitem[{\emph{{Sunshine} et~al.}(2006)\emph{{Sunshine}, {A'Hearn},
  {Groussin}, {Li}, {Belton}, {Delamere}, {Kissel}, {Klaasen}, {McFadden},
  {Meech}, {Melosh}, {Schultz}, {Thomas}, {Veverka}, {Yeomans}, {Busko},
  {Desnoyer}, {Farnham}, {Feaga}, {Hampton}, {Lindler}, {Lisse}, and
  {Wellnitz}}}]{Sunshine_DI_surfaceice}
{Sunshine} J.~M., {A'Hearn} M.~F., {Groussin} O. et~al. (2006) \emph{{Exposed
  Water Ice Deposits on the Surface of Comet 9P/Tempel 1}}, \emph{Science},
  \emph{311}, 1453--1455.

\bibitem[{\emph{{Sunshine} et~al.}(2009)\emph{{Sunshine}, {Farnham}, {Feaga},
  {Groussin}, {Merlin}, {Milliken}, and {A'Hearn}}}]{Sunshine_Moon}
{Sunshine} J.~M., {Farnham} T.~L., {Feaga} L.~M. et~al. (2009) \emph{{Temporal
  and Spatial Variability of Lunar Hydration As Observed by the Deep Impact
  Spacecraft}}, \emph{Science}, \emph{326}, 565.

\bibitem[{\emph{{Sunshine} and {Feaga}}(2021)}]{Sunshine_Feaga_hyper}
{Sunshine} J.~M. and {Feaga} L.~M. (2021) \emph{{All Comets are Somewhat
  Hyperactive and the Implications Thereof}}, \emph{\psj}, \emph{2}, 92.

\bibitem[{\emph{{Sunshine} et~al.}(2007)\emph{{Sunshine}, {Groussin},
  {Schultz}, {A'Hearn}, {Feaga}, {Farnham}, and
  {Klaasen}}}]{Sunshine_DI_ejectaice}
{Sunshine} J.~M., {Groussin} O., {Schultz} P.~H. et~al. (2007) \emph{{The
  distribution of water ice in the interior of Comet Tempel 1}},
  \emph{\icarus}, \emph{191}, 73--83.

\bibitem[{\emph{{Sunshine} et~al.}(2016)\emph{{Sunshine}, {Thomas},
  {El-Maarry}, and {Farnham}}}]{Sunshine_comet_processes}
{Sunshine} J.~M., {Thomas} N., {El-Maarry} M.~R. et~al. (2016) \emph{{Evidence
  for geologic processes on comets}}, \emph{Journal of Geophysical Research
  (Planets)}, \emph{121}, 2194--2210.

\bibitem[{\emph{{Swenson} et~al.}(1987)\emph{{Swenson}, {Squyres}, and
  {Knight}}}]{Swenson1987}
{Swenson} B.~L., {Squyres} S.~W., and {Knight} T.~C.~D. (1987) \emph{{A
  Proposed Comet Nucleus Penetrator for the Comet Rendezvous Asteroid Flyby
  Mission}}, \emph{Acta Astronautica}, \emph{15}, 471--479.

\bibitem[{\emph{{Tang} et~al.}(2021)\emph{{Tang}, {Xiao}, {Jiang}, {Zhang},
  {Chi}, {Su}, and {Hu}}}]{ZhengHe-penetrator}
{Tang} J., {Xiao} J., {Jiang} S. et~al. (2021) in \emph{43rd COSPAR Scientific
  Assembly. Held 28 January - 4 February}, vol.~43, p. 153.

\bibitem[{\emph{{Taylor} et~al.}(2017)\emph{{Taylor}, {Altobelli}, {Buratti},
  and {Choukroun}}}]{Taylor-Rosetta}
{Taylor} M.~G.~G.~T., {Altobelli} N., {Buratti} B.~J. et~al. (2017) \emph{{The
  Rosetta mission orbiter science overview: the comet phase}},
  \emph{Philosophical Transactions of the Royal Society of London Series A},
  \emph{375}, 20160262.

\bibitem[{\emph{{Thomas} et~al.}(2017)\emph{{Thomas}, {Cremonese}, {Ziethe},
  {Gerber}, {Br{\"a}ndli}, {Bruno}, {Erismann}, {Gambicorti}, {Gerber},
  {Ghose}, {Gruber}, {Gubler}, {Mischler}, {Jost}, {Piazza}, {Pommerol},
  {Rieder}, {Roloff}, {Servonet}, {Trottmann}, {Uthaicharoenpong},
  {Zimmermann}, {Vernani}, {Johnson}, {Pel{\`o}}, {Weigel}, {Viertl}, {De
  Roux}, {Lochmatter}, {Sutter}, {Casciello}, {Hausner}, {Ficai Veltroni}, {Da
  Deppo}, {Orleanski}, {Nowosielski}, {Zawistowski}, {Szalai}, {Sodor},
  {Tulyakov}, {Troznai}, {Banaskiewicz}, {Bridges}, {Byrne}, {Debei},
  {El-Maarry}, {Hauber}, {Hansen}, {Ivanov}, {Keszthelyi}, {Kirk}, {Kuzmin},
  {Mangold}, {Marinangeli}, {Markiewicz}, {Massironi}, {McEwen}, {Okubo},
  {Tornabene}, {Wajer}, and {Wray}}}]{CaSSIS}
{Thomas} N., {Cremonese} G., {Ziethe} R. et~al. (2017) \emph{{The Colour and
  Stereo Surface Imaging System (CaSSIS) for the ExoMars Trace Gas Orbiter}},
  \emph{\ssr}, \emph{212}, 1897--1944.

\bibitem[{\emph{{Thomas} et~al.}(2015{\natexlab{a}})\emph{{Thomas},
  {Davidsson}, {El-Maarry}, {Fornasier}, {Giacomini}, {Gracia-Bern{\'a}},
  {Hviid}, {Ip}, {Jorda}, {Keller}, {Knollenberg}, {K{\"u}hrt}, {La Forgia},
  {Lai}, {Liao}, {Marschall}, {Massironi}, {Mottola}, {Pajola}, {Poch},
  {Pommerol}, {Preusker}, {Scholten}, {Su}, {Wu}, {Vincent}, {Sierks},
  {Barbieri}, {Lamy}, {Rodrigo}, {Koschny}, {Rickman}, {A'Hearn}, {Barucci},
  {Bertaux}, {Bertini}, {Cremonese}, {Da Deppo}, {Debei}, {de Cecco}, {Fulle},
  {Groussin}, {Gutierrez}, {Kramm}, {K{\"u}ppers}, {Lara}, {Lazzarin}, {Lopez
  Moreno}, {Marzari}, {Michalik}, {Naletto}, {Agarwal}, {G{\"u}ttler}, {Oklay},
  and {Tubiana}}}]{Thomas2015-fallback}
{Thomas} N., {Davidsson} B., {El-Maarry} M.~R. et~al. (2015{\natexlab{a}})
  \emph{{Redistribution of particles across the nucleus of comet
  67P/Churyumov-Gerasimenko}}, \emph{\aap}, \emph{583}, A17.

\bibitem[{\emph{{Thomas} et~al.}(2015{\natexlab{b}})\emph{{Thomas}, {Sierks},
  {Barbieri}, {Lamy}, {Rodrigo}, {Rickman}, {Koschny}, {Keller}, {Agarwal},
  {A'Hearn}, {Angrilli}, {Auger}, {Barucci}, {Bertaux}, {Bertini}, {Besse},
  {Bodewits}, {Cremonese}, {Da Deppo}, {Davidsson}, {De Cecco}, {Debei},
  {El-Maarry}, {Ferri}, {Fornasier}, {Fulle}, {Giacomini}, {Groussin},
  {Gutierrez}, {G{\"u}ttler}, {Hviid}, {Ip}, {Jorda}, {Knollenberg}, {Kramm},
  {K{\"u}hrt}, {K{\"u}ppers}, {La Forgia}, {Lara}, {Lazzarin}, {Moreno},
  {Magrin}, {Marchi}, {Marzari}, {Massironi}, {Michalik}, {Moissl}, {Mottola},
  {Naletto}, {Oklay}, {Pajola}, {Pommerol}, {Preusker}, {Sabau}, {Scholten},
  {Snodgrass}, {Tubiana}, {Vincent}, and {Wenzel}}}]{Thomas2015}
{Thomas} N., {Sierks} H., {Barbieri} C. et~al. (2015{\natexlab{b}}) \emph{{The
  morphological diversity of comet 67P/Churyumov-Gerasimenko}}, \emph{Science},
  \emph{347}, aaa0440.

\bibitem[{\emph{{Thomas} et~al.}(2019)\emph{{Thomas}, {Ulamec}, {K{\"u}hrt},
  {Ciarletti}, {Gundlach}, {Yoldi}, {Schwehm}, {Snodgrass}, and
  {Green}}}]{Thomas2019SSRv}
{Thomas} N., {Ulamec} S., {K{\"u}hrt} E. et~al. (2019) \emph{{Towards New Comet
  Missions}}, \emph{\ssr}, \emph{215}, 47.

\bibitem[{\emph{{Thomas} et~al.}(2013{\natexlab{a}})\emph{{Thomas}, {A'Hearn},
  {Belton}, {Brownlee}, {Carcich}, {Hermalyn}, {Klaasen}, {Sackett}, {Schultz},
  {Veverka}, {Bhaskaran}, {Bodewits}, {Chesley}, {Clark}, {Farnham},
  {Groussin}, {Harris}, {Kissel}, {Li}, {Meech}, {Melosh}, {Quick},
  {Richardson}, {Sunshine}, and {Wellnitz}}}]{Thomas2013}
{Thomas} P., {A'Hearn} M., {Belton} M.~J.~S. et~al. (2013{\natexlab{a}})
  \emph{{The nucleus of Comet 9P/Tempel 1: Shape and geology from two flybys}},
  \emph{\icarus}, \emph{222}, 453--466.

\bibitem[{\emph{{Thomas} et~al.}(2013{\natexlab{b}})\emph{{Thomas}, {A'Hearn},
  {Veverka}, {Belton}, {Kissel}, {Klaasen}, {McFadden}, {Melosh}, {Schultz},
  {Besse}, {Carcich}, {Farnham}, {Groussin}, {Hermalyn}, {Li}, {Lindler},
  {Lisse}, {Meech}, and {Richardson}}}]{Thomas2013b_H2}
{Thomas} P.~C., {A'Hearn} M.~F., {Veverka} J. et~al. (2013{\natexlab{b}})
  \emph{{Shape, density, and geology of the nucleus of Comet 103P/Hartley 2}},
  \emph{\icarus}, \emph{222}, 550--558.

\bibitem[{\emph{{Thomas} et~al.}(2007)\emph{{Thomas}, {Veverka}, {Belton},
  {Hidy}, {A'Hearn}, {Farnham}, {Groussin}, {Li}, {McFadden}, {Sunshine},
  {Wellnitz}, {Lisse}, {Schultz}, {Meech}, and {Delamere}}}]{Thomas_DI_shape}
{Thomas} P.~C., {Veverka} J., {Belton} M. J.~S. et~al. (2007) \emph{{The shape,
  topography, and geology of Tempel 1 from Deep Impact observations}},
  \emph{\icarus}, \emph{191}, 51--62.

\bibitem[{\emph{{Thomas} et~al.}(2011)\emph{{Thomas}, {Makowski}, {Brown},
  {McCarthy}, {Bruno}, {Cardoso}, {Chiville}, {Meyer}, {Nelson}, {Pavri},
  {Termohlen}, {Violet}, and {Williams}}}]{Dawn}
{Thomas} V.~C., {Makowski} J.~M., {Brown} G.~M. et~al. (2011) \emph{{The Dawn
  Spacecraft}}, \emph{\ssr}, \emph{163}, 175--249.

\bibitem[{\emph{{Tran} et~al.}(1997)\emph{{Tran}, {Johnson}, {Rasky}, {Hui},
  {Hsu}, {Chen}, {Chen}, {Paragas}, and {Kobayashi}}}]{Stardust_heatshield}
{Tran} H.~K., {Johnson} C.~E., {Rasky} D.~J. et~al. (1997) \emph{{Phenolic
  Impregnated Carbon Ablators (PICA) as Thermal Protection Systems for
  Discovery Missions}}, \emph{NASA Technical Reports, NASA-TM-110440}.

\bibitem[{\emph{{Tsou} et~al.}(2004)\emph{{Tsou}, {Brownlee}, {Anderson},
  {Bhaskaran}, {Cheuvront}, {Clark}, {Duxbury}, {Economou}, {Green}, {Hanner},
  {H{\"o}rz}, {Kissel}, {McDonnell}, {Newburn}, {Ryan}, {Sandford}, {Sekanina},
  {Tuzzolino}, {Vellinga}, and {Zolensky}}}]{Tsouetal2004}
{Tsou} P., {Brownlee} D.~E., {Anderson} J.~D. et~al. (2004) \emph{{Stardust
  encounters comet 81P/Wild 2}}, \emph{Journal of Geophysical Research
  (Planets)}, \emph{109}, E12S01.

\bibitem[{\emph{{Tubiana} et~al.}(2015)\emph{{Tubiana}, {Snodgrass}, {Bertini},
  {Mottola}, {Vincent}, {Lara}, {Fornasier}, {Knollenberg}, {Thomas}, {Fulle},
  {Agarwal}, {Bodewits}, {Ferri}, {G{\"u}ttler}, {Gutierrez}, {La Forgia},
  {Lowry}, {Magrin}, {Oklay}, {Pajola}, {Rodrigo}, {Sierks}, {A'Hearn},
  {Angrilli}, {Barbieri}, {Barucci}, {Bertaux}, {Cremonese}, {Da Deppo},
  {Davidsson}, {De Cecco}, {Debei}, {Groussin}, {Hviid}, {Ip}, {Jorda},
  {Keller}, {Koschny}, {Kramm}, {K{\"u}hrt}, {K{\"u}ppers}, {Lazzarin}, {Lamy},
  {Lopez Moreno}, {Marzari}, {Michalik}, {Naletto}, {Rickman}, {Sabau}, and
  {Wenzel}}}]{Tubiana-67Papproach}
{Tubiana} C., {Snodgrass} C., {Bertini} I. et~al. (2015)
  \emph{{67P/Churyumov-Gerasimenko: Activity between March and June 2014 as
  observed from Rosetta/OSIRIS}}, \emph{\aap}, \emph{573}, A62.

\bibitem[{\emph{{Tuzzolino} et~al.}(2003)\emph{{Tuzzolino}, {Economou},
  {McKibben}, {Simpson}, {McDonnell}, {Burchell}, {Vaughan}, {Tsou}, {Hanner},
  {Clark}, and {Brownlee}}}]{Tuzzolino2003}
{Tuzzolino} A.~J., {Economou} T.~E., {McKibben} R.~B. et~al. (2003) \emph{{Dust
  Flux Monitor Instrument for the Stardust mission to comet Wild 2}},
  \emph{Journal of Geophysical Research (Planets)}, \emph{108}, 8115.

\bibitem[{\emph{{Ulamec} et~al.}(2015)\emph{{Ulamec}, {Biele}, {Blazquez},
  {Cozzoni}, {Delmas}, {Fantinati}, {Gaudon}, {Geurts}, {Jurado},
  {K{\"u}chemann}, {Lommatsch}, {Maibaum}, {Sierks}, and {Witte}}}]{Ulamec2015}
{Ulamec} S., {Biele} J., {Blazquez} A. et~al. (2015) \emph{{Rosetta Lander -
  Philae: Landing preparations}}, \emph{Acta Astronautica}, \emph{107}, 79--86.

\bibitem[{\emph{{Ulamec} et~al.}(2006)\emph{{Ulamec}, {Espinasse},
  {Feuerbacher}, {Hilchenbach}, {Moura}, {Rosenbauer}, {Scheuerle}, and
  {Willnecker}}}]{Ulamec2006}
{Ulamec} S., {Espinasse} S., {Feuerbacher} B. et~al. (2006) \emph{{Rosetta
  Lander{\textemdash}Philae: Implications of an alternative mission}},
  \emph{Acta Astronautica}, \emph{58}, 435--441.

\bibitem[{\emph{{Ulamec} et~al.}(2011)\emph{{Ulamec}, {Kucherenko}, {Biele},
  {Bogatchev}, {Makurin}, and {Matrossov}}}]{Ulamec2011}
{Ulamec} S., {Kucherenko} V., {Biele} J. et~al. (2011) \emph{{Hopper concepts
  for small body landers}}, \emph{Advances in Space Research}, \emph{47},
  428--439.

\bibitem[{\emph{{Usher} et~al.}(2020)\emph{{Usher}, {Snodgrass}, {Green},
  {Norton}, and {Roche}}}]{Usher2020}
{Usher} H., {Snodgrass} C., {Green} S.~F. et~al. (2020) \emph{{Seeing the
  Bigger Picture: Rosetta Mission Amateur Observing Campaign and Lessons for
  the Future}}, \emph{\psj}, \emph{1}, 84.

\bibitem[{\emph{{Vallat} et~al.}(2017)\emph{{Vallat}, {Altobelli}, {Geiger},
  {Grieger}, {Kueppers}, {Mu{\~n}oz Crego}, {Moissl}, {Taylor}, {Alexander},
  {Buratti}, {Choukroun}, and {the RSGS Liaison scientists
  Group}}}]{Vallat2017}
{Vallat} C., {Altobelli} N., {Geiger} B. et~al. (2017) \emph{{The science
  planning process on the Rosetta mission}}, \emph{Acta Astronautica},
  \emph{133}, 244--257.

\bibitem[{\emph{{Veverka} et~al.}(2013)\emph{{Veverka}, {Klaasen}, {A'Hearn},
  {Belton}, {Brownlee}, {Chesley}, {Clark}, {Economou}, {Farquhar}, {Green},
  {Groussin}, {Harris}, {Kissel}, {Li}, {Meech}, {Melosh}, {Richardson},
  {Schultz}, {Silen}, {Sunshine}, {Thomas}, {Bhaskaran}, {Bodewits}, {Carcich},
  {Cheuvront}, {Farnham}, {Sackett}, {Wellnitz}, and {Wolf}}}]{Veverka2013}
{Veverka} J., {Klaasen} K., {A'Hearn} M. et~al. (2013) \emph{{Return to Comet
  Tempel 1: Overview of Stardust-NExT results}}, \emph{\icarus}, \emph{222},
  424--435.

\bibitem[{\emph{{Veverka} et~al.}(2011)\emph{{Veverka}, {Klaasen}, {Kissel},
  {Silen}, {Carcich}, and {Sackett}}}]{VeverkaPDS}
{Veverka} J.~F., {Klaasen} K.~P., {Kissel} J. et~al. (2011) \emph{{Stardust RDR
  NAVCAM Images of 9P/Tempel 1, V1.0}}, \emph{NASA Planetary Data System},
  SDU-C/CAL-NAVCAM-3-NEXT-TEMPEL1-V1.0.

\bibitem[{\emph{{Vilnrotter} et~al.}(2008)\emph{{Vilnrotter}, {Tsao}, {Lee},
  {Cornish}, {Paal}, and {Jamnejad}}}]{DI_tech_demo}
{Vilnrotter} V., {Tsao} P.~C., {Lee} D.~K. et~al. (2008) \emph{{EPOXI Uplink
  Array Experiment of June 27, 2008}}, \emph{Interplanetary Network Progress
  Report}, \emph{42-174}, 1--25.

\bibitem[{\emph{{Vincent} et~al.}(2016)\emph{{Vincent}, {A'Hearn}, {Lin},
  {El-Maarry}, {Pajola}, {Sierks}, {Barbieri}, {Lamy}, {Rodrigo}, {Koschny},
  {Rickman}, {Keller}, {Agarwal}, {Barucci}, {Bertaux}, {Bertini}, {Besse},
  {Bodewits}, {Cremonese}, {Da Deppo}, {Davidsson}, {Debei}, {De Cecco},
  {Deller}, {Fornasier}, {Fulle}, {Gicquel}, {Groussin}, {Guti{\'e}rrez},
  {Guti{\'e}rrez-Marquez}, {G{\"u}ttler}, {H{\"o}fner}, {Hofmann}, {Hviid},
  {Ip}, {Jorda}, {Knollenberg}, {Kovacs}, {Kramm}, {K{\"u}hrt}, {K{\"u}ppers},
  {Lara}, {Lazzarin}, {Lopez Moreno}, {Marzari}, {Massironi}, {Mottola},
  {Naletto}, {Oklay}, {Preusker}, {Scholten}, {Shi}, {Thomas}, {Toth}, and
  {Tubiana}}}]{JB-summer-fireworks}
{Vincent} J.~B., {A'Hearn} M.~F., {Lin} Z.~Y. et~al. (2016) \emph{{Summer
  fireworks on comet 67P}}, \emph{\mnras}, \emph{462}, S184--S194.

\bibitem[{\emph{{Vincent} et~al.}(2013)\emph{{Vincent}, {Lara}, {Tozzi}, {Lin},
  and {Sierks}}}]{67Ppole3}
{Vincent} J.~B., {Lara} L.~M., {Tozzi} G.~P. et~al. (2013) \emph{{Spin and
  activity of comet 67P/Churyumov-Gerasimenko}}, \emph{\aap}, \emph{549}, A121.

\bibitem[{\emph{{Vincent} et~al.}(2015)\emph{{Vincent}, {Oklay}, {Marchi},
  {H{\"o}fner}, and {Sierks}}}]{Vincent2015}
{Vincent} J.-B., {Oklay} N., {Marchi} S. et~al. (2015) \emph{{Craters on
  comets}}, \emph{\planss}, \emph{107}, 53--63.

\bibitem[{\emph{{Wellnitz} et~al.}(2013)\emph{{Wellnitz}, {Collins}, {A'Hearn},
  {Deep Impact Mission Team}, and {Stardust-NExT Mission
  Team}}}]{Wellnitz_DI_impact}
{Wellnitz} D.~D., {Collins} S.~M., {A'Hearn} M.~F. et~al. (2013) \emph{{The
  location of the impact point of the Deep Impact Impactor on Comet 9P/Tempel
  1}}, \emph{\icarus}, \emph{222}, 487--491.

\bibitem[{\emph{Westphal et~al.}(2021)\emph{Westphal, Nittler, Stroud,
  Zolensky, Chabot, Dello~Russo, Elsila, Sandford, Glavin, Evans, Nuth,
  Sunshine, Vervack~Jr., and Weaver}}]{Westphal_cryo}
Westphal A., Nittler L.~R., Stroud R. et~al. (2021) \emph{Cryogenic comet
  sample return}, \emph{Bulletin of the AAS}, \emph{53},
  https://baas.aas.org/pub/2021n4i014.

\bibitem[{\emph{{Wright} et~al.}(2007)\emph{{Wright}, {Barber}, {Morgan},
  {Morse}, {Sheridan}, {Andrews}, {Maynard}, {Yau}, {Evans}, {Leese},
  {Zarnecki}, {Kent}, {Waltham}, {Whalley}, {Heys}, {Drummond}, {Edeson},
  {Sawyer}, {Turner}, and {Pillinger}}}]{RosIns-Ptolemy}
{Wright} I.~P., {Barber} S.~J., {Morgan} G.~H. et~al. (2007) \emph{{Ptolemy an
  Instrument to Measure Stable Isotopic Ratios of Key Volatiles on a Cometary
  Nucleus}}, \emph{\ssr}, \emph{128}, 363--381.

\bibitem[{\emph{{Young} et~al.}(2004)\emph{{Young}, {Crary}, {Nordholt},
  {Bagenal}, {Boice}, {Burch}, {Eviatar}, {Goldstein}, {Hanley}, {Lawrence},
  {McComas}, {Meier}, {Reisenfeld}, {Sauer}, and {Wiens}}}]{Young2004}
{Young} D.~T., {Crary} F.~J., {Nordholt} J.~E. et~al. (2004) \emph{{Solar wind
  interactions with Comet 19P/Borrelly}}, \emph{\icarus}, \emph{167}, 80--88.

\bibitem[{\emph{{Zacny} et~al.}(2021)\emph{{Zacny}, {Chu}, {Vendiola}, {Seto},
  {Quinn}, {Eichenbaum}, {J.}, {Kleinhenz}, {Colaprete}, and
  {Elphic}}}]{Zacny2021}
{Zacny} K., {Chu} P., {Vendiola} V. et~al. (2021) in \emph{52nd Lunar and
  Planetary Science Conference}, Lunar and Planetary Science Conference, p.
  2400.

\bibitem[{\emph{{Zhang} et~al.}(2021)\emph{{Zhang}, {Xu}, and
  {Ding}}}]{ZhengHe2}
{Zhang} T., {Xu} K., and {Ding} X. (2021) \emph{{China's ambitions and
  challenges for asteroid-comet exploration}}, \emph{Nature Astronomy},
  \emph{5}, 730--731.

\bibitem[{\emph{{Zhang} et~al.}(2019)\emph{{Zhang}, {Huang}, {Wang}, and
  {Huo}}}]{ZhengHe1}
{Zhang} X., {Huang} J., {Wang} T. et~al. (2019) in \emph{Lunar and Planetary
  Science Conference}, Lunar and Planetary Science Conference, p. 1045.

\bibitem[{\emph{Zhao et~al.}(2022)\emph{Zhao, Wang, Li, Li, Wang, Liu, Li,
  Chen, and Zhuang}}]{ZHAO2022}
Zhao C., Wang Y., Li D. et~al. (2022) \emph{A design of dust analyzer for
  future main belt comet exploration mission}, \emph{Advances in Space
  Research}.

\bibitem[{\emph{{Zolensky} et~al.}(2000)\emph{{Zolensky}, {Pieters}, {Clark},
  and {Papike}}}]{Zolenskyetal2000}
{Zolensky} M.~E., {Pieters} C., {Clark} B. et~al. (2000) \emph{{Invited Review
  Small is beautiful: The analysis of nanogram-sized astromaterials}},
  \emph{\maps}, \emph{35}, 9--29.

\bibitem[{\emph{{Zolensky} et~al.}(2006)\emph{{Zolensky}, {Zega}, {Yano},
  {Wirick}, {Westphal}, {Weisberg}, {Weber}, {Warren}, {Velbel}, {Tsuchiyama},
  {Tsou}, {Toppani}, {Tomioka}, {Tomeoka}, {Teslich}, {Taheri}, {Susini},
  {Stroud}, {Stephan}, {Stadermann}, {Snead}, {Simon}, {Simionovici}, {See},
  {Robert}, {Rietmeijer}, {Rao}, {Perronnet}, {Papanastassiou}, {Okudaira},
  {Ohsumi}, {Ohnishi}, {Nakamura-Messenger}, {Nakamura}, {Mostefaoui},
  {Mikouchi}, {Meibom}, {Matrajt}, {Marcus}, {Leroux}, {Lemelle}, {Le},
  {Lanzirotti}, {Langenhorst}, {Krot}, {Keller}, {Kearsley}, {Joswiak},
  {Jacob}, {Ishii}, {Harvey}, {Hagiya}, {Grossman}, {Grossman}, {Graham},
  {Gounelle}, {Gillet}, {Genge}, {Flynn}, {Ferroir}, {Fallon}, {Ebel}, {Dai},
  {Cordier}, {Clark}, {Chi}, {Butterworth}, {Brownlee}, {Bridges}, {Brennan},
  {Brearley}, {Bradley}, {Bleuet}, {Bland}, and {Bastien}}}]{Zolenskyetal2006}
{Zolensky} M.~E., {Zega} T.~J., {Yano} H. et~al. (2006) \emph{{Mineralogy and
  Petrology of Comet 81P/Wild 2 Nucleus Samples}}, \emph{Science}, \emph{314},
  1735.

\end{thebibliography}

\end{document}